\pdfoutput=1

\documentclass[3p]{elsarticle}

\usepackage{amsmath}
\usepackage{amssymb}
\usepackage{mathrsfs}
\usepackage{graphicx}
\usepackage{color}
\usepackage{subfigure}
\usepackage[hidelinks]{hyperref}
\usepackage[capitalise]{cleveref}

\definecolor {darkred}{rgb}{0.6,0,0}
\definecolor {orangered}{rgb}{0.85,0.325,0.098}
\definecolor {orange}{rgb}{1,0.5,0}
\definecolor {darkgreen}{rgb}{0,0.6,0.2}
\definecolor {lightgreen}{rgb}{0.5,0.8,0.35}
\definecolor {lightblue}{rgb}{0.686,0.776,0.914}
\definecolor {mediumblue}{rgb}{0,0.5,0.9}
\definecolor {purple}{rgb}{0.5,0,1}
\definecolor {darkpurple}{rgb}{0.251,0,0.502}
\definecolor {colModelD}{rgb}{0,0.549,0.278}
\definecolor {colModelJ}{rgb}{0.85,0.325,0.098}
\definecolor {lightgray}{rgb}{0.5,0.5,0.5}

\newcommand {\pa}[2]{\frac{\partial{#1}}{\partial{#2}}}

\newcommand {\tr}{\mathrm{tr}}

\newcommand {\fc}{\mathbf{f}_{\mathrm{c}}}
\newcommand {\fck}{\mathbf{f}_{\mathrm{c},k}}
\newcommand {\fn}{\mathbf{f}_{\mathrm{n}}}
\newcommand {\fnk}{\mathbf{f}_{\mathrm{n},k}}
\newcommand {\ft}{\mathbf{f}_{\mathrm{t}}}
\newcommand {\ftk}{\mathbf{f}_{\mathrm{t},k}}

\newcommand {\kc}{\mathbf{k}_{\mathrm{c}}}
\newcommand {\kckk}{\mathbf{k}_{\mathrm{c},kk}}
\newcommand {\kckl}{\mathbf{k}_{\mathrm{c},k\ell}}
\newcommand {\knkk}{\mathbf{k}_{\mathrm{n},kk}}
\newcommand {\knkl}{\mathbf{k}_{\mathrm{n},k\ell}}
\newcommand {\ktkk}{\mathbf{k}_{\mathrm{t},kk}}
\newcommand {\ktkl}{\mathbf{k}_{\mathrm{t},k\ell}}

\newcommand {\Nk}{\mathbf{N}_k}

\newcommand{\adh}{{\mathrm{adh}}}

\newcommand{\extn}{{\mathrm{ext}}}
\newcommand{\intn}{{\mathrm{int}}}

\newcommand{\Gamc}[1]{{\Gamma_{\mathrm{c}#1}^e}}

\newcommand{\dBc}[1]{{\partial_\mathrm{c}\mathcal{B}_{#1}}}

\newcommand{\zk}{{0k}}

\newcommand{\dA}{{\mathrm{d}A}}
\newcommand{\da}{{\mathrm{d}a}}

\newcommand{\AH}{{A_\mathrm{H}}}
\newcommand{\Ac}{{A_\mathrm{c}}}
\newcommand{\Acz}{{A_{\mathrm{c},0}}}
\newcommand{\Areal}{{A_\mathrm{real}}}

\newcommand{\epsT}{{\varepsilon_\mathrm{t}}}
\newcommand{\Fn}{F_\mathrm{n}}
\newcommand{\Ft}{F_\mathrm{t}}
\newcommand{\Lc}{{L_\mathrm{c}}}
\newcommand{\Lcz}{{L_{\mathrm{c},0}}}
\newcommand{\Lpeel}{{L_{\mathrm{peel}}}}
\newcommand{\Lslide}{{L_{\mathrm{slide}}}}

\newcommand{\Tn}{T_\mathrm{n}}
\newcommand{\tn}{t_\mathrm{n}}

\newcommand{\bTnk}{{\boldsymbol{T}_{\mathrm{n},k}}}

\newcommand{\bTtk}{{\boldsymbol{T}_{\mathrm{t},k}}}
\newcommand{\btt}{\boldsymbol{t}_\mathrm{t}}
\newcommand{\bttk}{{\boldsymbol{t}_{\mathrm{t},k}}}
\newcommand{\bTtrial}{\boldsymbol{T}_\mathrm{trial}}
\newcommand{\bttrial}{\boldsymbol{t}_\mathrm{trial}}
\newcommand{\Tt}{T_\mathrm{t}}

\newcommand{\Ttmax}{T_\mathrm{slide}}
\newcommand{\ttmax}{t_\mathrm{slide}}
\newcommand{\Tnmax}{T_\mathrm{max}}
\newcommand{\gequ}{{g_\mathrm{eq}}}
\newcommand{\garea}{{g_\mathrm{area}}}
\newcommand{\gn}{{g_\mathrm{n}}}
\newcommand{\bgn}{{\boldsymbol{g}_\mathrm{n}}}

\newcommand{\bgt}{{\boldsymbol{g}_\mathrm{t}}}
\newcommand{\bgs}{{\boldsymbol{g}_\mathrm{s}}}
\newcommand{\Lbgs}{{\mathcal{L}\boldsymbol{g}_\mathrm{s}}}
\newcommand{\Dbge}{{\Delta\boldsymbol{g}_\mathrm{e}}}
\newcommand{\xip}[1]{\xi_{\mathrm{p}}^{\alpha #1}}
\newcommand{\xis}[1]{\xi_{\mathrm{s}}^{\alpha #1}}
\newcommand{\Dxie}[1]{{\Delta\xi_{\mathrm{e} #1}^\alpha}}
\newcommand{\bxip}[1]{{\boldsymbol{\xi}_{\mathrm{p}}^{#1}}}
\newcommand{\bxis}[1]{{\boldsymbol{\xi}_{\mathrm{s}}^{#1}}}
\newcommand{\gmax}{{g_\mathrm{max}}}
\newcommand{\fs}{f_\mathrm{s}}

\newcommand{\gcut}{{g_\mathrm{cut}}}
\newcommand{\scut}{{s_\mathrm{cut}}}

\newcommand{\alphbet}{{\alpha\beta}}
\newcommand{\apco}[1]{{\boldsymbol{a}_{#1}^\mathrm{p}}}
\newcommand{\apcon}[1]{{\boldsymbol{a}^{#1}_\mathrm{p}}}
\newcommand{\Apco}[1]{{\boldsymbol{A}_{#1}^\mathrm{p}}}
\newcommand{\Apcon}[1]{{\boldsymbol{A}^{#1}_\mathrm{p}}}
\newcommand{\Jck}{J_{\mathrm{c}k}}
\newcommand{\Jcl}{J_{\mathrm{c}\ell}}
\newcommand{\Jcke}{J_{\mathrm{c}k}^e}
\newcommand{\Jcle}{J_{\mathrm{c}\ell}^e}
\newcommand{\np}{\boldsymbol{n}_\mathrm{p}}
\newcommand{\nt}{\boldsymbol{n}_\mathrm{t}}
\newcommand{\Nt}{\boldsymbol{N}_\mathrm{t}}
\newcommand{\xk}{{\boldsymbol{x}_k}}
\newcommand{\xl}{{\boldsymbol{x}_\ell}}
\newcommand{\xp}{{\boldsymbol{x}_\mathrm{p}}}
\newcommand{\uke}{{\mathbf{u}_k^e}}
\newcommand{\ule}{{\mathbf{u}_\ell^e}}

\newcommand {\mra}{\mathrm{a}}

\newcommand {\mrc}{\mathrm{c}}

\newcommand {\mre}{\mathrm{e}}

\newcommand {\mrm}{\mathrm{m}}
\newcommand {\mrn}{\mathrm{n}}

\newcommand {\mrp}{\mathrm{p}}

\newcommand {\mrs}{\mathrm{s}}
\newcommand {\mrt}{\mathrm{t}}

\newcommand {\mrA}{\mathrm{A}}

\newcommand {\mrD}{\mathrm{D}}
\newcommand {\mrE}{\mathrm{E}}

\newcommand {\mrG}{\mathrm{G}}

\newcommand {\mrI}{\mathrm{I}}
\newcommand {\mrJ}{\mathrm{J}}

\newcommand {\mrM}{\mathrm{M}}
\newcommand {\mrN}{\mathrm{N}}

\newcommand {\mrP}{\mathrm{P}}

\newcommand {\mrT}{\mathrm{T}}

\newcommand {\bzero}{\boldsymbol{0}}
\newcommand {\bone}{\boldsymbol{1}}

\newcommand {\ba}{\boldsymbol{a}}

\newcommand {\bg}{\boldsymbol{g}}

\newcommand {\bp}{\boldsymbol{p}}

\newcommand {\bx}{\boldsymbol{x}}

\newcommand {\bP}{\boldsymbol{P}}

\newcommand {\bX}{\boldsymbol{X}}

\newcommand {\bsig}{\mbox{\boldmath$\sigma$}}

\newcommand {\bxi}{\mbox{\boldmath$\xi$}}

\newcommand {\muu}{\mathbf{u}}

\newcommand {\mN}{\mathbf{N}}

\newcommand {\sB}{\mathcal{B}}

\newcommand {\sD}{\mathcal{D}}

\newcommand {\sL}{\mathcal{L}}

\newcommand{\EA}{EA}
\newcommand{\EAe}{{\mrE\mrA}}
\newcommand{\DI}{DI}
\newcommand{\DIe}{{\mrD\mrI}}

\newcommand\blfootnote[1]{%
	\begingroup
	\renewcommand\thefootnote{}\footnote{#1}%
	\addtocounter{footnote}{-1}%
	\endgroup
}

\Crefname{equation}{Eq.}{Eqs.}
\Crefname{figure}{Fig.}{Figs.}
\crefname{table}{Tab.}{Tabs.}
\Crefname{table}{Tab.}{Tabs.}

\journal{the Journal of the Mechanics and Physics of Solids}

\biboptions{authoryear}

\begin{document}


\begin{frontmatter}

\title{Contact with coupled adhesion and friction: \\
	Computational framework, applications, and new insights
	\blfootnote{\textcopyright\ 2020. This manuscript version is made
	available under the
	\href{https://creativecommons.org/licenses/by-nc-nd/4.0/}
	{CC-BY-NC-ND 4.0 license}.}}

\author[addressAICES]{Janine C.~Mergel\fnref{previous}\corref{corrauthorJCM}}
\fntext[previous]{Former member of Graduate School AICES}
\cortext[corrauthorJCM]{Corresponding author, janine.mergel@rwth-aachen.de}

\author[addressLTDS]{Julien Scheibert}

\author[addressAICES,addressIIT,addressGdansk]
	{Roger A.~Sauer\corref{corrauthorRAS}}
\cortext[corrauthorRAS]{Corresponding author, sauer@aices.rwth-aachen.de}

\address[addressAICES]{Graduate School AICES, RWTH Aachen University,
	Templergraben 55, 52056 Aachen, Germany}
\address[addressLTDS]{Univ Lyon, Ecole Centrale de Lyon, ENISE, ENTPE,
	CNRS, Laboratoire de Tribologie et Dynamique des Syst{\`e}mes LTDS,
	UMR 5513, F-69134, Ecully, France}
\address[addressIIT]{Department of Mechanical Engineering,
	Indian Institute of Technology Kanpur, UP 208016, India}
\address[addressGdansk]{Faculty of Civil and Environmental Engineering,
	Gda\'{n}sk University of Technology, ul.~Narutowicza 11/12, \\
	80-233 Gda\'{n}sk, Poland}
\address{\textnormal{Published in the \href{https://doi.org/10.1016/j.jmps.2020.104194}
	{Journal of the Mechanics and Physics of Solids 146:104194, 2021}}}

\begin{abstract}
Contact involving soft materials often combines dry adhesion, sliding friction,
and large deformations. At the local level, these three aspects are rarely
captured simultaneously, but included in the theoretical models by
\cite{mergel19jadh}. We here develop a corresponding finite element framework
that captures 3D finite-strain contact of two deformable bodies. This framework
is suitable to investigate sliding friction even under tensile normal loads.
First, we demonstrate the capabilities of our finite element model using both
2D and 3D test cases, which range from compliant tapes to structures with high
stiffness, and include deformable--rigid and deformable--deformable contact. We
then provide new results on the onset of sliding of smooth elastomer--glass
interfaces, a setup that couples nonlinear material behavior, adhesion, and
large frictional stresses. Our simulations not only agree well with both
experimental and theoretical findings, they also provide new insights into the
current debate on the shear-induced reduction of the contact area in
elastomeric contact.
\end{abstract}

\begin{keyword}
van der Waals interactions \sep computational contact mechanics \sep nonlinear
finite element methods \sep peeling \sep elastomer contact
\end{keyword}

\end{frontmatter}


\section{Introduction} \label{s:intro}

Soft materials like compliant tapes, elastomers, and biological adhesive pads
(appearing e.g.~in insects and lizards) play a major role in a large variety of
dry, solid contact. This kind of contact usually features large adhesive (or
tensile) stresses, large frictional stresses, and large deformations,
simultaneously. Present computational contact models do not appropriately
capture all these three features at once. However, this would be desirable in
order to reproduce and interpret a wide range of experimental observations for
such systems, including gecko pads \citep{autumn02}, tape peeling
\citep{dezotti19}, or rubber friction \citep{sahli18}.

Dry friction is often described with the classical Amontons-Coulomb law
of friction
\begin{equation}
	\Ft = \mu \, \Fn, \qquad \Fn > 0, \label{e:FtCoulomb}
\end{equation}
where~$\mu$ is a coefficient of friction relating the sliding friction
force~$\Ft$ to the normal load~$\Fn$. However, in applications dominated by
adhesion, this friction force is often found to be proportional to the real
contact area, $\Areal$, i.e., the total area of small microasperities in actual
contact \citep{carpick97, degrandicontraires12, yashima15, sahli18}. In this
case,
\begin{equation}
	\Ft = \tau_0 \, \Areal, \label{e:FtTabor}
\end{equation}
where $\tau_0$ denotes a frictional shear strength that depends on both
materials of the interacting surfaces. It was thus natural to propose the
following intermediate law~\citep{derjaguin34, bowden42}:
\begin{equation}
	\Ft = \mu \, \Fn + \tau_0 \, \Areal, \label{e:FtExtAmontons}
\end{equation}
which interpolates between the two limit cases of \cref{e:FtCoulomb,e:FtTabor}.
In the following, \cref{e:FtExtAmontons} will be referred to as the
``extended'' Amontons' law of friction. According to \cite{ruths05}, it is
suitable to describe the friction force between dry surfaces sliding over each
other in the presence of adhesion. As shown experimentally by \cite{homola90}
and others, it depends on the specific application which of these terms has
stronger influence. The transition between both terms is also discussed in
\cite{berman98}, \cite{gao04b}, and \cite{jagota11}. Sometimes using a slightly
different notation, relation~\labelcref{e:FtExtAmontons} was considered in the
context of microtribology~\citep{briscoe79}, molecular
dynamics~\citep{sivebaek08, mo09}, or adhesion and friction of biologic and
bio-mimetic systems~\citep{zeng09, hill11}. It was further used by
\cite{tabor81} to state a pressure-dependent, effective friction coefficient,
which was then incorporated into a computational model by~\cite{wriggers90}.
Moreover, \cref{e:FtExtAmontons} is also known as Mohr--Coulomb criterion in
soil mechanics. For a comprehensive review of tribological models, in general,
we refer to \cite{vakis18}.

Note that \crefrange{e:FtCoulomb}{e:FtExtAmontons} refer to the total
forces~$\Fn$ and~$\Ft$ applied at the interface. Locally, the normal contact
stress may vary between tension and compression within the same macroscopic
contact area (see e.g.~measurements by~\cite{eason15} for gecko toes). It is
thus desirable to formulate the two macroscopic friction
laws~\labelcref{e:FtTabor} and~\labelcref{e:FtExtAmontons} in terms of local
contact tractions, resulting in two general continuum models for dry adhesion
and friction. This is exactly what was recently proposed in~\cite{mergel19jadh}:
\begin{enumerate}[1)]
\item Model~\DI: In this model, the local sliding resistance is constant
everywhere along the contact interface, with a value equal to~$\tau_0$. This
yields~\cref{e:FtTabor} at the macroscale. The local sliding resistance itself
is independent of the (generally varying) contact pressure, which in turn is
expressed as function of the (small, but non-vanishing) local normal distance
between the contacting surfaces. The model is thus called~\DI, for
\underline{D}istance-\underline{I}ndependent.
\item Model~\EA: This corresponds to a local version of
\cref{e:FtExtAmontons}, and is denoted~\EA\ for \underline{E}xtended
\underline{A}montons. Its local sliding resistance is the sum of a constant
term (like in model~\DI) and a term that linearly depends on the (local) normal
pressure between the surfaces.
\end{enumerate}
As will be seen, both models are capable of capturing friction even for zero or
tensile contact pressures. This capability is an important feature of the class
of models investigated here, and is motivated by e.g.~soft and compliant
bio-adhesive pads, which are able to generate friction forces under tensile
normal loads.

The first aim of the present paper is to formulate a computational framework
for adhesive-friction models in general, and for the two above-mentioned models
in particular. For the sake of self-consistency, we first briefly review
models~\DI\ and \EA\ as well as the underlying assumptions. We then derive the
equations that are necessary to implement these models into a nonlinear 3D
finite element (FE) formulation based on large-deformation continuum mechanics.
Such formulations go back to \cite{laursen93}, and were extended subsequently
to consider e.g.~wear \citep{stromberg96}, irreversible adhesion
\citep{raous99}, reversible adhesion \citep{sauer07ijnme}, and multiscale
contact \citep{wriggers09}. Regarding non-adhesive 3D frictional contact,
important computational advances were made in the context of surface smoothing
\citep{padmanabhan01, krstulovic02}, mortar methods \citep{puso04, gitterle10,
dittmann14}, moving cone formulations \citep{wriggers04}, isogeometric analysis
\citep{delorenzis11, temizer12}, and unbiased friction algorithms
\citep{sauer15ijnme}. The latter work is used as a basis for the present
formulation. A literature survey on more broadly related models (including
e.g.~cohesive zone models) is given in \cite{mergel19jadh}.

The second aim of this paper is to provide various examples of application for
our framework. The first three especially serve to illustrate the wide range of
systems that can be studied using this framework, including various shapes and
dimensionalities. We then provide a detailed study of a fourth example, which
is currently a matter of scientific debate in the literature: the onset of
sliding of Hertzian elastomer--glass contact \citep{sahli18, sahli19, menga18,
menga19, mergel19jadh, khajehsalehani19, papangelo19, mcmeeking20, wang20,
lengiewicz20}. We here present new results that agree well with recent
experimental \citep{sahli19} and theoretical results \citep{chen08b,
papangelo19b}. These results lead us to suggest that the shear-induced
reduction of the contact area, discussed in those references, may exist even in
the absence of adhesion as well as for compressible materials.



The remainder of this paper is structured as follows. In \cref{s:model} we
state our general computational framework, and outline models~\DI\ and \EA.
\Cref{s:comput} contains the algorithmic treatment of adhesive friction as well
as the resulting finite element formulation. In \cref{s:results} we illustrate
the validity and applicability of our computational models by discussing 2D and
3D applications ranging from soft and compliant tapes to rather stiff
structures. A detailed study of the onset of sliding for smooth
elastomer--glass contact follows in \cref{s:cap}. \Cref{s:concl} concludes this
paper.


\section{Continuum modeling of dry adhesion and friction} \label{s:model}

The models for adhesive friction discussed here are expressed as functions of
the (generally varying) local normal distance between two surfaces. We thus
first provide the fundamentals to describe this distance mathematically. For
this purpose, we introduce a co-variant description for both surfaces, which
will be described in the following. As commonly done in large-deformation
continuum mechanics, we use uppercase letters for variables in the
\emph{reference} configuration of a body (denoted $\sB_0$), and small letters
for variables in the \emph{current} configuration ($\sB$).


\subsection{Contact kinematics} \label{s:model:kinem}

\begin{figure}[ht]
	\centering
	\includegraphics[width=0.45\textwidth]{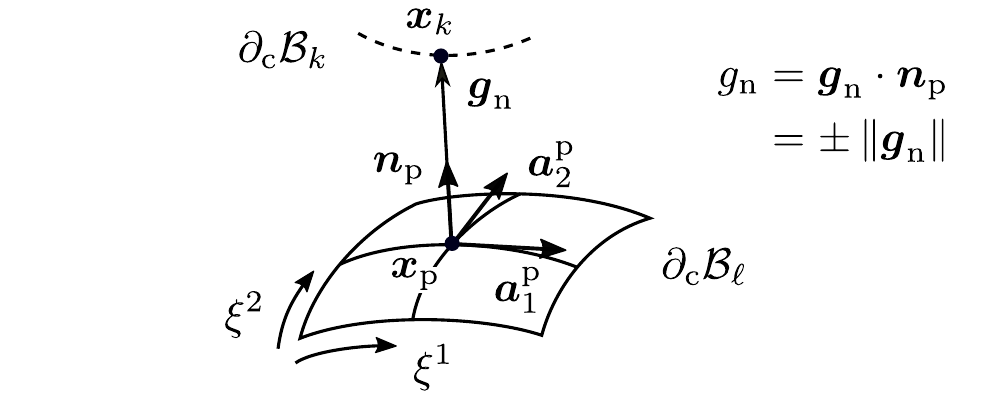}
	\caption{Closest projection point,~$\xp = \xl(\bxi_\mrp)$, of~$\xk$
		onto surface~$\dBc{\ell}$, and illustration of the tangent
		vectors~$\apco{\alpha}$ ($\alpha = 1,\,2$), surface normal~$\np$, and
		normal gap vector~$\bgn$.}
	\label{f:contkin}
\end{figure}

For a certain material point on the contact surface of one body, $\xk \in
\dBc{k}$ ($k = 1,\,2$), we need to determine its closest projection point, $\xp
\in \dBc{\ell}$ ($\ell = 2,\,1$), that minimizes the distance between~$\xk$ and
the surface~$\dBc{\ell}$ of the neighboring body (\cref{f:contkin}). To this
end, we assume that $\dBc{\ell}$ can be parametrized by two convective
coordinates $\bxi = \left\{ \xi^1,\xi^2\right\}$ such that $\xp =
\xl(\bxi_\mrp)$. At this (still unknown) point~$\xp$, $\dBc{\ell}$ is
characterized by its co-variant and contra-variant tangent vectors,
$\apco{\alpha}$ and $\apcon{\alpha}$ ($\alpha = 1,\,2$), and by its surface
normal $\np$. These vectors are defined as
\begin{align}
	\apco{\alpha}	& := \left.{ \pa{\xl(\bxi)}{\xi^\alpha}
						}\right|_{\boldsymbol{\xi}\,=\,
						\boldsymbol{\xi}_\mrp}\!, \label{e:apco} \\
	\apcon{\alpha}	& := a_\mrp^\alphbet \, \apco{\beta}, \qquad
			\big[a_\mrp^\alphbet\big] = {\big[a_\alphbet^\mrp\big]}^{-1},
			\qquad a_\alphbet^\mrp = \apco{\alpha} \cdot \apco{\beta},
			\label{e:apcontra} \\[1ex]
	\np		& := \frac{\apco{1} \times \apco{2}}{\|\apco{1} \times \apco{2}\|},
			\label{e:np}
\end{align}
where $\times$ denotes the cross product, and the definition of
$\apcon{\alpha}$ in \cref{e:apcontra} contains a summation over $\beta =
1,\,2$. The coordinates~$\bxi_\mrp$ of the projection point~$\xp$ can then be
determined by solving the (generally nonlinear) equations
\begin{equation}
	(\xp - \xk) \cdot \apco{\alpha} = 0, \qquad \alpha = 1,\,2;
\end{equation}
see e.g.~Appendix~B of \cite{sauer15ijnme}. Once this projection point is
found, one can define a normal gap vector, $\bgn$, and a scalar normal gap,
$\gn$, as
\begin{equation}
	\bgn := \xk - \xp, \qquad \gn := \bgn \cdot \np. \label{e:gn}
\end{equation}

In order to distinguish between tangential sticking and sliding, we (for now
only conceptually) introduce a tangential gap vector,~$\bgt$, which is
decomposed into 1)~a reversible (``elastic'') part,~$\Dbge$, associated with a
non-vanishing, tangential stiffness of the interface during sticking, and 2)~an
irreversible (``inelastic'') part,~$\bgs$, due to local sliding,
\begin{equation}
	\bgt = \Dbge + \bgs. \label{e:gt}
\end{equation}
The concept of an elastic, tangential gap is used for the algorithmic treatment
of friction, and explained in detail in \cref{s:comput:algo}.


\subsection{Modeling of normal tractions} \label{s:model:adh}

\begin{figure}[h]
	\centering
	\includegraphics[width=0.47\textwidth]{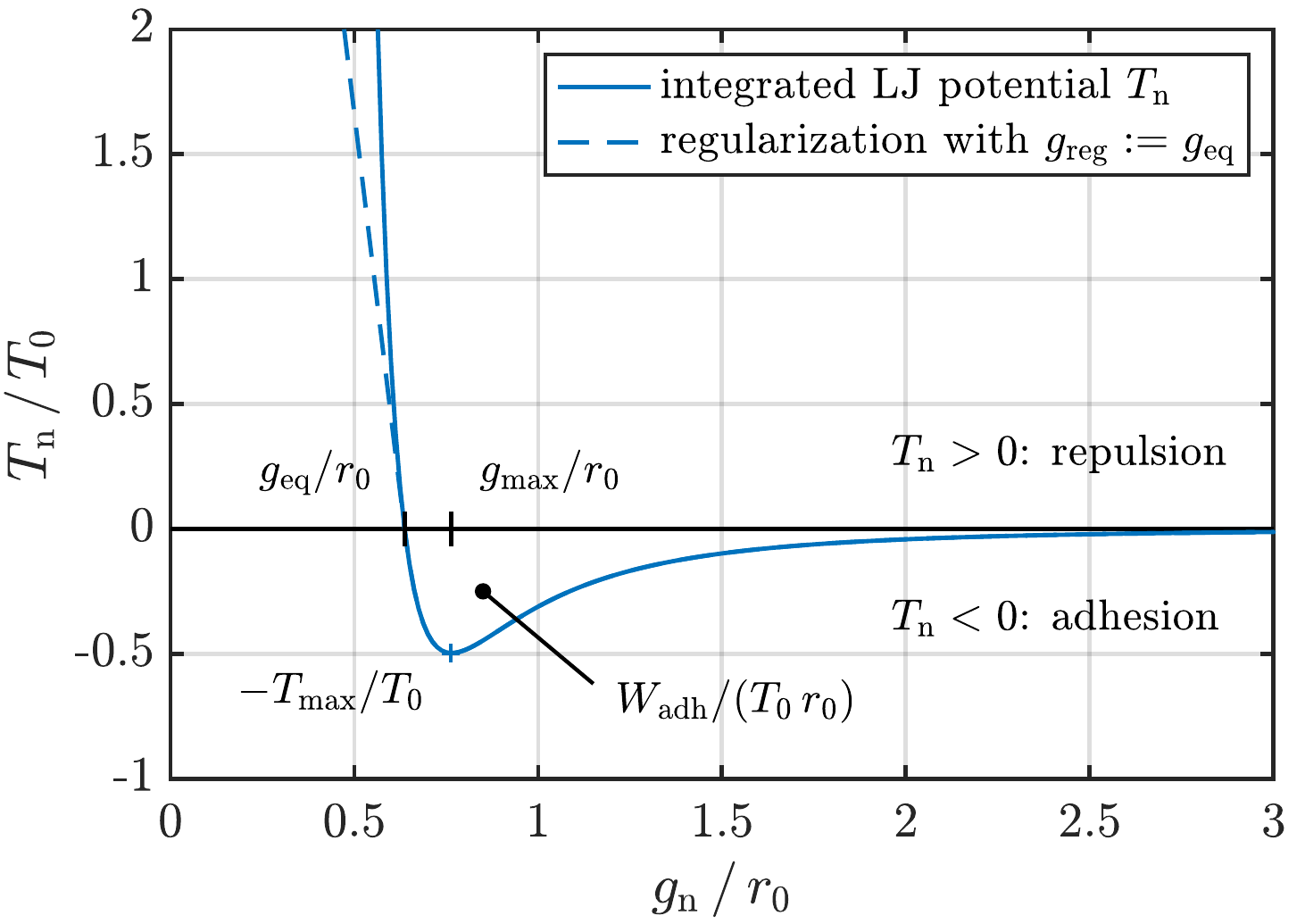}
	\hspace*{0.01\textwidth}
	\unitlength\textwidth
	\begin{picture}(0.47,0.30)
		\put(0,0.295){\textbf{Characteristic parameters:}}
		\put(0,0.14){\begin{tabular}{l@{\ \ }r@{\ }c@{\ }l}
			$\bullet$\ \ Equilibrium distance:
				& $\gequ$	& $=$	& $r_0 / \sqrt[6]{15}$ \\ \\[-2ex]
			$\bullet$\ \ Max.~adhesive traction:
				& $\Tnmax$ & $=$	& $\displaystyle
				\frac{\sqrt{5}\,\AH}{9\pi \, r_0^3}$ \\ \\[-2ex]
			$\bullet$\ \ Location of $-\Tnmax$:
				& $\gmax$	& $=$	& $r_0 / \sqrt[6]{5}$ \\ \\[-2ex]
			$\bullet$\ \ Work of adhesion:
				& $W_\adh$	& $=$	& $\displaystyle
				\frac{\sqrt[3]{15}\,\AH}{16\pi \, r_0^2}$ \\ \\[-2ex]
		\end{tabular}}
	\end{picture}
	\caption{Normal contact traction in the model of \cite{sauer09cmame} for
		frictionless adhesion and repulsion as well as characteristic
		parameters; $T_0 = \AH\,/\,\big(2\pi r_0^3\big)$; the blue dashed line
		shows a regularization of~$\Tn$ based on linear extrapolation at
		$\gn = \gequ$ \citep{mergel19jadh}.}
	\label{f:Tn}
\end{figure}

To allow for both tensile (adhesive) and compressive (repulsive) normal
tractions at the contact interface between~$\sB_k$ and $\sB_\ell$, we use a
model based on an integrated Lennard-Jones (LJ) potential \citep{sauer07ijnme,
sauer09cmame}. According to this model, the contact traction (force per area)
at point $\xk \in \dBc{k}$, due to~$\sB_\ell$, is given by
\begin{equation}
	\bTnk = \frac{\theta_k}{J_\ell} \, \Tn(\gn) \, \np, \qquad
	\Tn(\gn) = \frac{\AH}{2\pi r_0^3} \left[{ \frac{1}{45}
	{\Big(\frac{r_0}{\gn}\Big)}^9 - \frac{1}{3}
	{\Big(\frac{r_0}{\gn}\Big)}^3}\right]. \label{e:Tn}
\end{equation}
Note that this contact law is formulated with respect to the undeformed
\emph{reference} configuration. The function~$\Tn$ is shown in \cref{f:Tn},
together with several characteristic parameters.

We here consider a surface-to-surface contact formulation in which we iterate
over both contact surfaces (see also \cref{s:comput:FE}). The index~$k$
(or~$\ell$) thus switches from~1 to~2 (or vice versa). According to
\cref{e:Tn}, $\bTnk$ depends on the normal vector and normal gap from
\cref{e:np,e:gn} as well as on the following quantities: the Hamaker
constant~$\AH$, the characteristic length~$r_0$ in the Lennard-Jones potential,
the current volume change~$J_\ell$ around $\xp \in \dBc{\ell}$, and a
scalar~$\theta_k$ involving the inclination and deformation of the two bodies.
In the following we use two assumptions discussed in full detail in
\cite{mergel19jadh}: $\theta_k \approx 1$ and $J_\ell \approx \Jcl$, where
$\Jcl$ is the local stretch of the surface~$\dBc{\ell}$, given by
\begin{equation}
	\Jcl := \frac{j_{a\ell}}{J_{A\ell}}, \qquad
	j_{a\ell} := \sqrt{\det \big[a_\alphbet^\mrp\big]}, \qquad
	J_{A\ell} := \sqrt{\det \big[A_\alphbet^\mrp\big]}. \label{e:Jcl}
\end{equation}
The entries of~$\big[a_\alphbet^\mrp\big]$ and $\big[A_\alphbet^\mrp\big]$
($\alpha,\beta = 1,2$) are determined from \cref{e:apcontra}, inserting the
co-variant tangent vectors from the current and the reference configurations,
$\apco{\alpha}$ and $\Apco{\alpha} = {\partial\bX_\ell(\bxi) /
\partial\xi^\alpha}|_{\boldsymbol{\xi}\,=\,{\boldsymbol{\xi}}_\mrp}$,
respectively.

\Cref{f:Tn} further shows a regularization of~$\Tn$ based on linear
extrapolation for small normal gaps, $\gn < g_\mathrm{reg}$. This
regularization is also shown in the following figures
(\cref{f:modelDI,f:modelEA}), and increases the robustness of the computational
model by preventing infinitely large slope magnitudes. We apply it in all our
examples of \cref{s:results,s:cap}, using $g_\mathrm{reg} = \gequ$.

Due to the strong repulsion in~$\Tn$ (\cref{e:Tn,f:Tn}), the two contacting
surfaces are mainly (without regularization: always) separated by a small, but
positive, normal gap. This directly affects the definition of the real contact
area. Although this area is not explicitly required for our computational
framework, it is important for the interpretation of the results. We thus
define a gap, $\garea$, to determine whether or not two neighboring surface
points belong to the contact area. It is reasonable to set~$\garea$ somewhere
between~$\gequ$ (below which the normal traction is compressive) and~$\gmax$
(above which tensile tractions decay). We found the sensitivity of the contact
area to be large for~$\garea \approx \gequ$, and minimal at~$\gmax$. In this
work, we thus consider the real area of contact as that part of the interface
that is separated by less than $\garea = \gmax$.


\subsection{Modeling of tangential tractions} \label{s:model:frict}

We now present a general framework for combined adhesion, repulsion, and
sliding friction. As specific examples we outline the two models~\DI\ and \EA\
mentioned in the introduction. The equations stated in the remaining section,
as well as in the algorithm description (\cref{s:comput:algo}), all refer to
friction between a surface point~$\xk \in \dBc{k}$ ($k = 1,\,2$) and the
neighboring surface~$\dBc{\ell}$ ($\ell = 2,\,1$). For a shorter notation, we
omit the indices~$k$ and~$\ell$ in the following.

In our framework we assume that the vector of tangential tractions, due to
sticking and sliding friction, depends on both the normal gap from \cref{e:gn}
and the tangential slip vector introduced in \cref{e:gt}, i.e.~$\btt =
\btt(\gn,\bgt)$. After defining a non-negative function $\ttmax(\gn)$ for the
(magnitude of the) frictional resistance during sliding, we consider friction
laws of the general form
\begin{equation}
	\|\btt(\gn,\bgt)\|\ \begin{cases}
	\ <\ \ttmax(\gn)	& \text{during\ sticking}, \\
	\ =\ \ttmax(\gn)	& \text{during\ sliding}.
	\end{cases} \label{e:ttorig}
\end{equation}
In \cref{e:ttorig} we assume that, for equal~$\gn$, the tangential traction
required to initiate sliding (after sticking) is equal to the traction in the
final sliding state. The motivation, validity, and restrictions of this
assumption are addressed in detail in \cite{mergel19jadh}.


\subsubsection{Model~\DI: Distance-independent sliding friction in the contact
area} \label{s:model:DI}

\begin{figure}[h]
	\centering
	\subfigure[Sliding threshold for varying normal gap $\gn$.]{
		\includegraphics[width=0.47\textwidth]{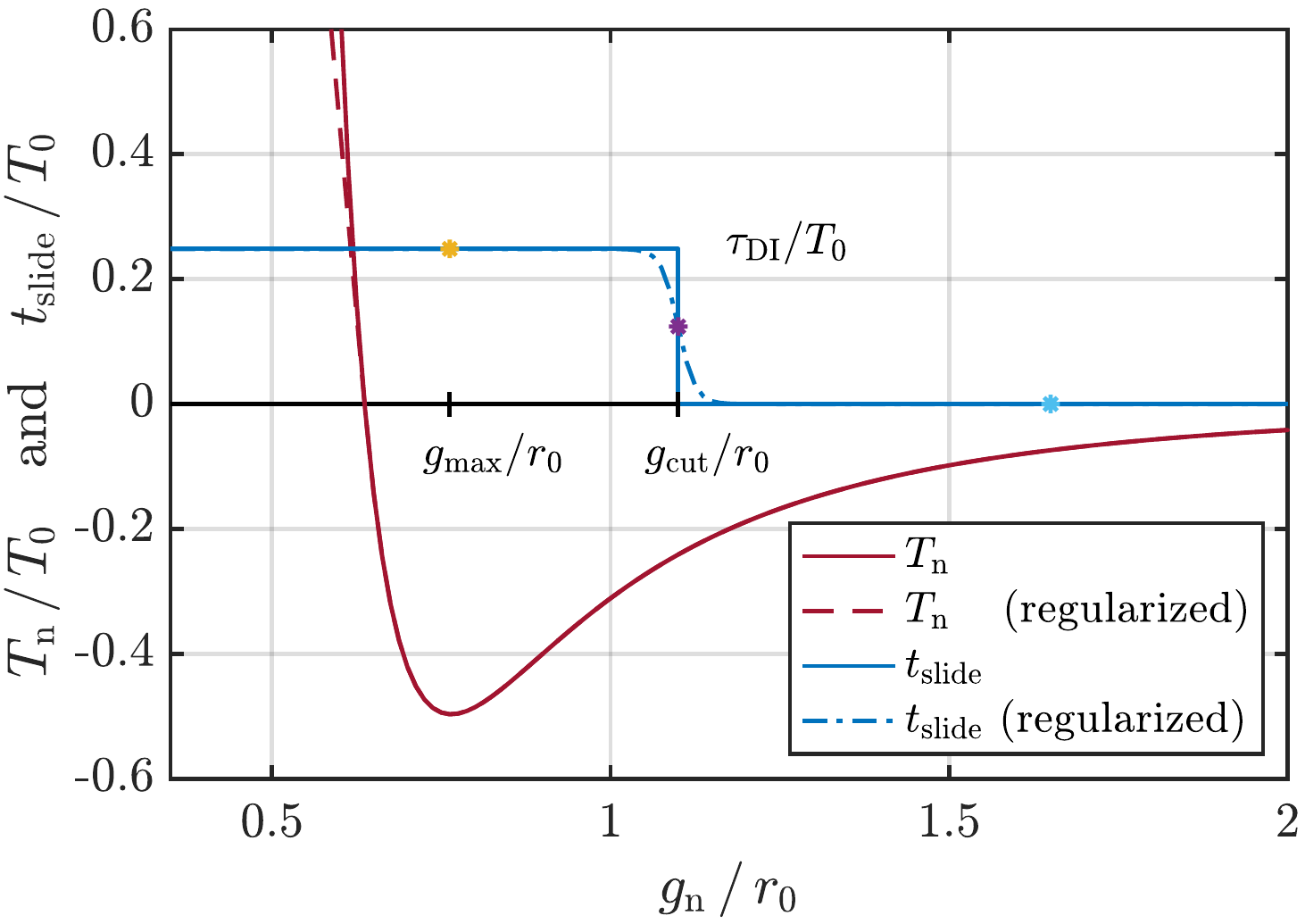}
		\label{f:tslideDI}
	}\hspace*{1ex}
	\subfigure[Friction law in 2D for different but fixed $\gn$.]{
		\includegraphics[width=0.47\textwidth]{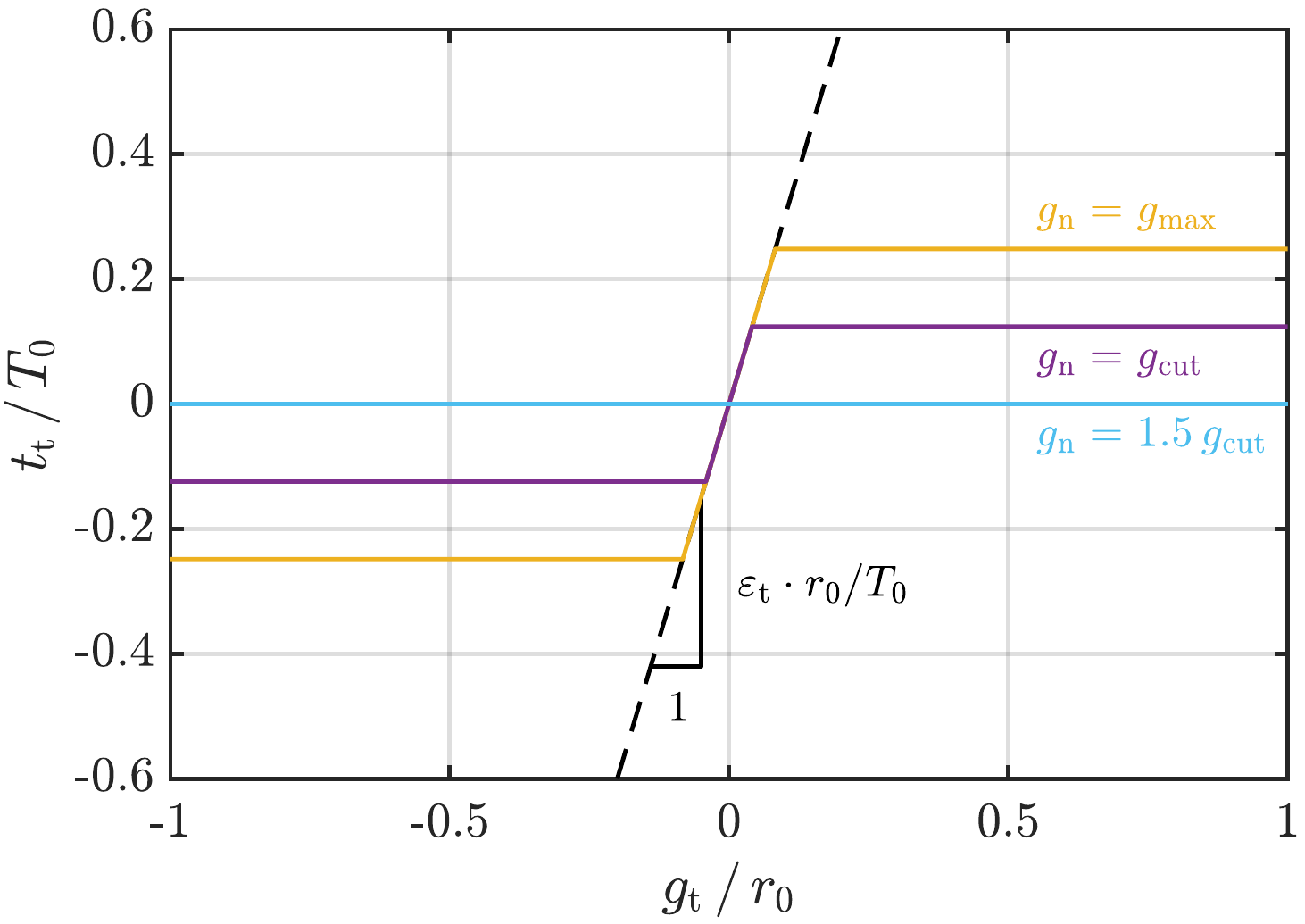}
		\label{f:ttlawDI}
	}
	\caption{Model~\DI: Constant sliding traction within the cutoff
		distance~$\gcut$ for $\tau_\DIe = 0.5 \, \Tnmax$, $\gcut = 1.1\,r_0$,
		and $k_\DIe = 80/r_0$; here, $T_0 = \AH \,/\, \big(2\pi r_0^3\big)$;
		the regularization of~$\Tn$ is discussed in \cref{s:model:adh};
		for~$\epsT$ see \cref{s:comput:algo}; the asterisks mark the
		distances~$\gmax$ (where $|\Tn| = \Tnmax$, yellow), $\gcut$ (purple),
		and $1.5\,\gcut$ (light blue).}
	\label{f:modelDI}
\end{figure}

The first model, denoted~\DI\ (for
\underline{D}istance-\underline{I}ndependent), is the local extension of
\cref{e:FtTabor}. It includes a constant sliding threshold for normal gaps
smaller than a certain cutoff distance~$\gcut$:
\begin{equation}
	\ttmax(\gn) = \begin{cases}
		\tau_\DIe,	& \gn \le \gcut, \\
		0,			& \gn > \gcut,
	\end{cases} \label{e:tslideDIunreg}
\end{equation}
where $\tau_\DIe > 0$ is a constant interface parameter. Note that the
adhesive-friction model~\DI\ is stated in the \emph{current} configuration of
the bodies, which means that the sliding threshold~$\tau_\DIe$ refers to the
\emph{actual} area of the contact interface. To overcome the discontinuity in
\cref{e:tslideDIunreg} at $\gn = \gcut$, for our computational framework we use
the following regularization,
\begin{equation}
	\ttmax(\gn) = \frac{\tau_\DIe}{1 + \mre^{\, k_\DIe (\gn - \gcut)}},
	\label{e:tslideDI}
\end{equation}
where $k_\DIe > 0$ is a sufficiently large regularization parameter for the
transition between $\tau_\DIe$ and zero. Its inverse, $1/k_\DIe$, can be
regarded as a characteristic decay length. Both the original and the
regularized versions of model~\DI\ are shown in \cref{f:tslideDI}, together
with the adhesion model from \cref{s:model:adh}.

\Cref{f:ttlawDI} shows the resulting friction law, i.e., the
traction--separation law in tangential direction for arbitrary, but fixed,
normal distances. In this friction law, we regularize the transition between
tangential sticking ($\|\btt\| < \ttmax$, see \cref{e:ttorig}) and sliding
($\|\btt\| = \ttmax$) by introducing a penalty parameter~$\epsT$. This
parameter represents the finite tangential stiffness of the interface during
sticking, and allows for a small, but reversible, tangential motion ($\Dbge$ in
\cref{e:gt}). Further details are provided in \cref{s:comput:algo}.

The sliding traction of model~\DI, $\tau_\DIe$, is independent from the normal
gap, and thus from the normal traction. If the same value is chosen for
both~$\gcut$ and~$\garea$ (see \cref{s:model:adh}), then the sliding threshold
will be equal to~$\tau_\DIe$ everywhere within the real contact area~$\Areal$.
Under such a condition, model~\DI\ yields \cref{e:FtTabor} at the macroscopic
scale, with $\tau_\DIe = \tau_0$. In this work, we always use model~\DI\ like
this, with $\gcut = \garea = \gmax$. For future comparison with model~\EA\
(\cref{s:results}), we further introduce a friction parameter\footnote{To avoid
any confusion with the classical coefficient of friction (defined as the ratio
between the frictional resistance and the pressure), for our two models we use
the terminology ``friction parameter'' instead.} $\mu_\DIe := \tau_\DIe \,/\,
\Tnmax$ to relate the sliding traction to the maximum adhesive traction from
\cref{f:Tn}.

A constant sliding resistance has also been considered in previous studies,
e.g.~in the context of sliding graphene sheets~\citep{deng12}, adhesive (and
small-deformation) contact between a rigid sphere and an elastic half-space
\citep{menga18, menga19, wang20}, or for non-adhesive friction of hyperelastic
materials \citep{lengiewicz20}. The strength of model~\DI\ is that it combines
adhesion and friction with large deformations and arbitrary-shaped geometries.
It includes the model of~\cite{deng12} as a special case with $\gcut = \gequ$,
while providing an additional regularization to overcome the discontinuity
at~$\gcut$. Note that (the unregularized version of) model~\DI\ is not to be
confused with the classical Maugis--Dugdale model. That model includes a
constant \emph{normal} traction for pure debonding, while the blue curve in
\cref{f:tslideDI} concerns the \emph{tangential} traction for sliding friction.


\subsubsection{Model~\EA: Extended Amontons' law in local form}
\label{s:model:EA}

The second adhesive-friction law, model~\EA\ (for \underline{E}xtended
\underline{A}montons) is intended as a local form of \cref{e:FtExtAmontons}.
Unlike model~\DI, this model explicitly depends on the (local) normal contact
tractions of our adhesion model, and thus on $\Tn(\gn)$ from \cref{e:Tn}. Let
us first choose a cutoff distance somewhere between the equilibrium
distance~$\gequ$ of~$\Tn$ and the location~$\gmax$ of~$-\Tnmax$ (see
\cref{f:Tn}):
\begin{equation}
	\gcut = \scut \, \gmax + (1 - \scut) \, \gequ, \qquad \scut \in [0,1].
	\label{e:gcutEA}
\end{equation}
We then consider a sliding threshold that is proportional to the shifted curve
$\Tn(\gn) + |\Tn(\gcut)|$, where the function value~$\Tn(\gcut)$ is either zero
or negative (because $\gcut \ge \gequ$):
\begin{equation}
	\Ttmax(\gn) = \begin{cases}
		\displaystyle
		\frac{\mu_\EAe}{\Jcl} \, \big[\Tn(\gn) - \Tn(\gcut)\big],
			& \gn < \gcut, \\[0.2ex]
		0,
			& \gn \ge \gcut.
	\end{cases} \label{e:TslideEA}
\end{equation}
The scalar $\mu_\EAe > 0$ denotes the friction parameter for model~\EA\ (see
also footnote~2). The first term in \cref{e:TslideEA} (top) expresses the
dependence on the normal traction that is formulated, by choice, w.r.t.~the
\emph{reference} configuration. If desired, both the normal and tangential
tractions can be mapped to the current configuration using~$\tn~\da = \Tn~\dA$.
The second term corresponds to an additional constant representing the effect
of adhesion. Since~$\gcut \le \gmax$ (\cref{e:gcutEA}), the sliding resistance
in~\cref{e:TslideEA} is always positive.
\begin{figure}[p]
	\centering
	\subfigure[Sliding threshold for varying $\gn$; $\scut = 0$.]{
		\includegraphics[width=0.47\textwidth]{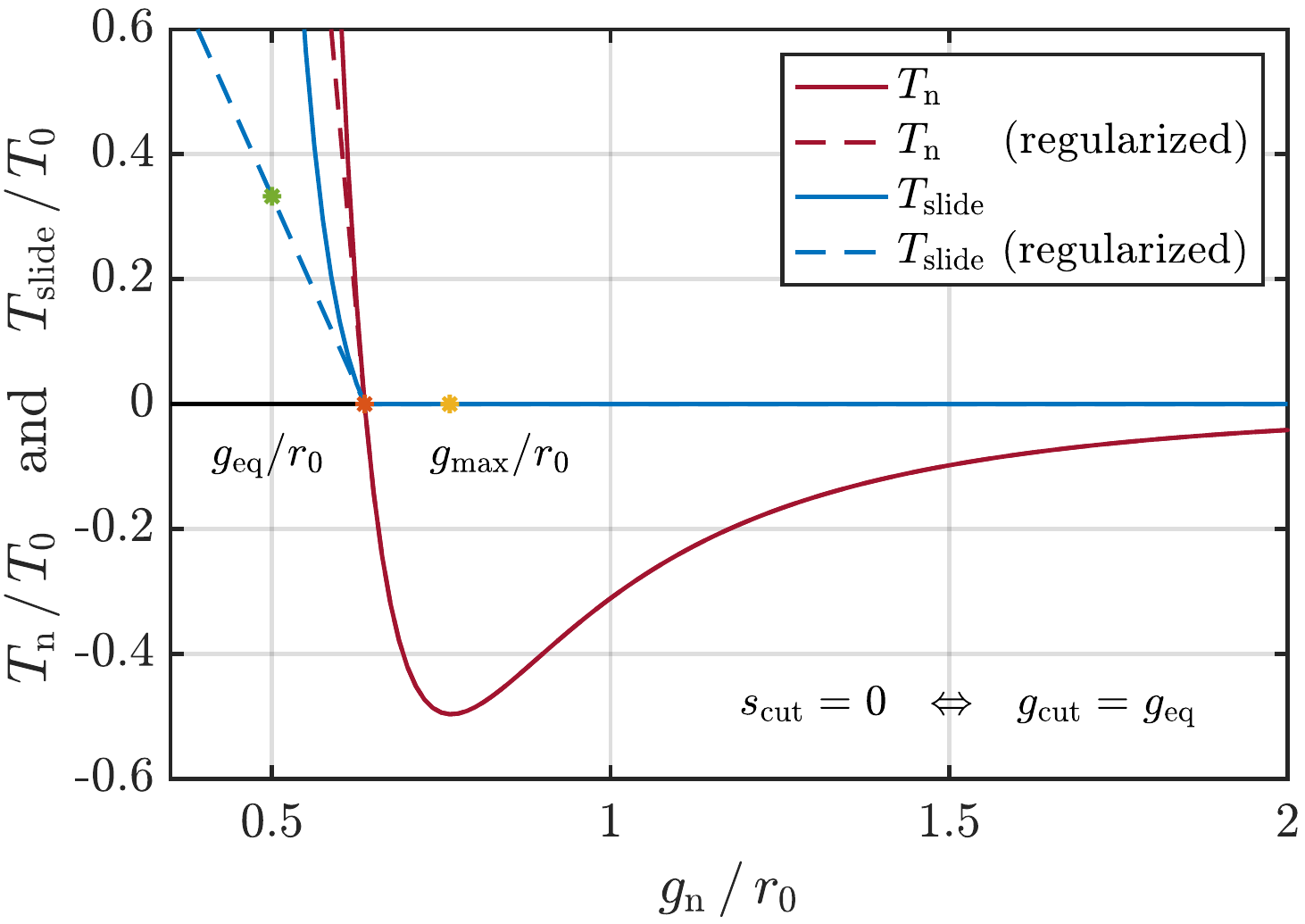}
		\label{f:TslideEA:s0}
	}\hspace*{1ex}
	\subfigure[2D friction law for different but fixed $\gn$; $\scut = 0$.]{
		\includegraphics[width=0.47\textwidth]{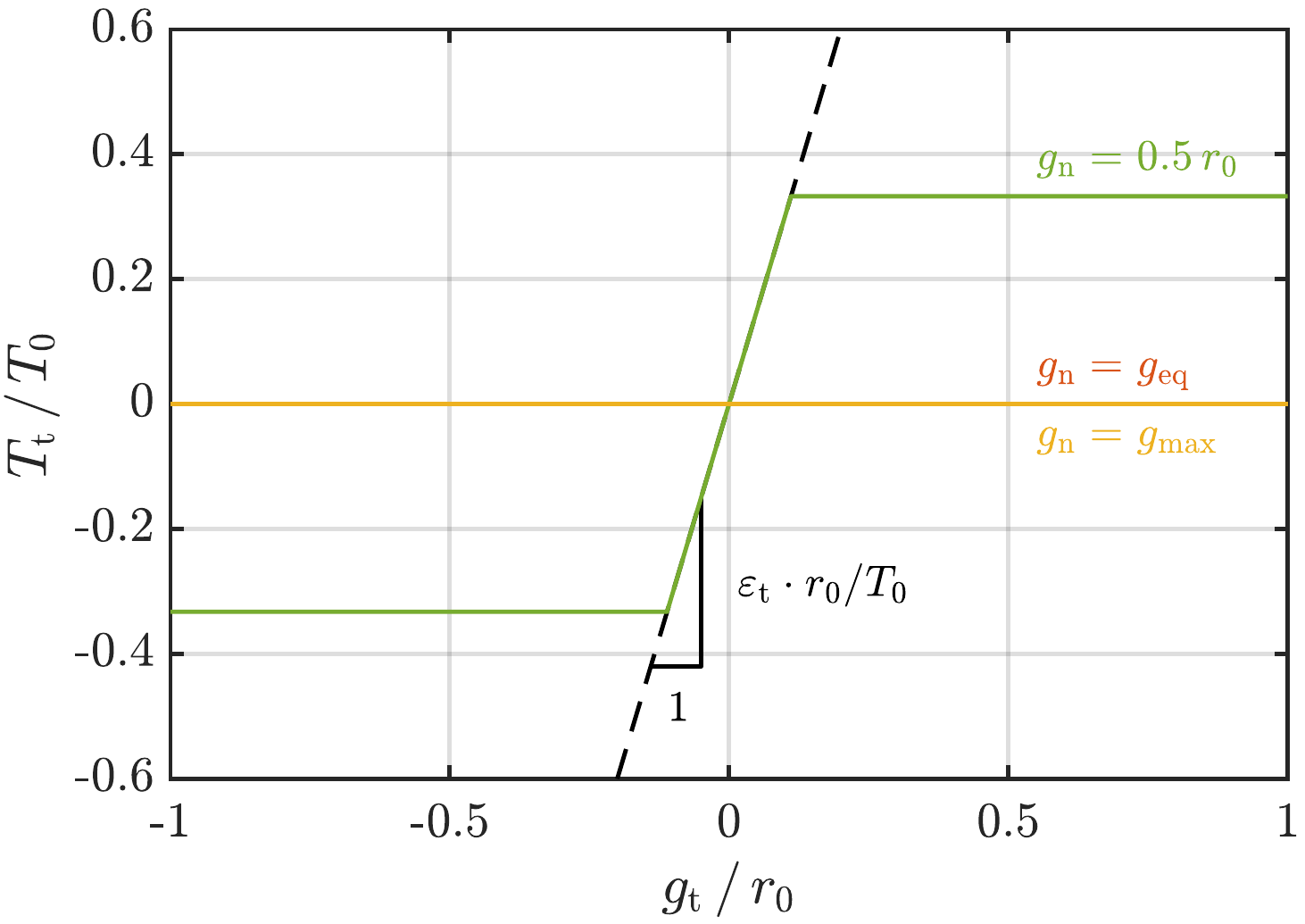}
		\label{f:TtlawEA:s0}
	}
	\subfigure[Sliding threshold for varying $\gn$; $\scut = 0.5$.]{
		\includegraphics[width=0.47\textwidth]{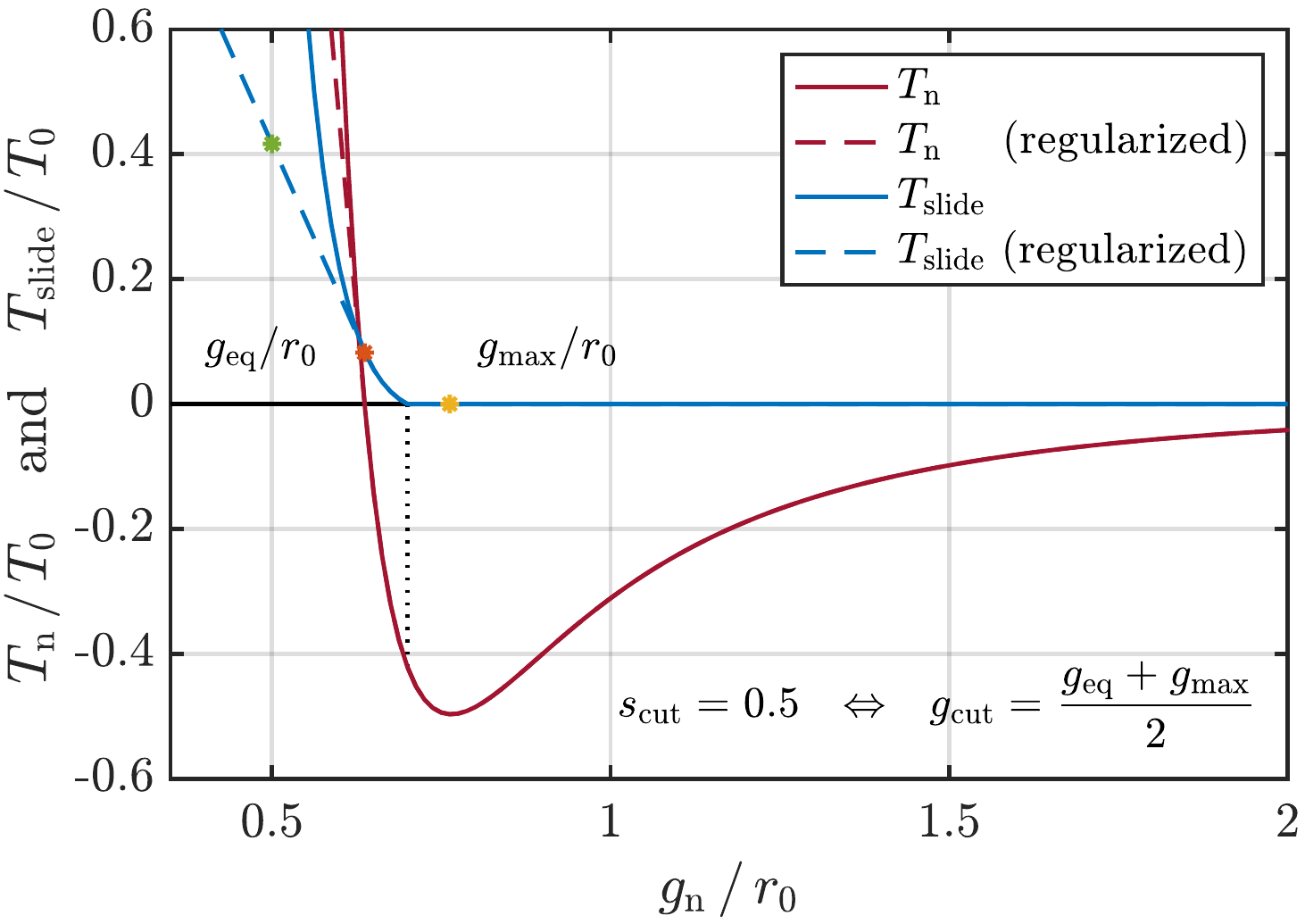}
		\label{f:TslideEA:s05}
	}\hspace*{1ex}
	\subfigure[2D friction law for different but fixed $\gn$; $\scut = 0.5$.]{
		\includegraphics[width=0.47\textwidth]{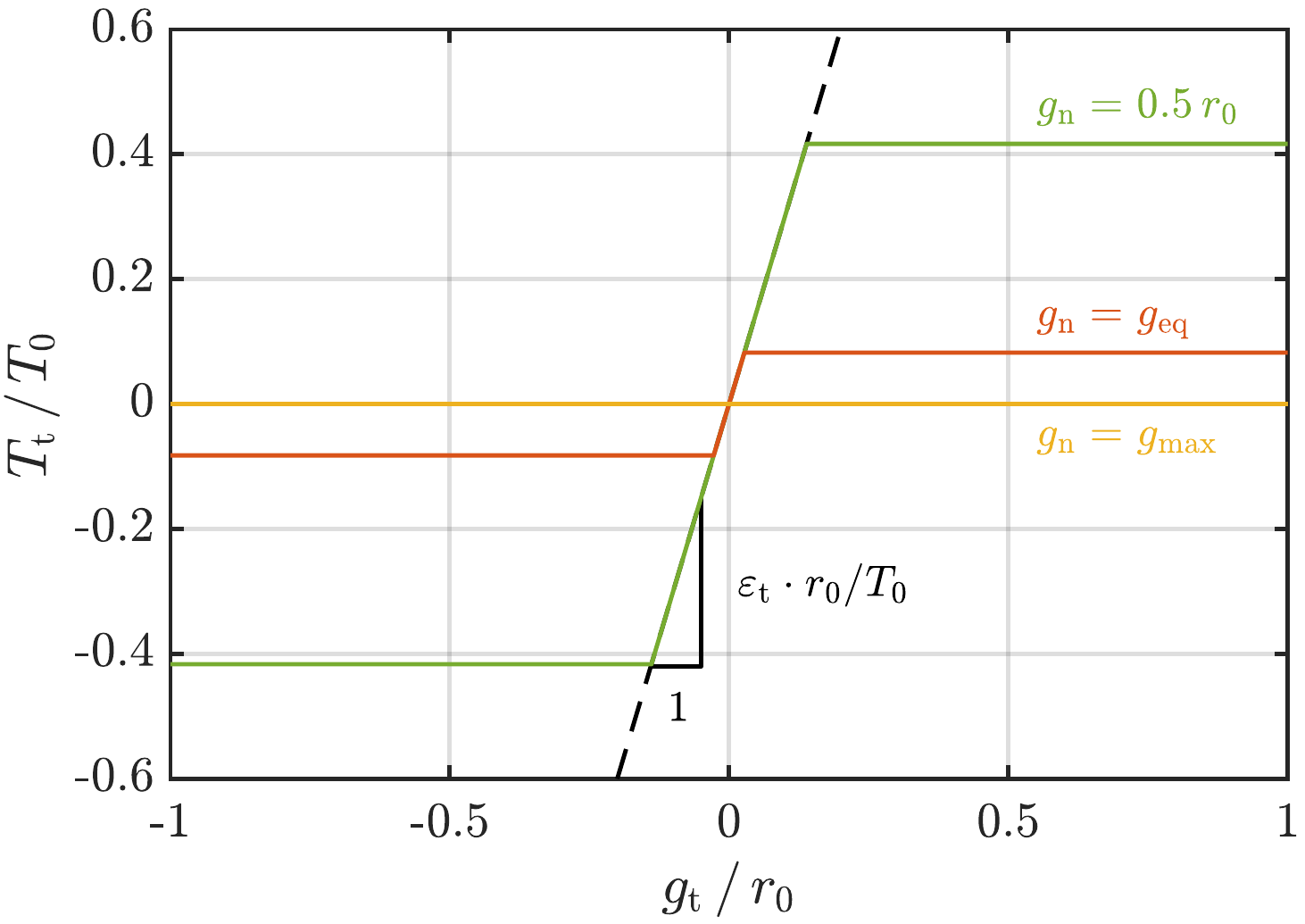}
		\label{f:TtlawEA:s05}
	}
	\subfigure[Sliding threshold for varying $\gn$; $\scut = 1$.]{
		\includegraphics[width=0.47\textwidth]{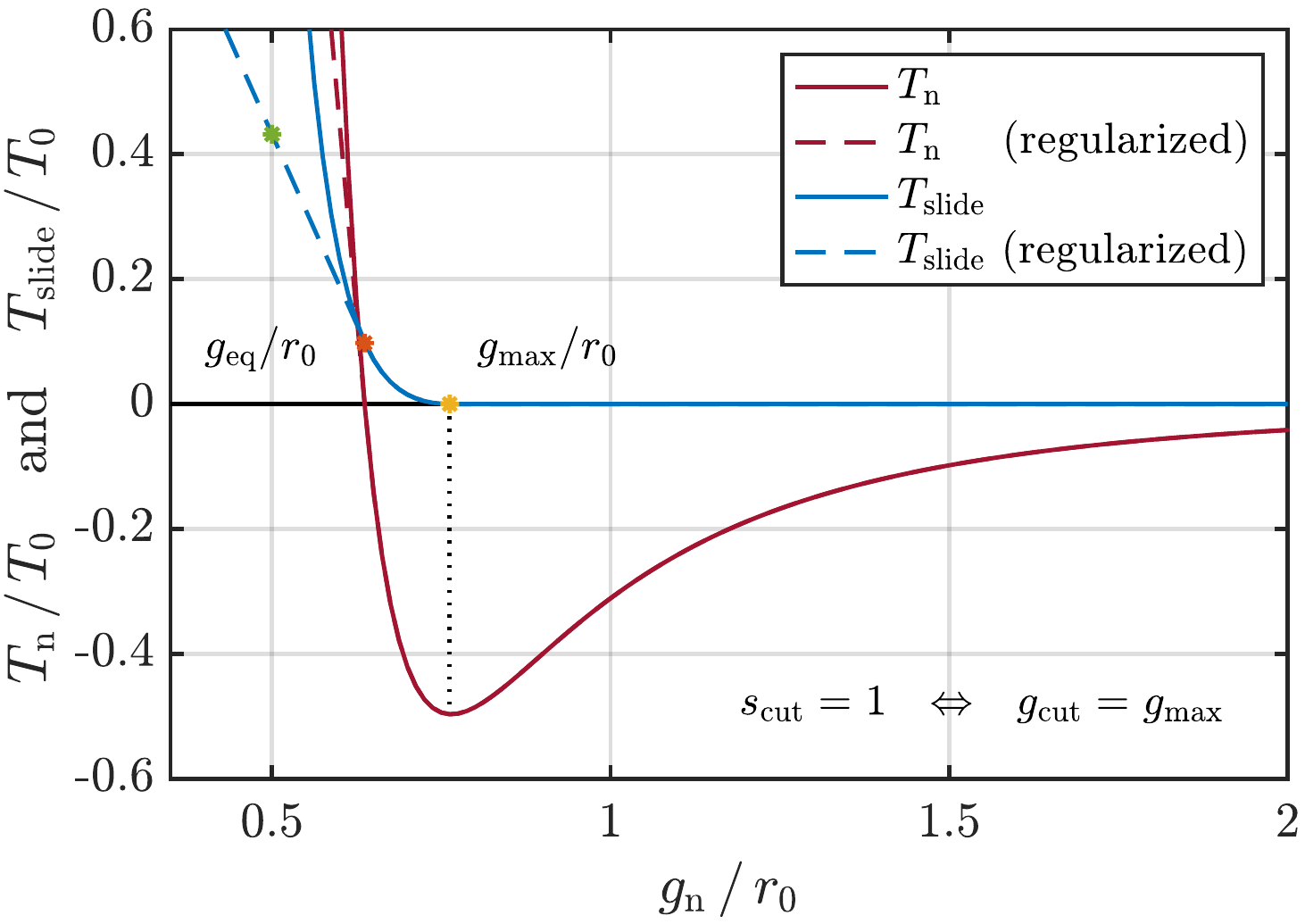}
		\label{f:TslideEA:s1}
	}\hspace*{1ex}
	\subfigure[2D friction law for different but fixed $\gn$; $\scut = 1$.]{
		\includegraphics[width=0.47\textwidth]{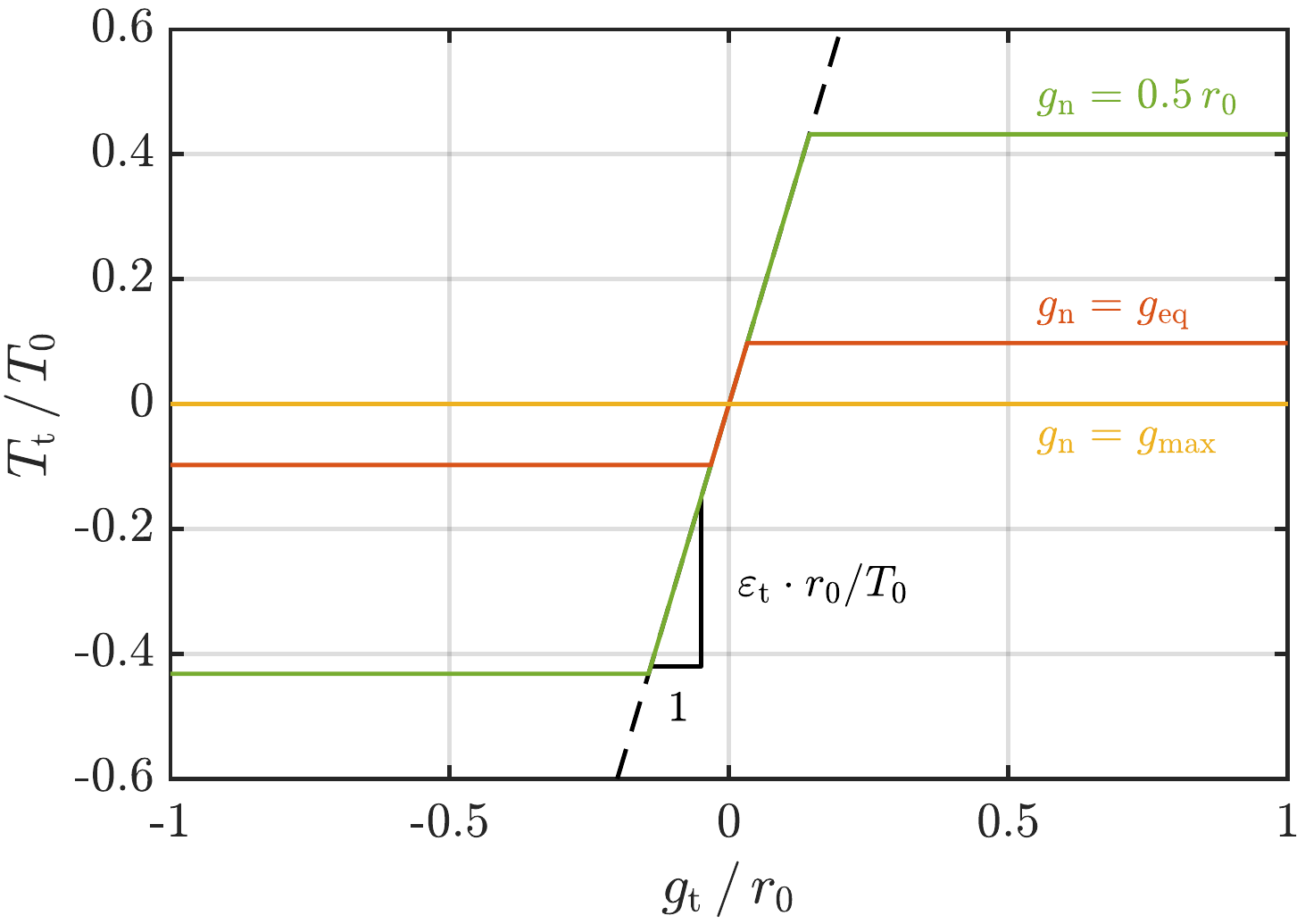}
		\label{f:TtlawEA:s1}
	}
	\caption{Model~\EA: Extended Amontons' law in \emph{local} form
		illustrated for $\mu_\EAe = 0.2$, $\Jcl \equiv 1$, and three different
		values of~$\scut$; here, $T_0 = \AH \,/\, \big(2\pi r_0^3\big)$; the
		regularization of~$\Tn$ is discussed in \cref{s:model:adh}; for~$\epsT$
		see \cref{s:comput:algo}; the colored asterisks mark the distances
		$0.5\,r_0$ (green), $\gequ$ (where $\Tn = 0$, orange), and $\gmax$
		(yellow).}
	\label{f:modelEA}
\end{figure}

\Cref{f:modelEA} illustrates model~\EA\ for the parameter values $\scut = 0$,
$0.5$, and $1$. In analogy to \cref{f:modelDI}, the left-hand side of
\cref{f:modelEA} depicts the relation between the normal gap,~$\gn$, and the
sliding threshold,~$\Ttmax$, while the right-hand side shows the resulting
friction laws. Note that the function $\Ttmax(\gn)$ is $C^1$-continuous only if
$\scut = 1$ (\cref{f:TslideEA:s1}). In any other case, a kink occurs at $\gn =
\gcut$, which requires a special algorithmic treatment in the solution strategy
(\cref{s:comput:actset}).

For $\gcut = \gequ$ (i.e., $\scut = 0$, see \cref{f:TslideEA:s0,f:TtlawEA:s0}),
model~\EA\ provides a frictional resistance (i.e., $\Ttmax > 0$) only for
positive, and thus compressive, normal tractions. This corresponds to the local
equivalent of the classical Coulomb--Amontons law for non-adhesive contact
(\cref{e:FtCoulomb}). It also follows the approach of many cohesive zone
models~\citep{chaboche97b, raous99, delpiero10}, which consider both local
tension and compression, but include sliding friction only where the normal
traction is compressive. Note that, in particular for adhesion-dominated setups
the model behavior may become very sensitive in this case. This is demonstrated
in \cite{mergelPhD} for the setup investigated in \cref{s:strip}. It also
agrees with the finding stated in \cref{s:model:adh}, namely that the real
contact area is very sensitive to~$\garea$ for $\garea \approx \gequ$.

As soon as~$\gcut > \gequ$, a frictional resistance can arise also in regions
where the normal traction is slightly tensile (up to the value~$\Tn(\gcut)$).
Note that, since in the present work we use $\garea = \gmax$
(\cref{s:model:adh}), and since $\gcut \le \gmax$, the tangential tractions are
always zero outside the real contact area.


\section{Computational framework} \label{s:comput}

This section contains the algorithmic treatment of adhesive friction and its
implementation into a nonlinear finite element code. We also address the
validity and restrictions of our framework.


\subsection{Algorithmic treatment of adhesive friction} \label{s:comput:algo}

In computational friction algorithms, sticking and sliding are often realized
by means of a predictor--corrector approach similar to that used for
elastoplasticity, see, e.g., \cite{simo98}: One first predicts tangential
sticking, checks whether the criterion for sliding is satisfied, and if so,
applies a return map to determine the sliding distance.

\begin{figure}[h]
	\centering
	\subfigure[Model~\DI.]{
		\includegraphics[width=0.43\textwidth]{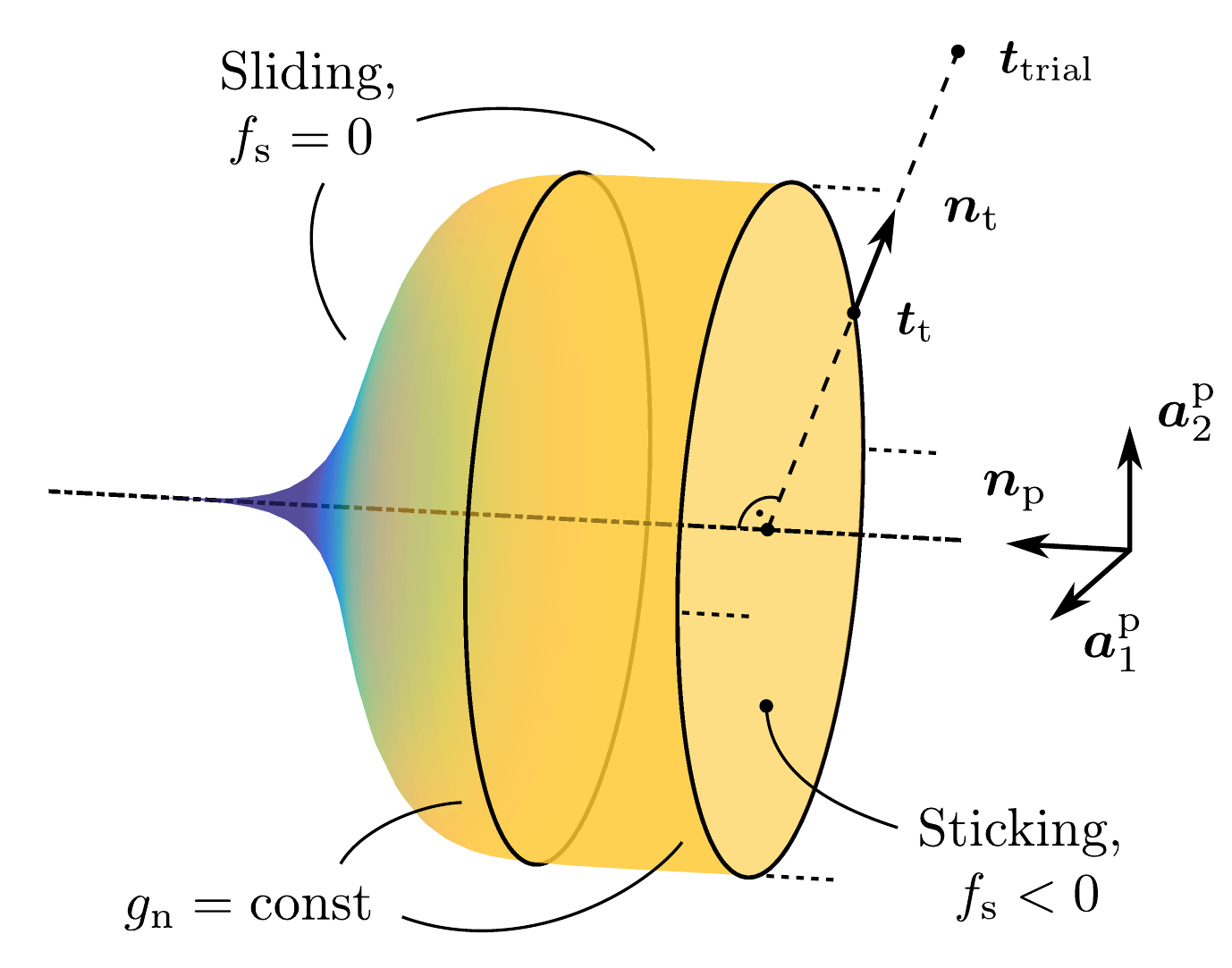}
		\label{f:returnmap:DI}
	}\hspace*{1ex}
	\subfigure[Model~\EA\ with $\gcut = \gmax$ (i.e., $\scut = 1$).]{
		\includegraphics[width=0.43\textwidth]{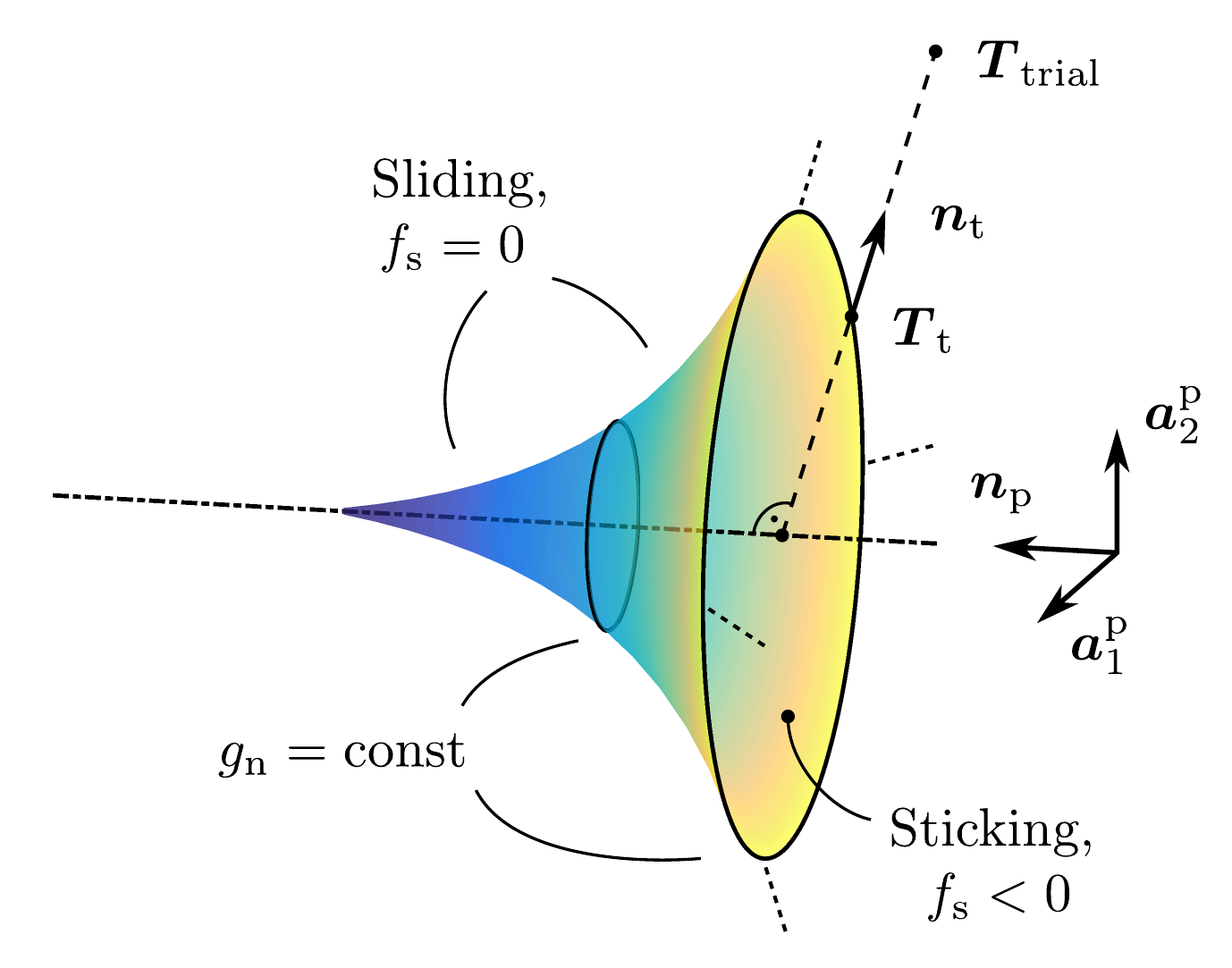}
		\label{f:returnmap:EA}
	}
	\caption{Illustration of the return map performed in the corrector step for
		tangential sliding.}
	\label{f:returnmap}
\end{figure}

The following algorithm is formulated in the current configuration, i.e., in
terms of the variables~$\btt$ and~$\ttmax$. However, as shown in
\cite{mergelPhD}, it is also possible to analogously state the algorithm with
respect to~$\Tt$ and~$\Ttmax$, the variables with which model~\EA\ is
formulated in \cref{s:model:EA}. Let us first define the sliding criterion
\begin{equation}
	\fs(\btt,\ttmax) := \|\btt\| - \ttmax, \label{e:fs}
\end{equation}
which must satisfy \citep{wriggers06}
\begin{equation}
	\fs \, \begin{cases}
	\ <\ 0	& \text{for\ sticking}, \\
	\ =\ 0	& \text{for\ sliding}.
	\end{cases} \label{e:slipcrit}
\end{equation}
These two cases are illustrated in \cref{f:returnmap:DI} for model~\DI. Here,
the colored surface marks $\fs = 0$, for which $\|\btt\|$ must be equal to the
sliding threshold~$\ttmax$ from \cref{e:tslideDI}. As a consequence, this
surface corresponds to nothing else but~$\ttmax$ revolved around the
$\gn$-axis, cf.~\cref{f:tslideDI,f:returnmap:DI}. In analogy,
\cref{f:returnmap:EA} indicates sticking and sliding for model~\EA\ with $\gcut
= \gmax$. The remaining quantities in \cref{f:returnmap} will be explained in
the following.

During sliding we must enforce $\fs(\btt,\ttmax) = 0$, and determine both the
direction and the magnitude of the resulting tangential gap. Since this
projection is carried out at a certain (pseudo-)time step, the normal gap from
\cref{e:gn}, $\gn$, is considered to be arbitrary but fixed. As shown in
\cite{mergelPhD}, both models~\EA\ and \DI\ fulfill convexity (i)~of the domain
of feasible tractions~$\btt$ satisfying $\fs(\btt,\ttmax) \le 0$, and (ii)~of
the function~$\fs$ within this domain. These properties ensure that the mapping
performed as a corrective step is unique. Let us now define the non-negative
dissipation during sliding,
\begin{equation}
	\sD_\mrs (\btt;\Lbgs) := \btt \cdot \sL\bg_\mrs,
\end{equation}
which represents the energy loss per time and area. The term
\begin{equation}
	\Lbgs := \dot{\xi}_\mrs^\alpha \, \ba_\alpha^\mrs, \qquad \ba_\alpha^\mrs
	:= \left.{ \pa{\xl(\bxi)}{\xi^\alpha}
	}\right|_{\boldsymbol{\xi}\,=\,\boldsymbol{\xi}_\mrs} \!, \label{e:dgs}
\end{equation}
denotes the relative slip velocity, i.e., the Lie derivative of the slip
vector~$\bgs$ associated with sliding (\cref{e:gt}, see also
\cite{sauer15ijnme}). Note that the dot denotes the material time derivative
of~$\xi_\mrs^\alpha$ (keeping material point~$\bX_\mrp$ fixed). According to
the principle of maximum dissipation \citep{simo98, wriggers06}, for a
\emph{given} slip velocity~$\Lbgs$, the actual (physically true)
traction~$\btt$ maximizes the dissipation among all feasible
tractions~$\btt^*$. By keeping $\Lbgs$ arbitrary but fixed, and by introducing
the Lagrange multiplier~$\gamma \ge 0$, we formulate the following constrained
minimization problem
\begin{equation}
	\left.{ \pa{\mathscr{L} \big(\btt^*,\gamma;\Lbgs\big)}{\btt^*}
	}\right|_{\btt^*\,=\,\btt} = \bzero
	\label{e:minLagrangian}
\end{equation}
with Lagrangian
\begin{equation}
	\mathscr{L} \big(\btt^*,\gamma;\Lbgs\big)
	:= - \sD_\mrs \big(\btt^*;\Lbgs\big)
	+ \gamma \cdot \fs\big(\btt^*,\ttmax\big),
\end{equation}
and the Karush--Kuhn--Tucker (KKT) conditions for optimality,
\begin{equation}
	\fs \le 0, \qquad \gamma \ge 0, \qquad \fs \cdot \gamma = 0.
\end{equation}
From \cref{e:minLagrangian} we then obtain the evolution law for the slip
velocity during sliding
\begin{equation}
	\Lbgs = \gamma \, \nt, \qquad \nt := \pa{\fs}{\btt}
	= \frac{\btt}{\|\btt\|}. \label{e:dgsslide}
\end{equation}
According to \cref{e:dgsslide}, the multiplier~$\gamma$ corresponds to the
magnitude of the tangential slip velocity. Since during sliding $\fs = 0$, the
traction satisfies
\begin{equation}
	\btt = \ttmax \, \nt, \label{e:ttslide}
\end{equation}
as required. The time derivatives~$\dot{\xi}_\mrs^\alpha$ in $\Lbgs$ can
finally be determined by inserting \cref{e:dgsslide} into \cref{e:dgs}, and
contracting the result with the tangent vectors~$\ba_\mrs^\alpha$, giving
\begin{equation}
	\dot{\xi}_\mrs^\alpha = \gamma \, \nt \cdot \ba_\mrs^\alpha, \qquad
	\alpha = 1,\,2. \label{e:dxis}
\end{equation}

The following procedure corresponds to the friction algorithm proposed by
\cite{sauer15ijnme}, but extended to a general sliding threshold~$\ttmax$. For
the derivation of the resulting equations we refer to the above-mentioned
paper. The key idea is to decompose the convective coordinates, $\xip{}$, of
the projection point into components that are either related to irreversible
sliding, $\xis{}$, or to reversible (``elastic'') sticking, $\Dxie{}$
(cf.~\cref{e:gt}). As described in detail in the following, for sliding the
traction is determined such that it satisfies \cref{e:slipcrit}; see also
\cref{f:returnmap}.

Given all quantities at a pseudo-time step~$t_n$, one proceeds to the next step
$t_{n+1}$ as follows:
\vspace*{1ex}

\hrule
\begin{enumerate}[1)]
\item
	Assume tangential sticking and compute a corresponding trial
	traction~(\cref{f:returnmap})
	\begin{equation}
		\bttrial^{n+1} = \epsT \left[{ \bx_\ell^{n+1}\big(\bxip{n+1}\big)
		- \bx_\ell^{n+1} (\bxis{n}) }\right]. \label{e:ttrial}
	\end{equation}
	The large, but finite, penalty parameter~$\epsT$ is used for
	regularization, i.e., to allow for a small, reversible tangential gap,
	given by $\bx_\ell^{n+1}\big(\bxip{n+1}\big) - \bx_\ell^{n+1} (\bxis{n})$.
	$\epsT$ is visualized in \cref{f:ttlawDI} for model~\DI, and e.g.~in
	\cref{f:TtlawEA:s05} for model~\EA.
\item Insert~$\bttrial^{n+1}$ into the criterion
	\begin{equation}
		\fs\big(\bttrial^{n+1},\ttmax^{n+1}\big) = \big\|\bttrial^{n+1}\big\|
		- \ttmax^{n+1}
	\end{equation}
	to check for sticking or sliding:
	\begin{enumerate}
	\item If $\fs < 0$, the point~$\xp$ is sticking tangentially; thus
		\begin{equation}
			\btt^{n+1} = \bttrial^{n+1}, \qquad
			\xis{\,n+1} = \xis{\,n}, \qquad
			\alpha = 1,\,2. \label{e:ttnewstick}
		\end{equation}
	\item If $\fs \ge 0$, the point~$\xp$ is sliding, and an additional
		correction is required to determine the actual traction~$\btt^{n+1}$
		such that it satisfies $\fs = 0$. Regarding \cref{f:returnmap}, in this
		case the trial traction lies outside the domain enclosed by the colored
		surface, and must be mapped back onto this surface. First compute the
		incremental update of the Lagrange multiplier in \cref{e:dgsslide},
		\begin{equation}
			\Delta\gamma^{n+1} =
			\frac{\fs\big(\bttrial^{n+1},\ttmax^{n+1}\big)}{\epsT},
		\end{equation}
		and then determine the updated coordinates~$\xis{\,n+1}$
		($\alpha = 1,\,2$) associated with sliding as well as the corresponding
		tangential traction,
		\begin{align}
			\xis{\,n+1} & = \xis{\,n} + \Delta\gamma^{n+1} \, \nt^{n+1} \cdot
							\ba_\mrp^{\alpha\,n+1}, \qquad
							\nt^{n+1} = \frac{\bttrial^{n+1}}
							{\left\|{\bttrial^{n+1}}\right\|},
							\label{e:ttslidealgo1} \\
			\btt^{n+1}	& = \bttrial^{n+1} - \epsT \, \Delta\gamma^{n+1}
							\, \nt^{n+1}. \label{e:ttslidealgo2}
		\end{align}
	\end{enumerate}
\end{enumerate}
\hrule
\vspace*{1ex}

\Cref{e:ttslidealgo1} is based on the approximation $\ba_\mrs^{\alpha\,n+1}
\approx \ba_\mrp^{\alpha\,n+1}$ (see \cite{sauer15ijnme})\footnote{This
approximation simplifies the linearization of the computational formulation
considerably, while introducing a negligible error proportional to $1/\epsT$.},
and on the discretization of \cref{e:dxis} by means of the implicit Euler
method. The traction~$\btt^{n+1}$ in \cref{e:ttslidealgo2} satisfies
\cref{e:ttslide}. In the case of frictionless sliding (i.e., $\ttmax^{n+1} =
0$), one simply sets $\btt^{n+1} = \bzero$ and $\xis{\,n+1} = \xip{\,n+1}$.


\subsection{Contact contributions to the finite element equations}
\label{s:comput:FE}

In order to discretize the corresponding model equations in space, we use a
standard Galerkin finite element (FE) method. For the sake of brevity we only
discuss the contact contributions here. To this end, we introduce a global FE
contact force vector~$\fc(\muu)$, which depends on the global displacement
vector~$\muu$ for all finite element nodes. Following the sign convention by
\cite{laursen02}, we now decompose $\fc(\muu)$ into the normal force~$\fn$ due
to adhesion and repulsion and the tangential force~$\ft$ due to friction,
\begin{equation}
	\fc(\muu) := \fn(\muu) - \ft(\muu). \label{e:fc}
\end{equation}
The vectors~$\fn$ and~$\ft$ are obtained by assembling the contributions from
each finite element on the contact surfaces. For contact between two deformable
bodies, we consider the \emph{two-half-pass} approach by \cite{sauer13cmame,
sauer15ijnme}. We thus iterate over the elements of \emph{both} contact
surfaces ($k = 1,\,2$), and compute the elemental contribution to the normal
forces from
\begin{equation}
	\fnk^e = - \int_{\Gamc{\zk}} \Nk^\mrT \, \bTnk~\dA_k. \label{e:fnke}
\end{equation}
Here, $\Gamc{\zk} \subset \dBc{}^h_\zk$, $\bTnk$ is given by \cref{e:Tn}, and
the array~$\Nk$ contains the nodal shape functions associated with the current
element. The tangential forces are determined similarly,
\begin{equation}
	\ftk^e = - \int_{\Gamc{k}} \Nk^\mrT \, \bttk~\da_k = -\int_{\Gamc{\zk}}
	\Nk^\mrT \, \bTtk~\dA_k, \label{e:ftke}
\end{equation}
depending on whether the model is defined in the current (model~\DI,
\cref{e:tslideDI}) or the reference configuration (model~\EA,
\cref{e:TslideEA}).

Since the vector~$\fc$ is nonlinear with respect to~$\muu$, we linearize it to
solve the governing equations with Newton's method. To this end, we derive the
tangent matrix associated with~$\fc$, denoted $\kc :=
\partial\fc/\partial\muu$. Like the global contact forces~$\fn$ and $\ft$,
$\kc$ is assembled from the contributions of each single surface element, see
\labelcref{a:kc}. Finally, we evaluate the integrals in \cref{e:fnke,e:ftke}
and in~\labelcref{a:kc} by means of Gaussian quadrature.


\subsection{Active set strategy} \label{s:comput:actset}

As seen in \cref{f:TslideEA:s0}, \labelcref{f:TslideEA:s05}, and
\labelcref{f:TslideEA:s1}, the curve $\Ttmax(\gn)$ of model~\EA\ is
$C^1$-continuous at the cutoff distance~$\gcut$ only if $\gcut = \gmax$ (i.e.,
$\scut = 1$). Otherwise, it is $C^0$-continuous at~$\gcut$. Moreover, we must
generally distinguish between discrete states of contact like tangential
sticking and sliding. Following \cite{sauer15ijnme}, we here consider an active
set strategy. To this end, we slightly modify the two active sets that are
often used for unilateral and frictional contact: (i)~``(compressive) contact
vs.~no contact'' and (ii)~``sticking vs.~sliding''. Since the function
$\Tn(\gn)$ in \cref{e:Tn} is continuous anyway (and thus does not require any
active set at all), we simply replace active set~(i) by ``frictional
vs.~frictionless sliding''. Note that it is also reasonable to define certain
cutoff values for which both the normal and tangential contact stresses are
considered to be negligible. This on one hand increases the efficiency, and on
the other hand circumvents computational problems in the sliding algorithm or
closest point projection (due to large separations of the bodies).

For changes in the (discrete) contact states, the FE force vector~$\fc(\muu)$
actually becomes non-differentiable, cf.~\labelcref{a:kc:ktstick,a:kc:ktslide}.
If the applied load step is too large, or if large parts of the contact surface
change their state at once, this may lead to a phenomenon called cycling, or
also zig-zagging \citep{wriggers06}. That is, the solution algorithm alternates
between two states in consecutive iterations. This can be generally overcome
e.g.~by means of semi-smooth Newton methods. Instead, we here pursue a strategy
that is based on a temporary freezing of the active sets; see also
\cite{wriggers06}. Accordingly, we keep the active sets fixed at the beginning
of the iterations, until the residual falls below a certain tolerance. We then
assume that the changes in the state are sufficiently small to prevent cycling.
Further improvement could be made by adjusting the load step adaptively when
the body starts or stops (full) sliding, in particular when it changes its
direction of sliding.

A special (and careful) treatment may be required for model~\EA\ with $\scut =
0$, i.e., $\gcut = \gequ$ (see also the end of \cref{s:model:EA}). In this
specific case, the switch from a non-zero to zero frictional resistance is
directly located at the equilibrium distance, $\gequ$, where the normal
traction is zero (\cref{f:TslideEA:s0}). Thus, for setups in which large parts
of the contact area are close to this distance (like those in
\cref{s:strip,s:tape}), the tangential traction can then alternate back and
forth between the states of frictional and frictionless sliding. This may
either require unnecessarily small load steps, or even lead to a failure of the
freezing strategy. As demonstrated in \cite{mergelPhD}, however, this issue is
successfully overcome by choosing a cutoff distance~$\gcut$ that is slightly
\emph{smaller} than $\gequ$. For such a case we recommend considering
\cref{e:gcutEA} with a parameter~$\scut$ that is negative but close to zero
(e.g., $\scut = -0.001$). Thus, the contact pressure must first exceed a
certain threshold to generate a frictional resistance.


\subsection{Validity, restrictions, and possible extensions of our models}
\label{s:comput:restrict}

Let us now discuss the validity and restrictions of our models, in particular
regarding computational aspects. For physical assumptions see
\cite{mergel19jadh}. The general properties of the applied two-half-pass
algorithm are discussed in \cite{sauer13cmame, sauer15ijnme}. A possible
extension to incorporate sticking friction larger than the sliding resistance
$\ttmax$ or $\Ttmax$ is briefly addressed in \cite{mergelPhD}.


\subsubsection{Jump-off and jump-to contact}

Strong adhesion combined with soft materials may cause sudden jump-off or
jump-to contact, see e.g.~\cite{chaboche01} and \cite{sauerPhD}. In that case
the (de-)bonding process inherently becomes physically unstable, and the
assumption of quasi-static conditions (as considered in this work) is not valid
anymore. According to \cite{raous11}, this is caused by the non-convexity and
the softening character (i.e., a decreasing slope with increasing distance) of
traction--separation laws like that in \cref{f:Tn}. A similar effect can occur
also for sliding friction \citep{mroz02}.

The issues mentioned above can be generally avoided by solving the model
equations with an arc length or continuation method \citep{sauerPhD}. Another
possibility would be to account for a viscous (velocity-dependent)
regularization as done in several cohesive zone models involving friction
\citep{raous99, chaboche01, delpiero10}. This approach is reasonable because
viscous effects often occur not only in the bulk material but also at the
interface. Besides, it would also be possible to account for inertial effects
and to consider the problem to be dynamic. Since the study of dynamic and
viscous effects lies outside the scope of this paper, in the following
application examples we focus on the attachment or detachment process itself,
but not on sudden jump-to or jump-off contact, respectively.


\subsubsection{Going from the nano- to the macroscale} \label{s:comput:scales}

As shown in \cref{f:Tn}, the normal contact traction~$\Tn$ from the adhesion
model by \cite{sauer09cmame} decays rapidly for increasing distances~$\gn$.
Depending on the involved materials, $\Tn$~typically acts only within the range
of several (tens of) nanometers. In addition, if $\gn$ approaches zero, the
traction increases to infinity while its slope approaches minus infinity. These
issues can cause numerical problems on one hand, and require a fine spatial
resolution (finite element size and applied displacements) on the other hand.

We thus recommend the following strategy to avoid such problems. First, a
sufficiently accurate surface discretization is crucial to reduce artificial
oscillations \citep{sauer11ijnme}, which could cause convergence problems in
the numerical solution procedure. In \cref{s:results,s:cap} we thus use
specially surface-enriched finite elements based either on Hermite
\citep{sauer11ijnme} or Non-Uniform Rational B-Spline (NURBS) shape functions
\citep{corbett14, corbett15}. Second, we recommend regularizing the normal
tractions as discussed in \cref{s:model:adh}. Third, it is generally possible
to calibrate the model parameters from \cref{s:model:adh} ($\AH$, $\Tnmax$, and
$W_\adh$) such that they match experimental data. This leads to a larger length
parameter~$r_0$, which automatically regularizes the curve~$\Tn$ in
\cref{e:Tn}. For further comments and restrictions see \cite{mergel19jadh}.


\section{Application examples} \label{s:results}

With the three following examples we illustrate the properties and qualitative
behavior of our two models, and demonstrate the generality of our method. We
consider not only different dimensions (2D and 3D), but also different contact
types (rigid/deformable and deformable/deformable). A fourth, and more
detailed, example follows in \cref{s:cap}.


\subsection{2D peeling of a thick strip from a rigid surface} \label{s:strip}

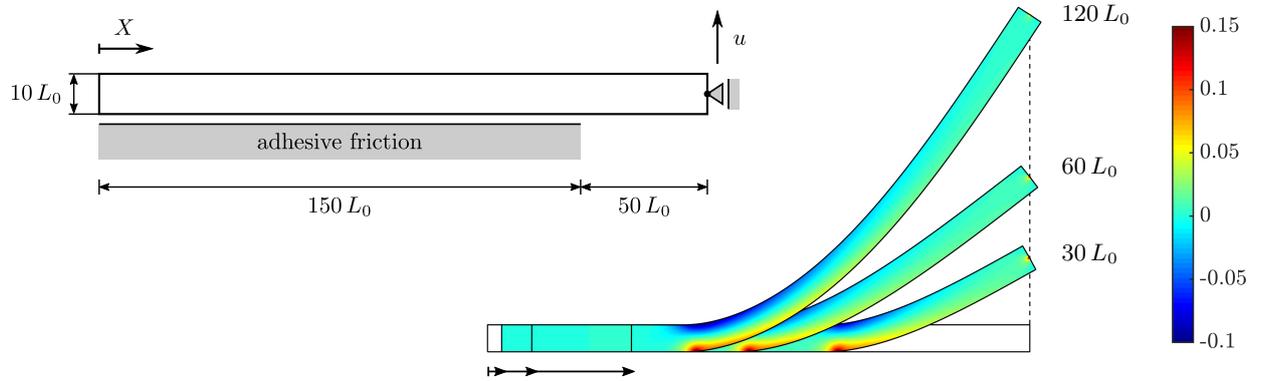
\begin{figure}[h]
	\unitlength\textwidth
	\begin{picture}(1,0.30)
		\put(0.38,0.02){\includegraphics[width=0.45\textwidth]
			{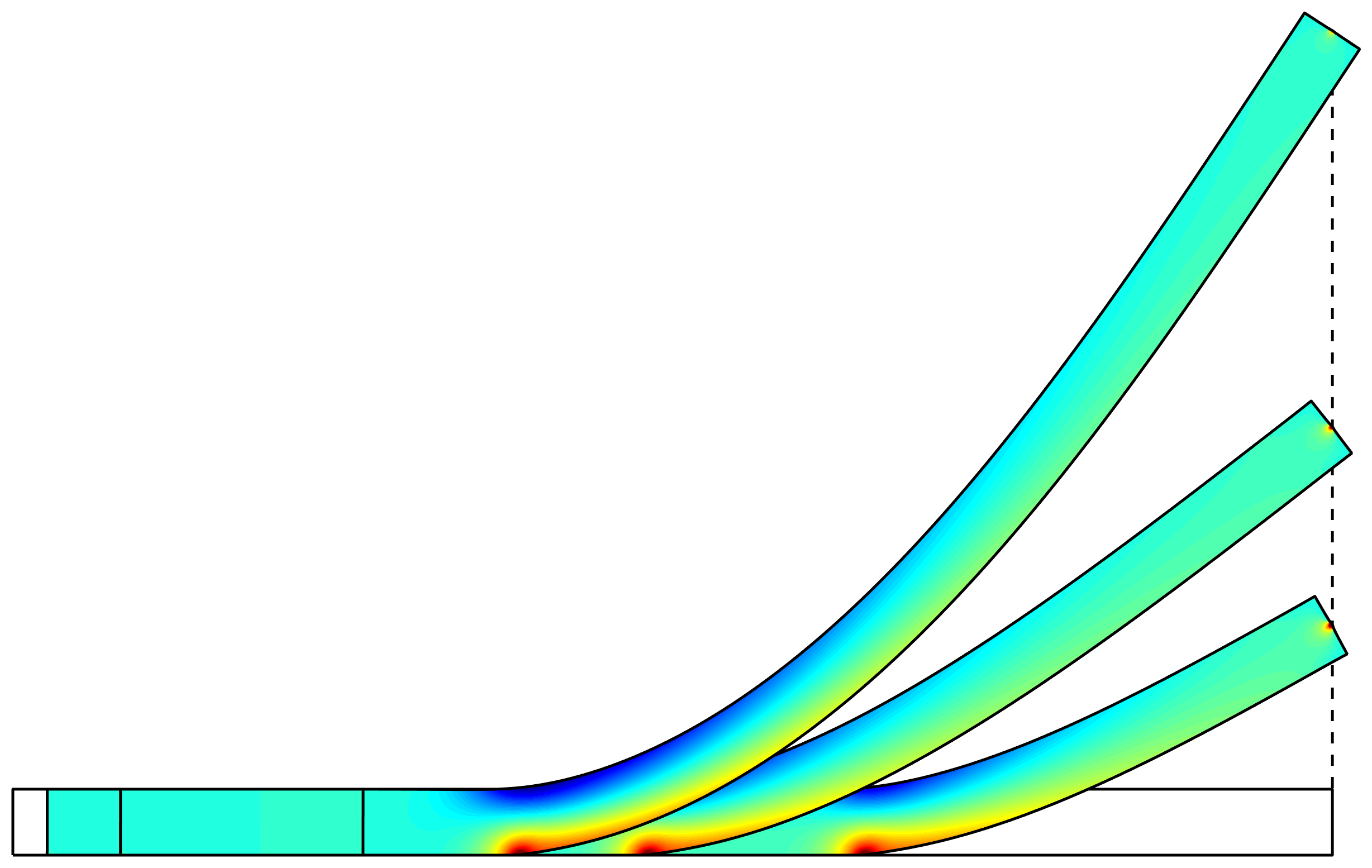}}
		\put(0.38,0){\includegraphics[width=0.45\textwidth]
			{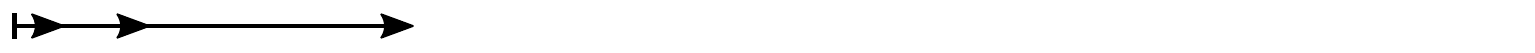}}
		\put(0.93,0.025){\includegraphics[height=0.27\textwidth]
			{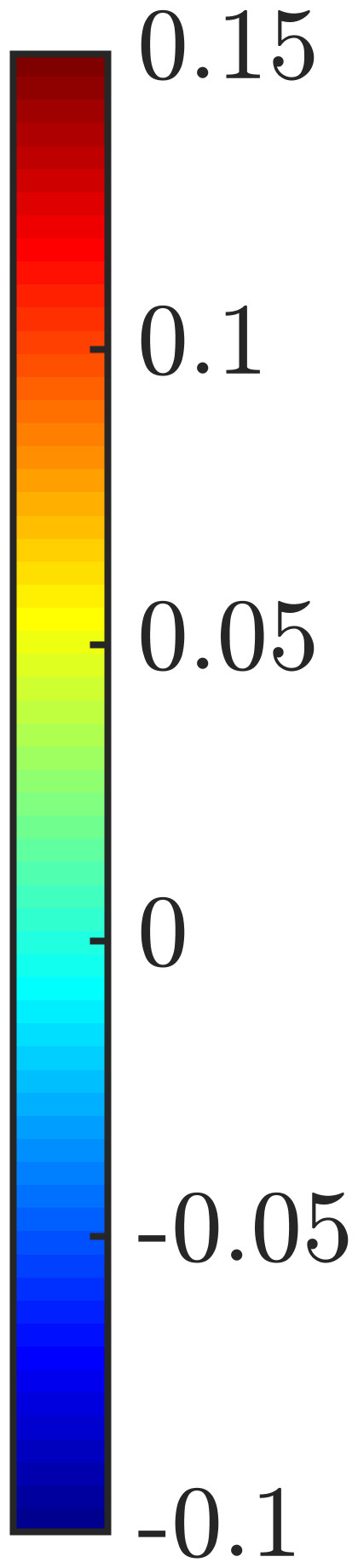}}
		\put(0,0.13){\includegraphics[width=0.6\textwidth]{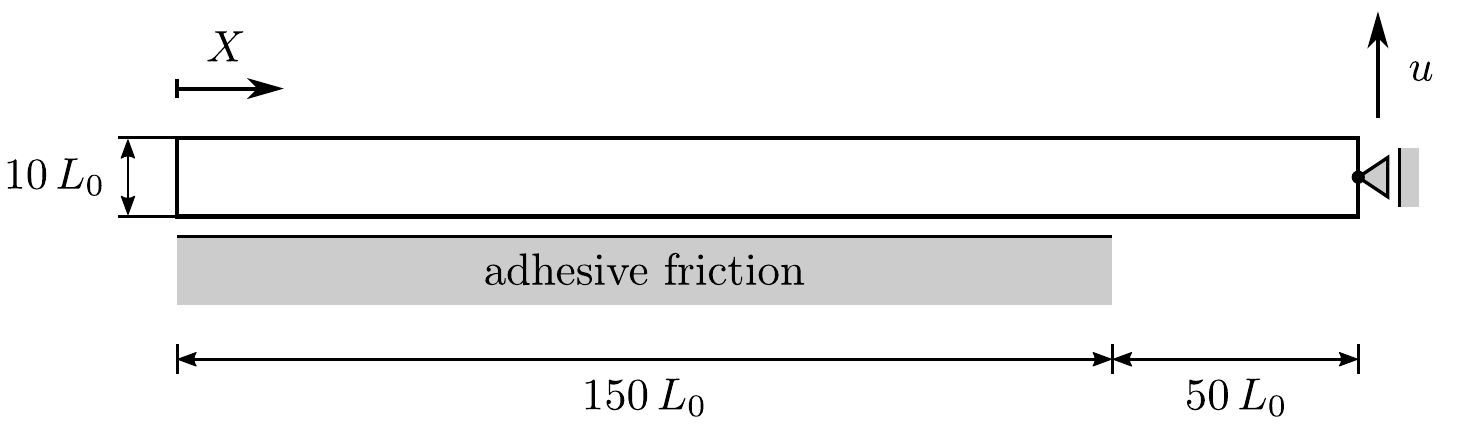}}
		\put(0.843,0.29){\small $120\,L_0$}
		\put(0.843,0.167){\small $60\,L_0$}
		\put(0.843,0.097){\small $30\,L_0$}
	\end{picture}
	\caption{2D strip peeling: Setup (left) and three peeling states for
		model~\EA\ with $\gcut = \gmax$ and $\mu_\EAe = 0.01$ (right); the
		colors show the first invariant of the Cauchy stress, $\tr\,\bsig$,
		divided by~$E$.}
	\label{f:strip:x}
\end{figure}

We first consider a massless, beam-like strip that initially adheres to a flat
and rigid substrate (\cref{f:strip:x}). Along this interface, the normal
traction is initially zero, so that the gap between the strip and the substrate
is equal to the equilibrium distance~($\gn = \gequ$) from \cref{f:Tn}. The
strip is then peeled off the substrate, moving the midpoint of its right
boundary by the vertical displacement~$u$. Such a setup is representative
e.g.~for tape peeling.

To avoid shear locking due to bending, we discretize the bulk of the strip with
$320\,\times\,8$ quadratic (Q2) finite elements. The contact surface is further
enriched by quadratic NURBS shape functions (Q2N2.1 elements)
\citep{corbett14}. In this example, as well as in the following subsection, we
consider the Neo-Hookean material model stated in \cite{bonet97}. For the
geometry, material, and contact parameters see \cref{t:strip:para}. All
results, shown here and in the following, are normalized by Young's modulus~$E$
and the unit length~$L_0$.
\begin{table}[h]
	\centering
	\begin{tabular}{c@{\qquad}c@{\qquad}c@{\qquad}c@{\qquad}c}
		\hline \noalign{\smallskip}
		$E$ & $\nu$ & $\AH$ & $r_0$ & $L_0$ \\
		\noalign{\smallskip} \hline \noalign{\smallskip}
		$2\,\mrG\mrP\mra$ & $0.2$ & $10^{-19}\,\mrJ$ & $0.4\,\mrn\mrm$
			& $1\,\mrn\mrm$ \\
		\noalign{\smallskip} \hline
	\end{tabular}
	\caption{2D strip peeling: Model parameters.}
	\label{t:strip:para}
\end{table}


\subsubsection{Contact tractions}

\begin{figure}[h]
	\centering
	\subfigure[Normal traction for model~\EA.]{
		\unitlength\textwidth
		\includegraphics[width=0.47\textwidth]{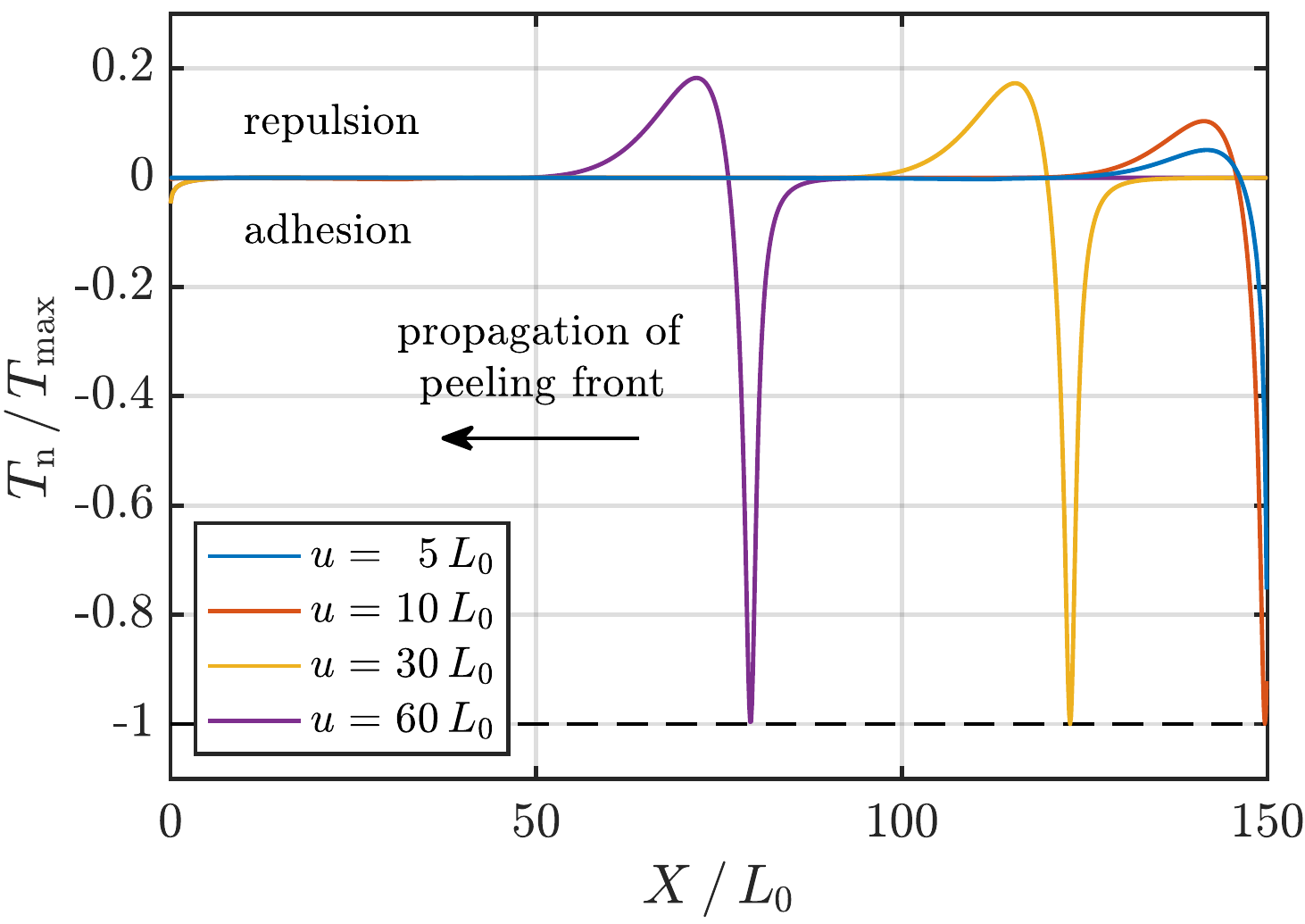}
		\label{f:strip:TnD}
	}\hspace*{1ex}
	\subfigure[Tangential traction for model~\EA.]{
		\includegraphics[width=0.47\textwidth]{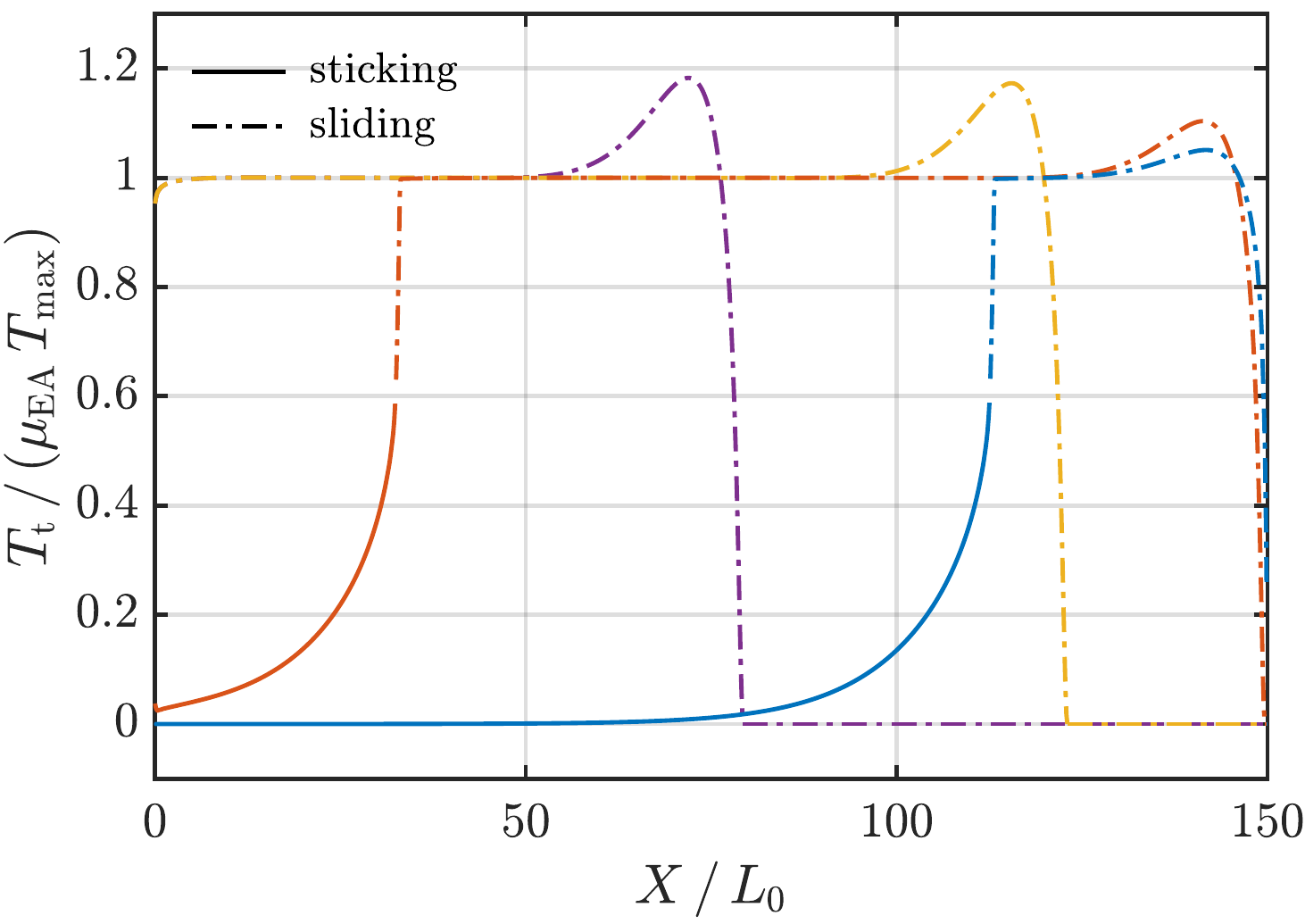}
		\label{f:strip:TtD}
	}
	\caption{2D strip peeling: Traction at the contact area (in dependence of
		the horizontal position~$X$ normalized by $L_0$) for model~\EA\ with
		$\mu_\EAe = 0.01$ and $\gcut = \gmax$; the dash-dotted lines in
		Fig.~(b) indicate sliding.}
	\label{f:strip:TnTt}
\end{figure}

\Cref{f:strip:x} illustrates the peeling of the strip for model~\EA, $\gcut =
\gmax$, and a small friction parameter ($\mu_\EAe = 0.01$).
\Cref{f:strip:TnD,f:strip:TtD} show the normal and tangential contact tractions
directly before and after the onset of peeling ($u = 5\,L_0$ and $10\,L_0$,
respectively), as well as for two of the configurations shown in
\cref{f:strip:x}. As seen in \cref{f:strip:TnD}, for small $u$ (like $5\,L_0$),
adhesion induces large tensile normal tractions at the right boundary of the
adhesive zone. When these tractions reach~$\Tnmax$ (at $u \approx 7\,L_0$),
first detachment occurs, and a peeling front nucleates. This front then
propagates to the left, visible as a sharp peak in \cref{f:strip:TnD}. Some
distance ahead of the peeling front, the normal tractions are repulsive due to
the finite bending stiffness of the strip. Together with the peeling front,
also a sliding front nucleates and propagates to the left (\cref{f:strip:TtD}).
This front marks the boundary between the region still sticking to the
substrate and the region already sliding. Note that a sliding region ahead of
the peeling front is indeed observed in experiments (see e.g.~\cite{newby97}).
When the sliding front has reached the left boundary of the strip, the
(remaining part of the) contact interface is fully sliding. As expected, for
model~\EA\ the tangential traction distribution looks similar to the normal
traction distribution within the sliding region.

When repeating the simulation from \cref{f:strip:TnTt} with model~\DI, a
friction parameter of $\mu_\DIe = \mu_\EAe = 0.01$, and the same cutoff
distance~($\gcut = \gmax$), we find the following behavior (not shown): The
normal tractions are almost identical to those from model~\EA, while the
tangential tractions remain constant at~$\tau_\DIe = \mu_\DIe\,\Tnmax$ within
the entire sliding region. In addition, the sliding front evolves very
similarly in both models.


\subsubsection{Peeling forces} \label{s:strip:forces}

\begin{figure}[ht]
	\centering
	\subfigure[Horizontal force$^*$ vs.~displacement.]{
		\includegraphics[width=0.47\textwidth]{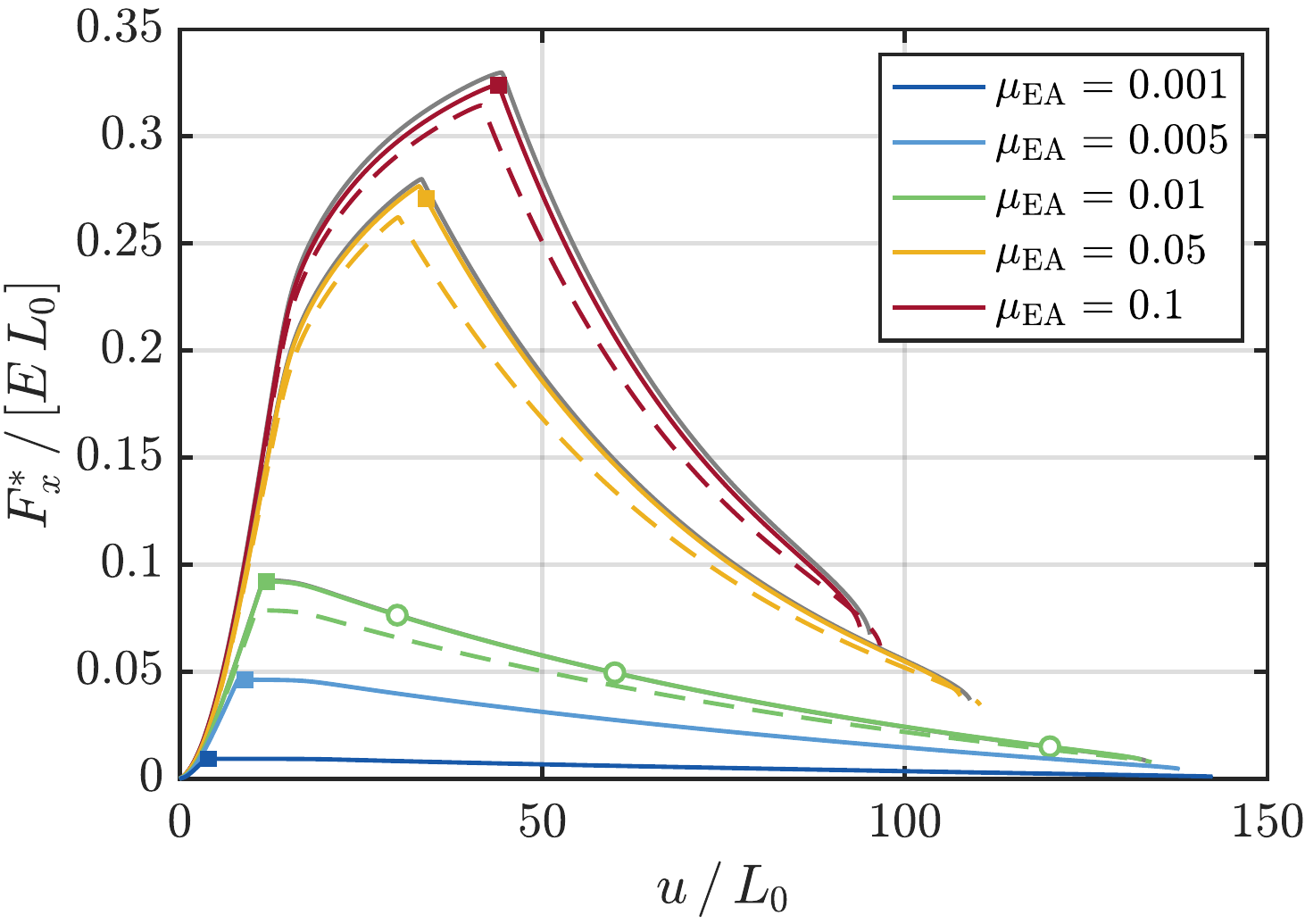}
		\label{f:strip:FDx}
	}\hspace*{1ex}
	\subfigure[Vertical force$^*$ vs.~displacement.]{
		\includegraphics[width=0.47\textwidth]{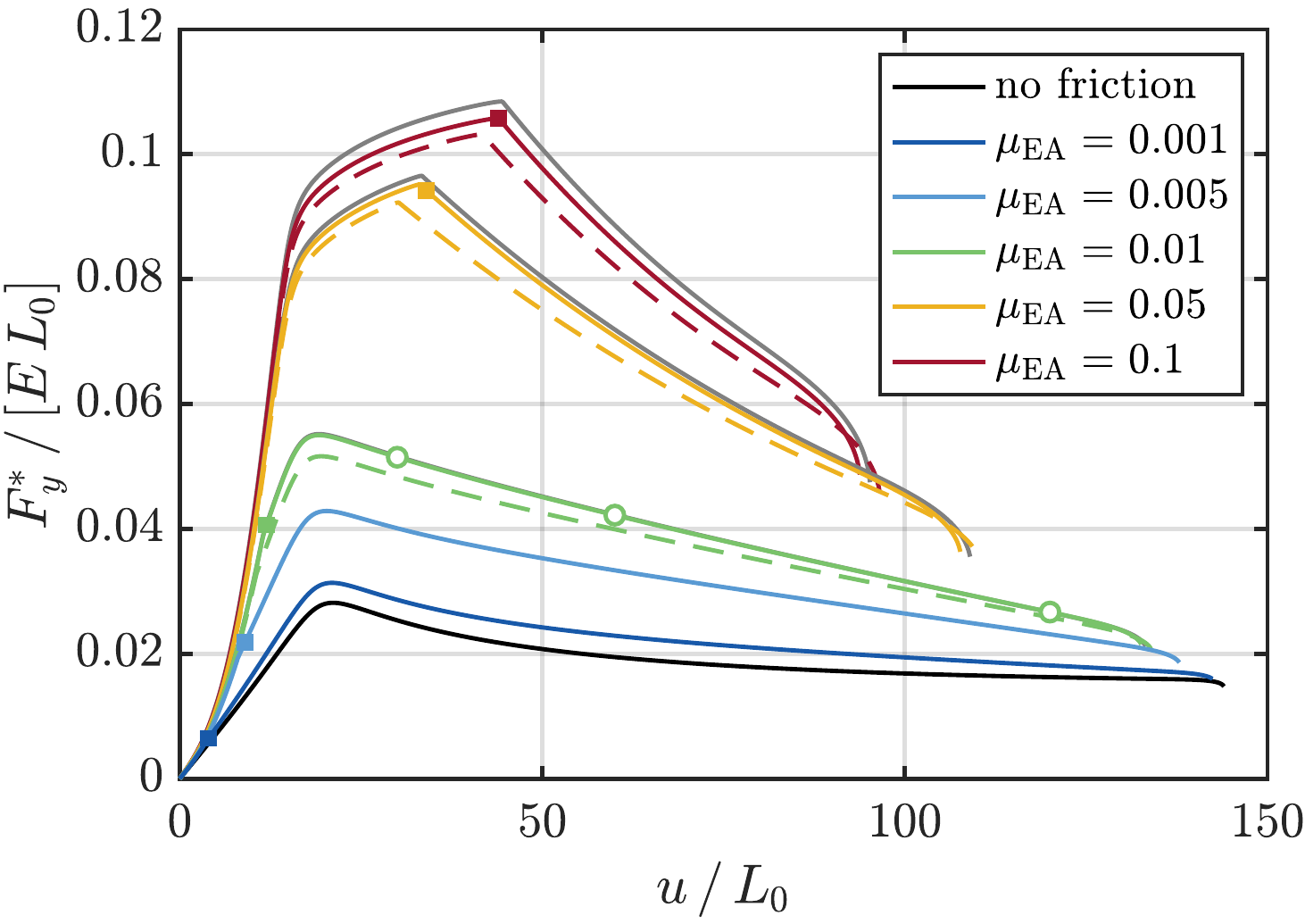}
		\label{f:strip:FDy}
	}
	\subfigure[Peeling and sliding lengths vs.~displacement.]{
		\includegraphics[width=0.47\textwidth]{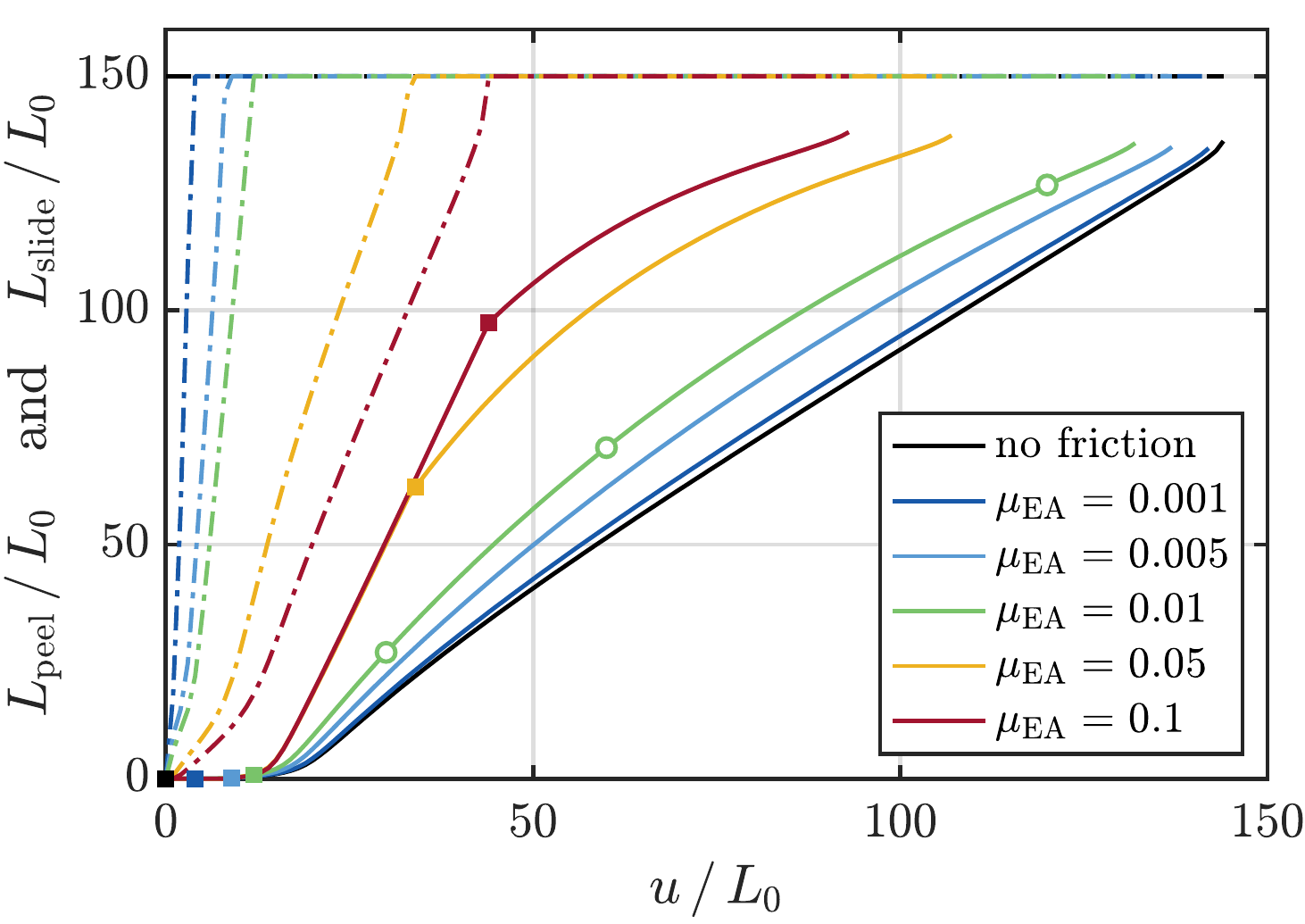}
		\label{f:strip:Lpeelslide}
	}\hspace*{1ex}
	\subfigure[Fracture energy$^*$ vs.~peeling length.]{
		\includegraphics[width=0.47\textwidth]{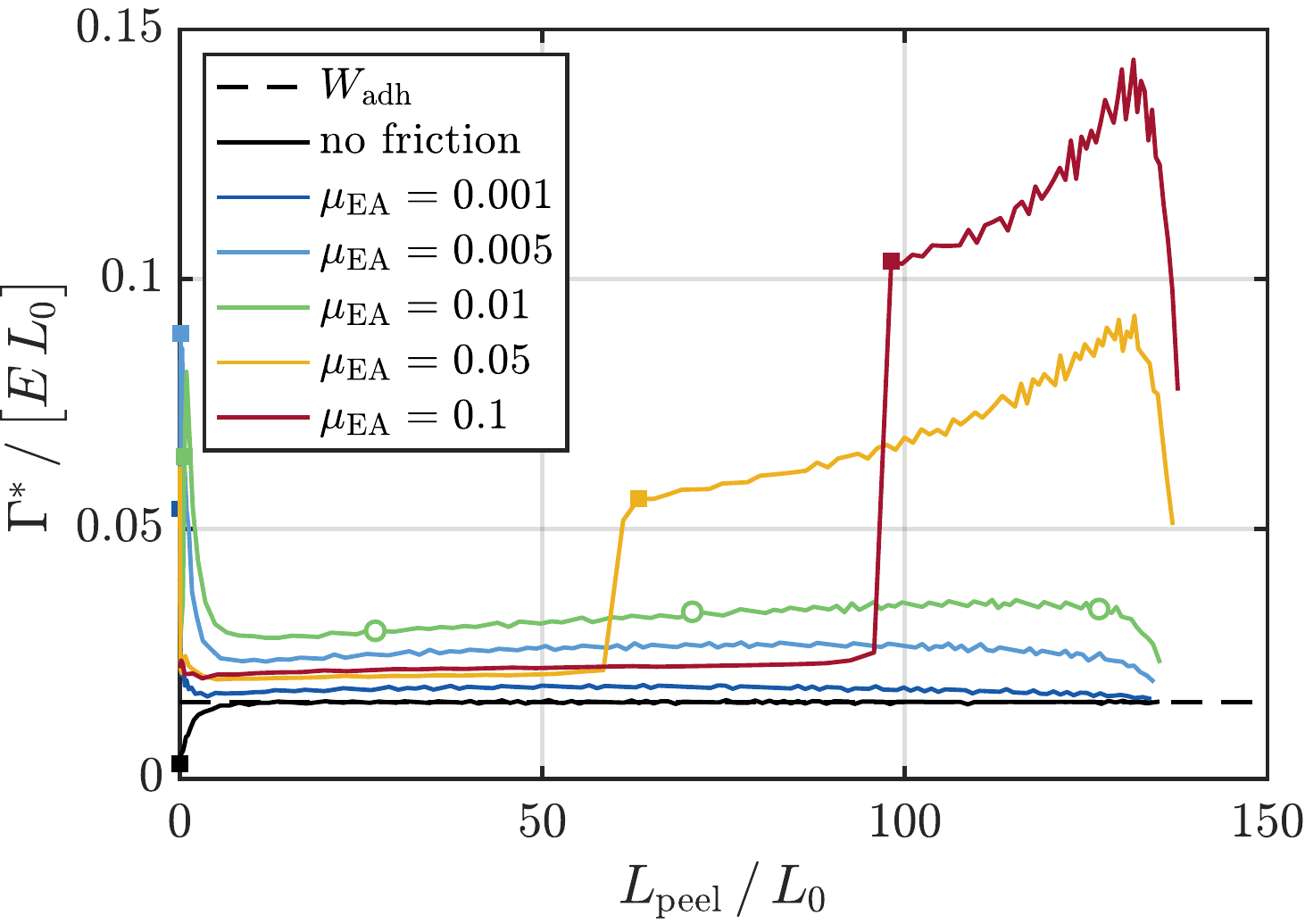}
		\label{f:strip:Gamma}
	}
	\caption{2D strip peeling: (a) \& (b)~Horizontal and vertical forces$^*$
		($^*$: per out-of-plane width); solid colored lines: model~\EA\ 
		with $\gcut = \gmax$; dashed colored lines: model~\EA\ with $\gcut
		= (\gequ + \gmax)/2$; solid gray lines: model~\DI\ with $\mu_\DIe
		= \mu_\EAe$ and $\gcut = \gmax$; (c)~nominal peeling length~$\Lpeel$
		(solid) and nominal sliding length~$\Lslide$ (dash-dotted);
		(d)~fracture energy$^*$; the white dots in Figs.~(a) -- (d) mark the
		configurations plotted in \cref{f:strip:x}; the filled squares indicate
		the onset of full sliding rounded to next integer value of~$u$.}
\end{figure}

Let us now examine the horizontal and vertical pull-off forces for different
friction parameters, focusing on model~\EA\ (\cref{f:strip:FDx,f:strip:FDy}).

With increasing friction parameter, large parts of the strip initially remain
sticking to the substrate, leading to both large horizontal and vertical
forces. As illustrated in \cref{f:strip:FDx,f:strip:FDy}, the force curves
slightly change for a different value for~$\gcut$ (sufficiently larger
than~$\gequ$, see \cref{s:model:EA}). In analogy to the contact tractions,
using model~\DI\ instead (with $\mu_\DIe = \mu_\EAe$ and $\gcut = \gmax$) does
not cause any significant changes in the forces. The good agreement between
models~\EA\ and \DI\ is generally expected if (i)~the surface stretches are
small (such that~$\Jck,\,\Jcl \approx 1$, and thus $\bttk \approx \bTtk$),
(ii)~the contacting interfaces are mainly separated by the equilibrium
distance~$\gequ$, and (iii)~if the model parameters are chosen as
\begin{equation}
	\gequ < \gcut \le \gmax, \qquad
	\mu_\DIe := \mu_\EAe \, \frac{|\Tn(\gcut)|}{\Tnmax}. \label{e:strip:EJeq}
\end{equation}

Note that models~\EA\ and \DI\ will behave differently from each other if major
parts of the contact area are in tension or compression, or if the surface
stretch has a strong influence. This is discussed in the following two
examples. Further deviation may be caused by the regularization parameter
$k_\DIe$ of \cref{e:tslideDI}. For model~\DI\ the regularized traction law in
\cref{e:tslideDI} is $C^1$-continuous for arbitrary cutoff distances~$\gcut$.
In contrast, \cref{e:TslideEA} for model~\EA\ has a kink at $\gcut$ if $\gcut
\neq \gmax$ (see e.g.~\cref{f:TslideEA:s05}). Nevertheless, due to our active
set and freezing strategies (\cref{s:comput:actset}), our algorithm works
robustly for all investigated cases.

To understand both the shape of the force curves in
\cref{f:strip:FDx,f:strip:FDy} and the evolution of their maximum when varying
the friction parameter~$\mu_\EAe$, we next examine the evolution of the peeling
and sliding fronts during peeling. \Cref{f:strip:Lpeelslide} hence shows the
nominal length~$\Lpeel$ of the part that is peeled off the substrate already.
It further illustrates the nominal sliding length, $\Lslide$, which we define
as the initial contact length minus the nominal length of the currently
sticking part. The sliding front starts propagating directly after imposing a
small, vertical displacement. Larger friction parameters result in later
propagation, because the local tangential tractions must be larger to overcome
the sliding threshold,~$\Ttmax$ (\cref{e:TslideEA}). In contrast, the peeling
front nucleates (i.e., $\Lpeel > 0$) only after a finite vertical displacement,
$u \approx 20\,L_0$, where the precise value slightly varies with~$\mu_\EAe$
(\cref{f:strip:FDy,f:strip:Lpeelslide}). At the nucleation of the peeling
front, the vertical force decreases afterwards for small~$\mu_\EAe$, but
continues to increase (although with a smaller slope) for the two largest
values of~$\mu_\EAe$.

In the range of~$\mu_\EAe$ considered here, we can distinguish between two
types of evolution for~$\Lpeel$: (i)~For~$\mu_\EAe \lesssim 0.02$, the sliding
front reaches the boundary of the strip ($\Lslide = 150\,L_0$) before the
peeling front nucleates ($\Lpeel > 0$). In this case, $\Lpeel$ increases almost
linearly, and the strip detaches slightly earlier for increasing~$\mu_\EAe$.
(ii)~For~$\mu_\EAe \gtrsim 0.02$, the sliding length has not reached the limit
of~$150\,L_0$ before the peeling front nucleates. In this case, $\Lpeel$
initially evolves independently from~$\mu_\EAe$. Then, when the entire strip
starts sliding, $\Lpeel$ suddenly propagates with a different slope, which
abruptly causes the horizontal and vertical forces to decrease again
(\cref{f:strip:FDx,f:strip:FDy}).

In the field of adhesive peeling, one is often interested in the fracture
energy that is required to separate a unit area from the interface
\citep{creton16}. To evaluate this energy, we first compute the difference
between the external work done during peeling and the internal energy due to
elastic deformation ($^*$ per out-of-plane width): $\Delta \Pi^* = \Pi^*_\extn
- \Pi^*_\intn$. The fracture energy $\Gamma^*$, shown in \cref{f:strip:Gamma},
then corresponds to the derivative of $\Delta\Pi^*$ with respect to the peeling
length, $\Gamma^* = \partial(\Delta\Pi^*)/\partial\Lpeel$. Again, the behavior
is found to be different for friction parameters smaller and larger than
about~$0.02$. Below this value, $\Gamma^*$ is essentially independent from the
peeling length. It is minimal for the frictionless case, for which $\Gamma^*$
approaches~$W_\adh$ from \cref{f:Tn}. This is expected because there is no
dissipation due to friction, so that the fracture energy corresponds to the
work of adhesion per unit area. For increasing values of~$\mu_\EAe$, $\Gamma^*$
then increases, because an additional energy is necessary to overcome
frictional dissipation within the sliding region, as discussed in \cite{lu07}.
For friction parameters beyond $0.02$, the fracture energy strongly increases
(by a factor of~2 to 5) at full sliding ($\Lslide = 150\,L_0$). Afterwards,
large sliding distances are associated with each increment of~$u$, inducing a
much larger frictional dissipation, and thus fracture energy.

Our results suggest that the evolution of the peeling forces and the fracture
energy can significantly depend on both the friction parameter and the finite
size of the strip. For the two highest friction parameters, the forces evolve
very similarly at the beginning, because the sliding region is still small
compared to the initial contact length. This behavior is more related to the
behavior expected for peeling of adhesive tapes, for which the friction
parameter is probably considerably larger. Interestingly, our results in
\cref{f:strip:Gamma} predict that for small peeling lengths (about~$20 -
60\,L_0$), the fracture energy reaches its maximum for a finite value of
$\mu_\EAe \approx 0.02$. Such a result is potentially of practical interest for
the design of optimal adhesives. This could also be tested experimentally,
although we anticipate that the viscoelastic nature of real adhesives, not
accounted for here, may affect the fracture behavior significantly.


\subsection{2D contact of two deformable cylinders} \label{s:cyls}

In this second example, we consider friction between two Hertzian solids. This
setup is particularly important for the field of tribology, because such solids
are frequently used to represent either micro-contact within rough surfaces or
single contact between smooth solids. Although Hertzian geometries have been
investigated in numerous studies (see \cite{vakis18}), their combination with
adhesion, friction, finite deformations, and hyperelasticity (as done here) is
much rarer.

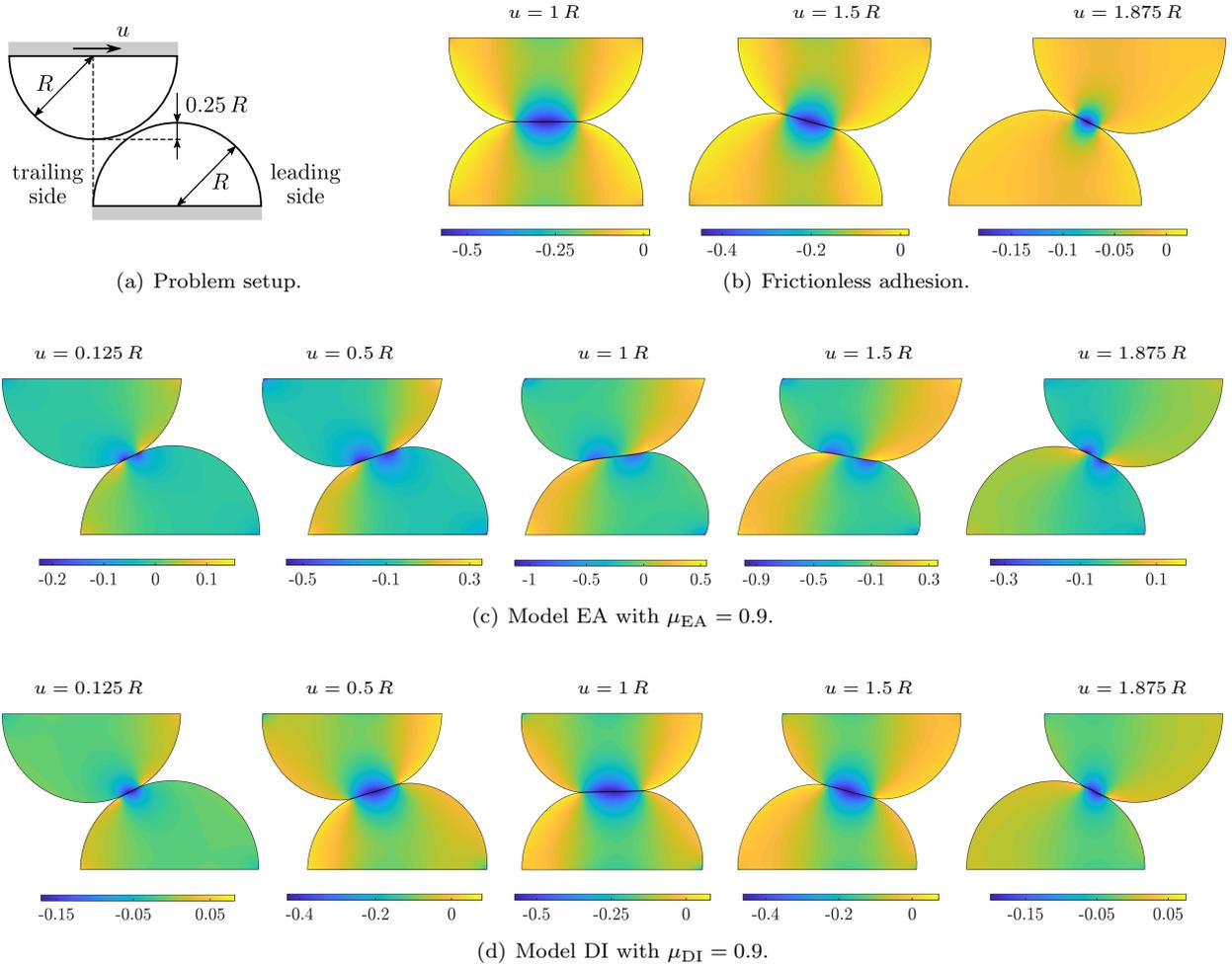
\begin{figure}[ht]
	\unitlength\textwidth
	\subfigure[Problem setup.]{
		\begin{picture}(0.33,0.22)
			\put(0,0.025){\includegraphics[width=0.28\textwidth]
				{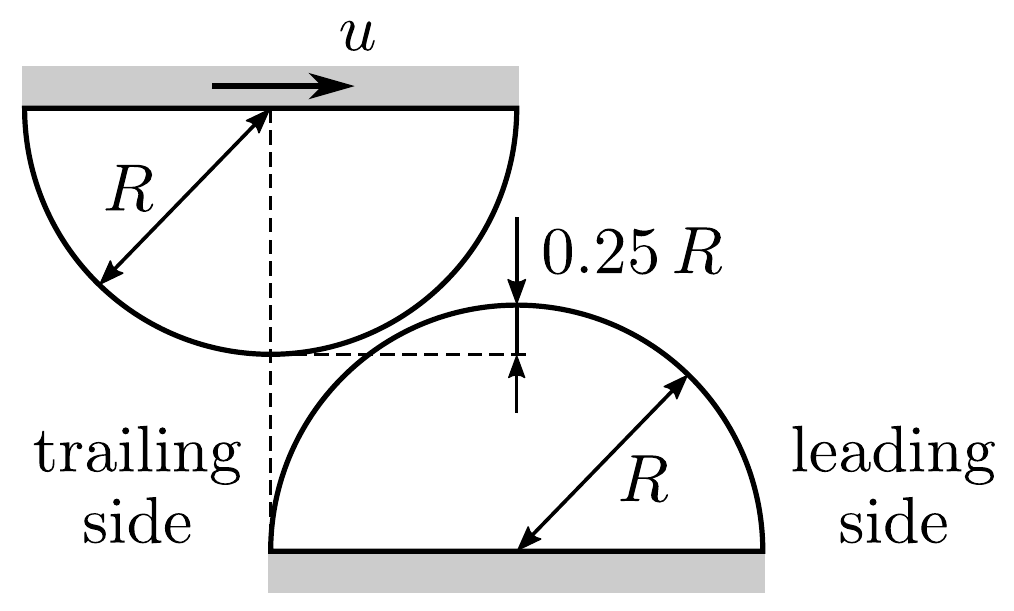}}
		\end{picture}
		\label{f:spheres:setup}
	}
	\subfigure[Frictionless adhesion.]{
		\begin{picture}(0.65,0.22)
			\put(0.005,0.04){\includegraphics[height=0.14\textwidth]
				{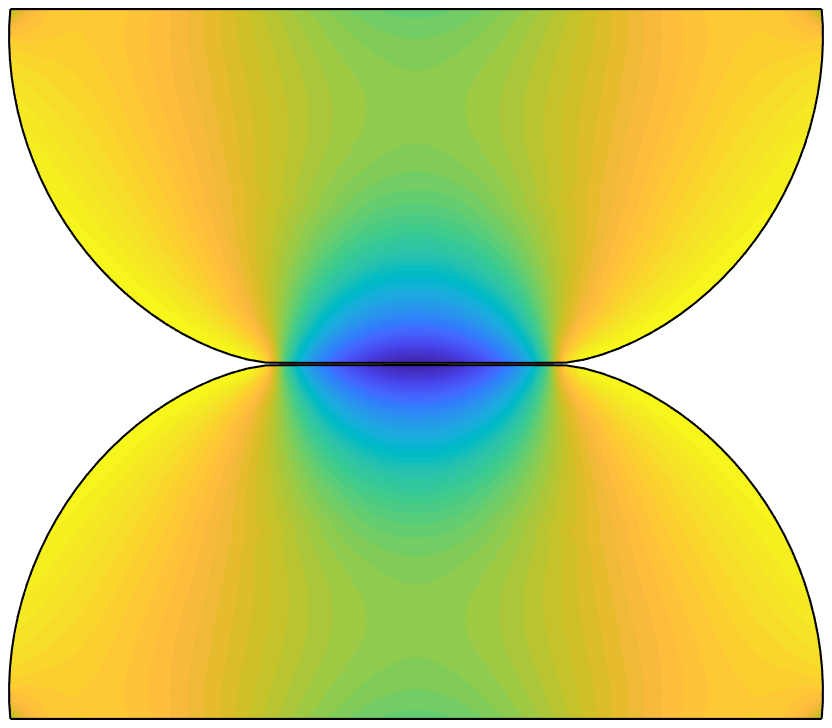}}
			\put(0.2,0.04){\includegraphics[height=0.14\textwidth]
				{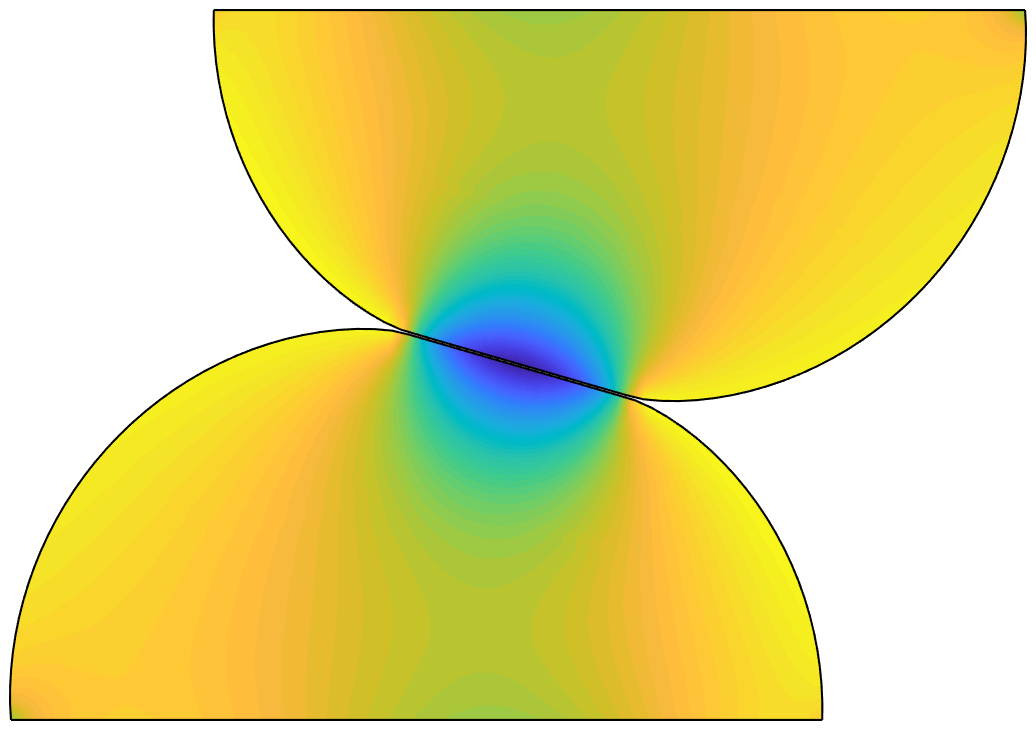}}
			\put(0.41,0.04){\includegraphics[height=0.14\textwidth]
				{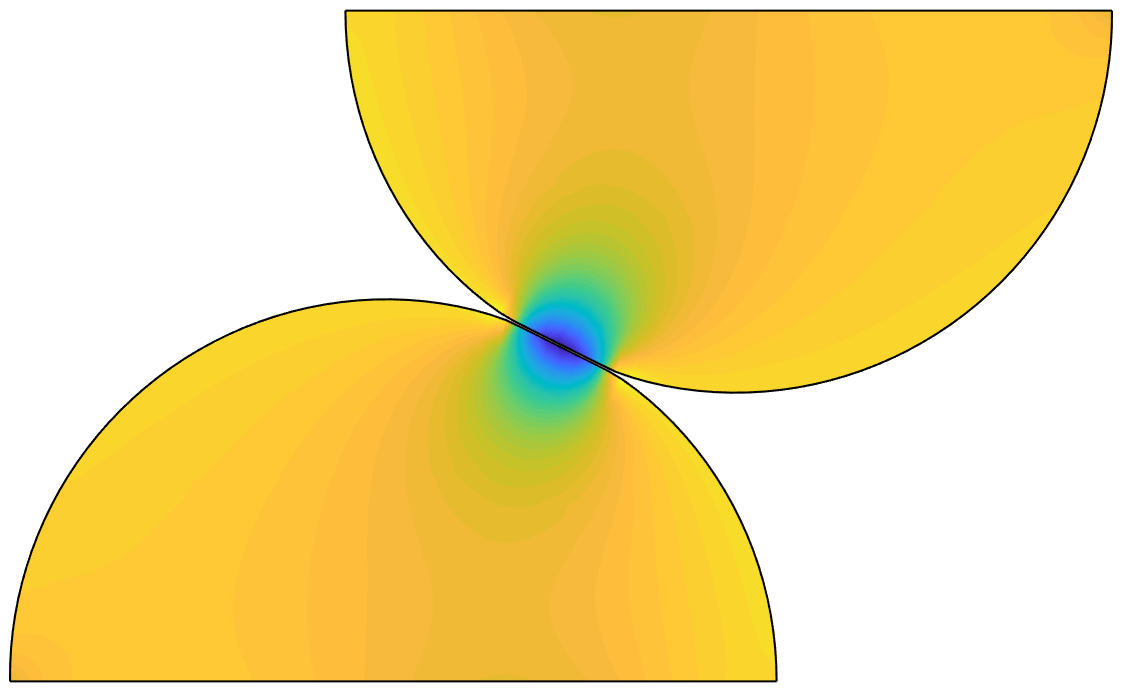}}
			\put(0,0){\includegraphics[width=0.17\textwidth]
				{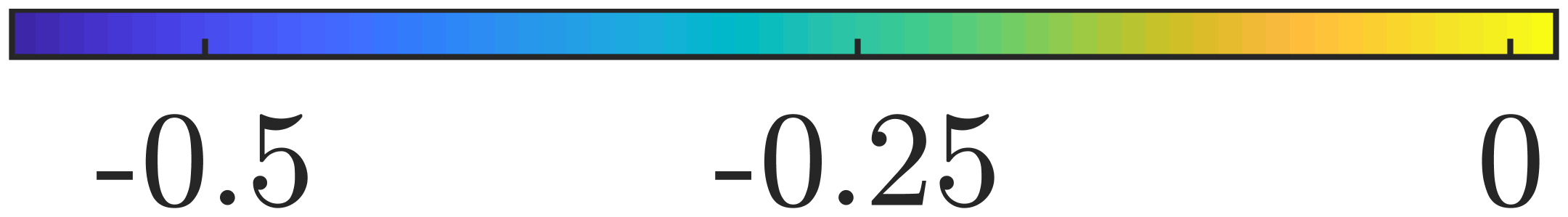}}
			\put(0.21,0){\includegraphics[width=0.17\textwidth]
				{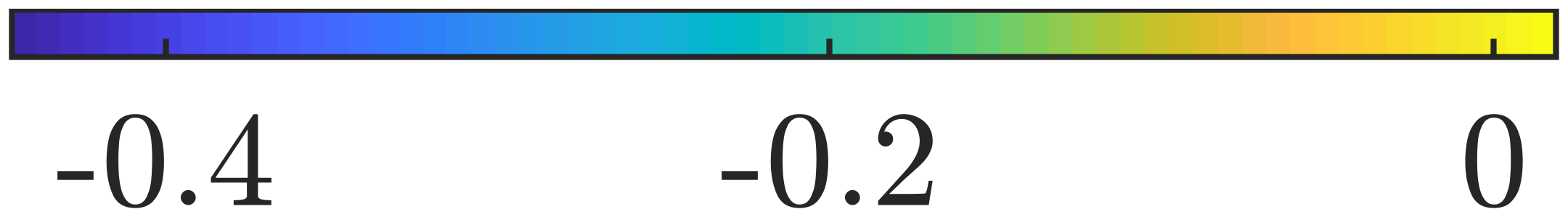}}
			\put(0.435,0){\includegraphics[width=0.17\textwidth]
				{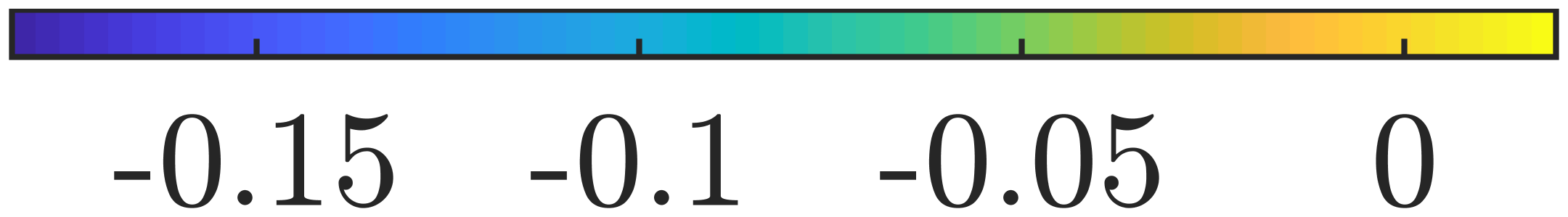}}
			\put(0.055,0.195){\scriptsize $u = 1\,R$}
			\put(0.285,0.195){\scriptsize $u = 1.5\,R$}
			\put(0.512,0.195){\scriptsize $u = 1.875\,R$}
		\end{picture}
		\label{f:spheres:xn0}
	}
	\subfigure[Model~\EA\ with $\mu_\EAe = 0.9$.]{
		\begin{picture}(1,0.22)
			\put(0,0.045){\includegraphics[height=0.13\textwidth]
				{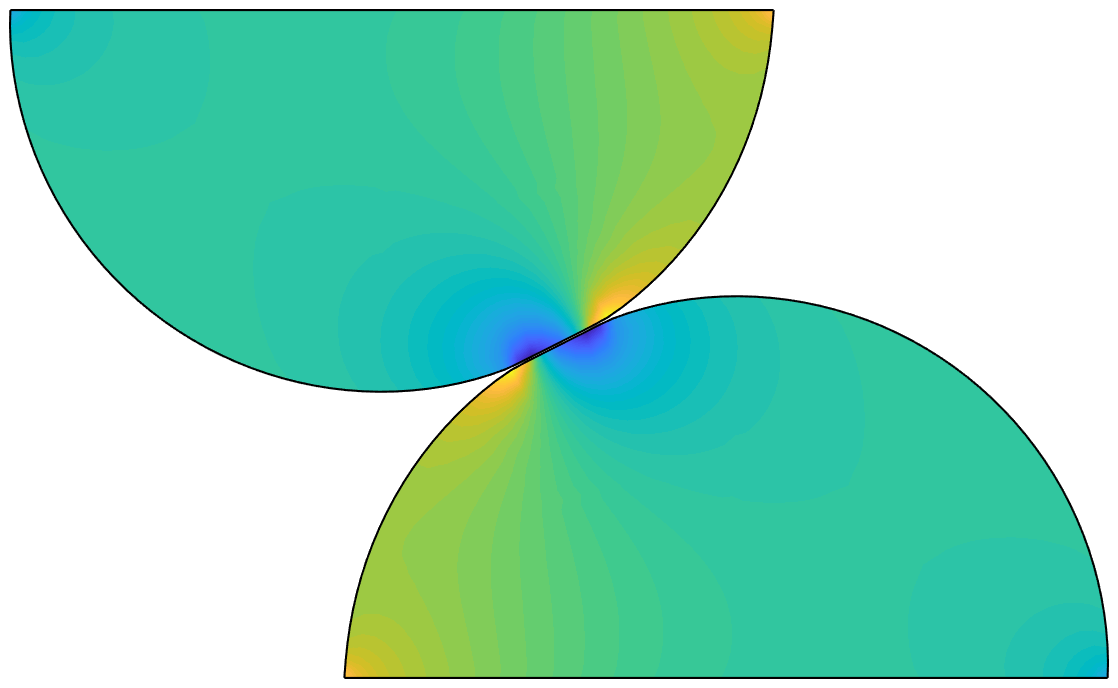}}
			\put(0.21,0.045){\includegraphics[height=0.13\textwidth]
				{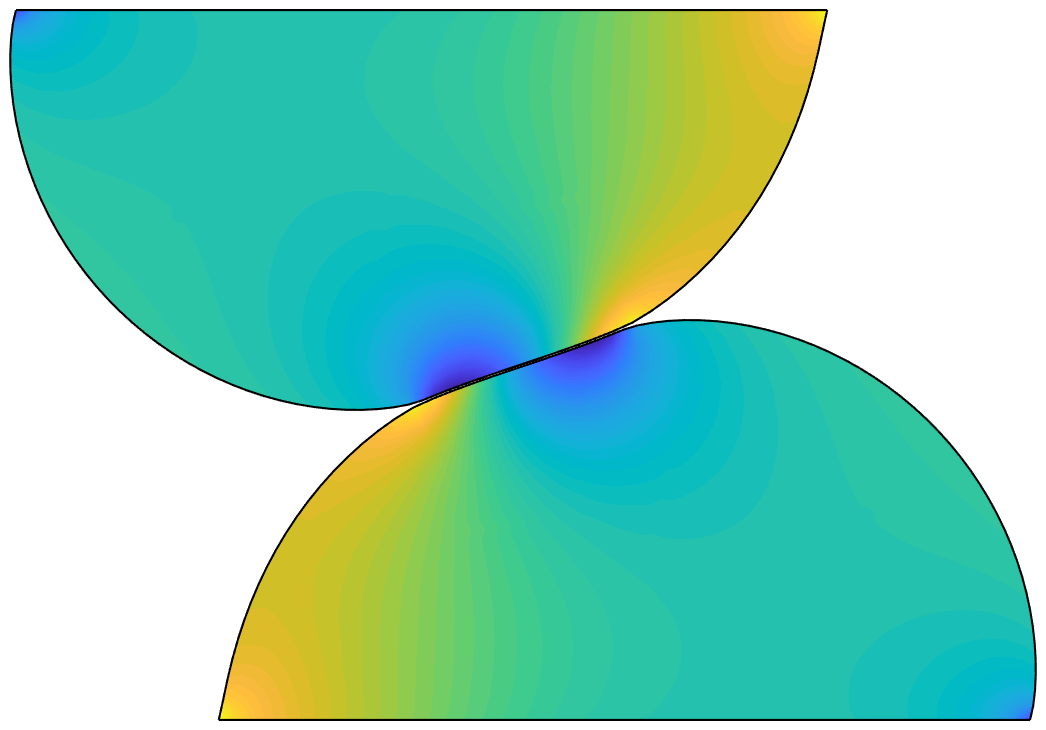}}
			\put(0.42,0.045){\includegraphics[height=0.13\textwidth]
				{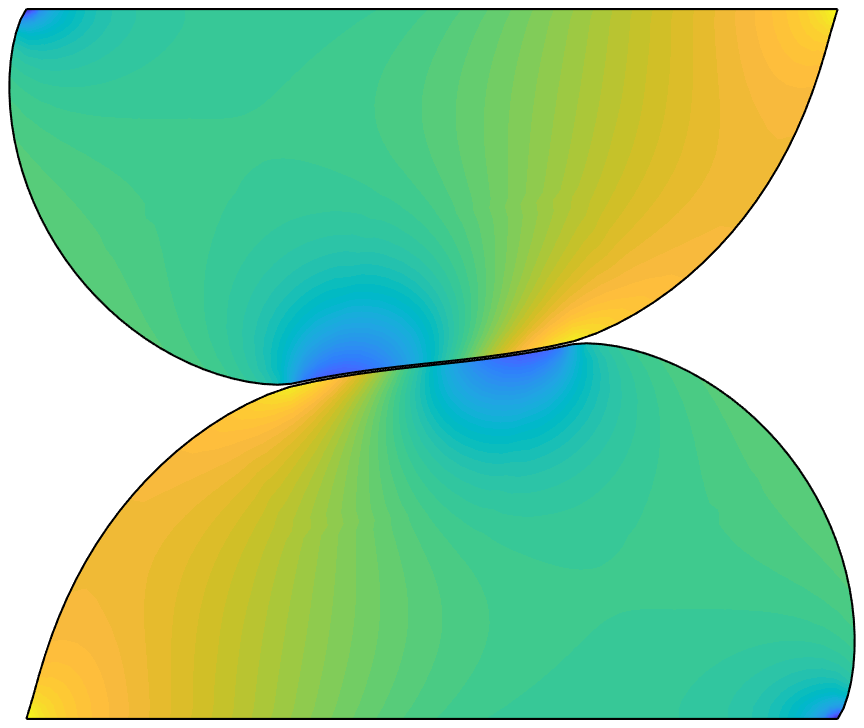}}
			\put(0.595,0.045){\includegraphics[height=0.13\textwidth]
				{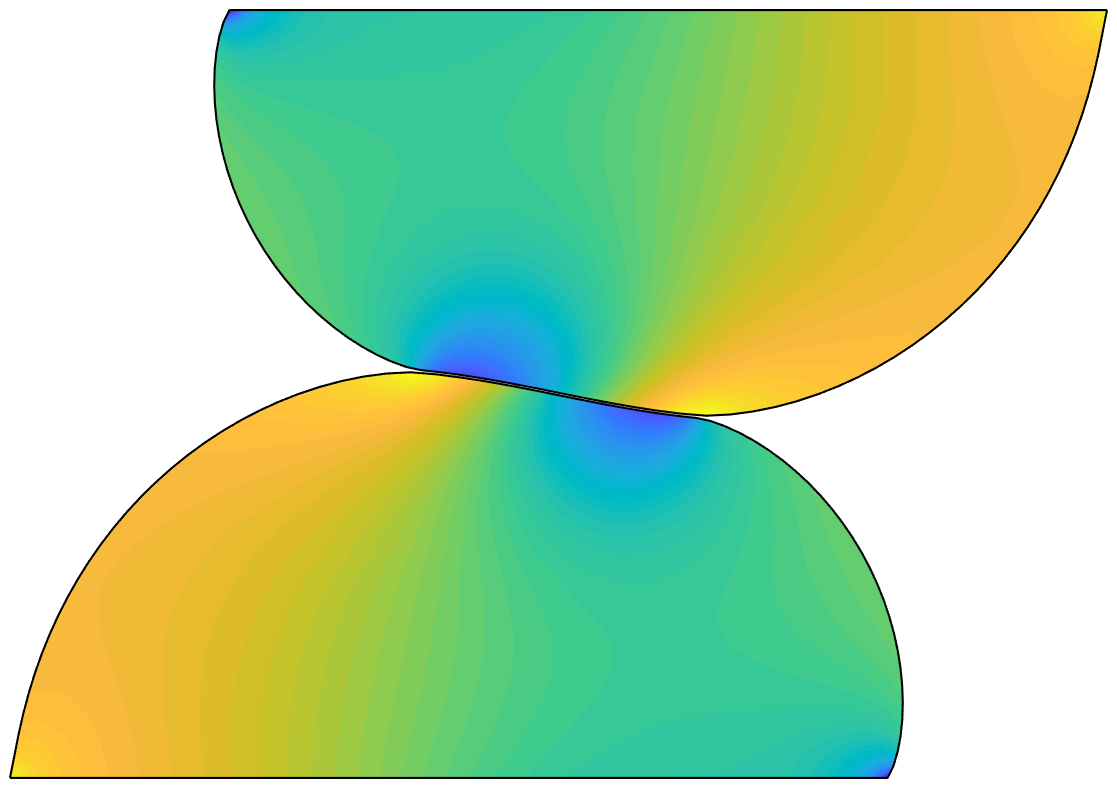}}
			\put(0.78,0.045){\includegraphics[height=0.13\textwidth]
				{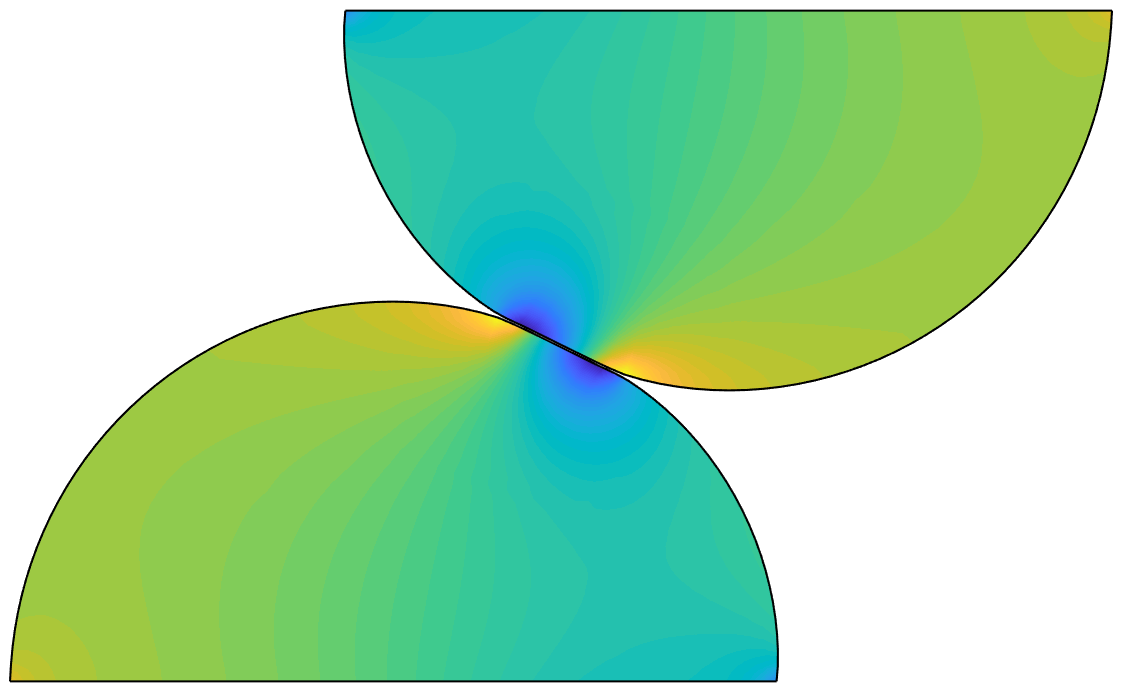}}
			\put(0.03,0.005){\includegraphics[width=0.16\textwidth]
				{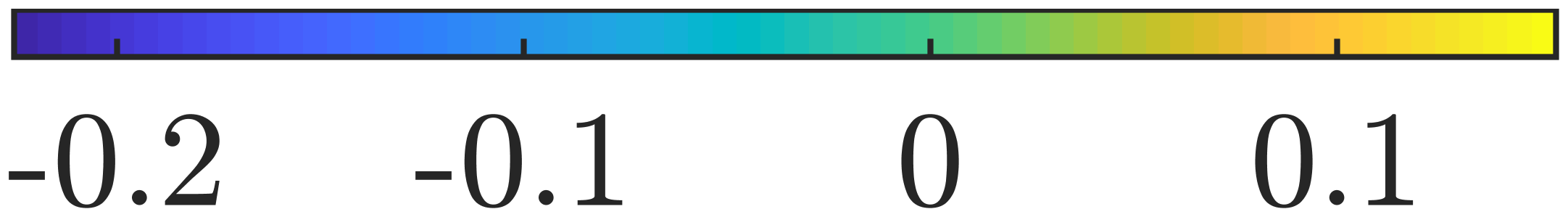}}
			\put(0.23,0.005){\includegraphics[width=0.16\textwidth]
				{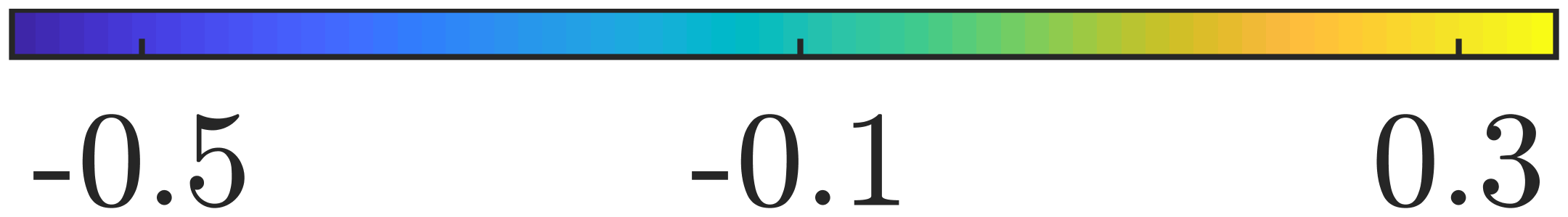}}
			\put(0.415,0.005){\includegraphics[width=0.16\textwidth]
				{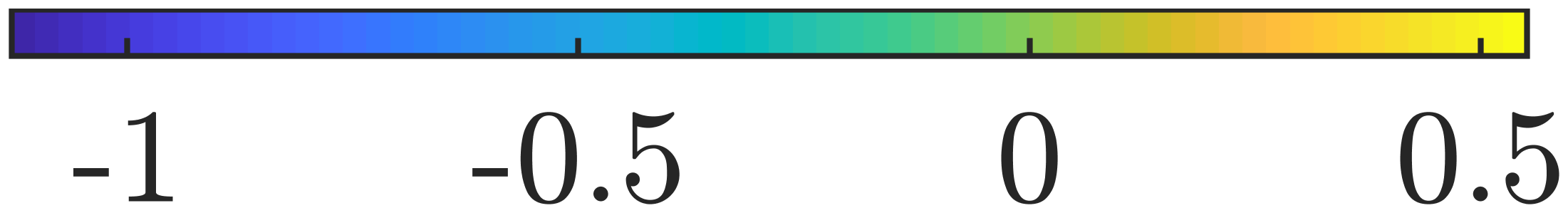}}
			\put(0.60,0.005){\includegraphics[width=0.16\textwidth]
				{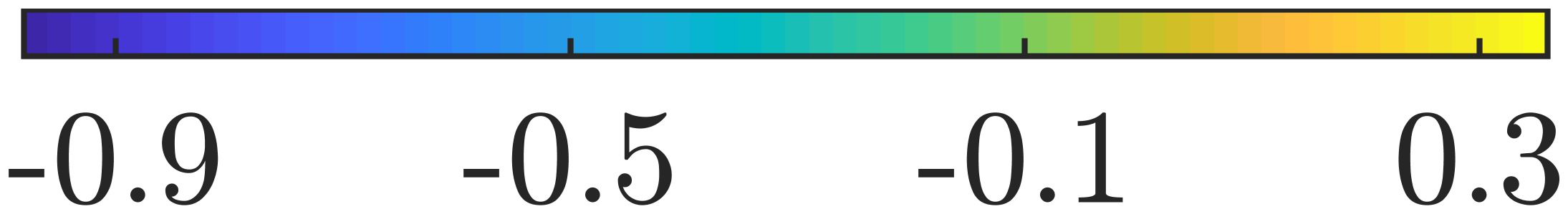}}
			\put(0.80,0.005){\includegraphics[width=0.16\textwidth]
				{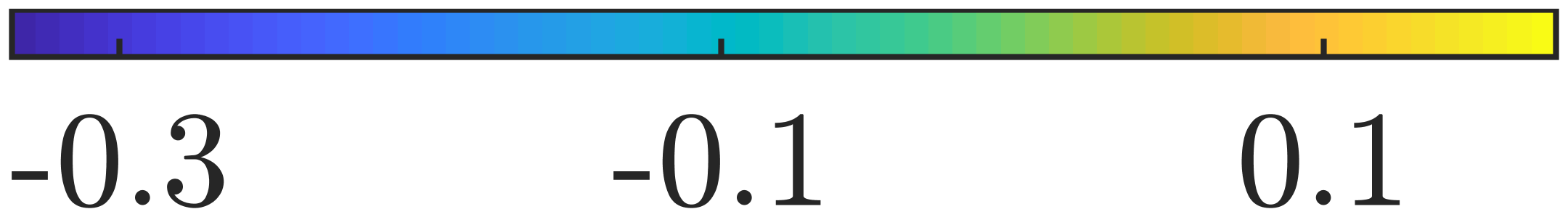}}
			\put(0.027,0.19){\scriptsize $u = 0.125\,R$}
			\put(0.247,0.19){\scriptsize $u = 0.5\,R$}
			\put(0.467,0.19){\scriptsize $u = 1\,R$}
			\put(0.667,0.19){\scriptsize $u = 1.5\,R$}
			\put(0.872,0.19){\scriptsize $u = 1.875\,R$}
		\end{picture}
		\label{f:spheres:xnD}
	}
	\subfigure[Model~\DI\ with $\mu_\DIe = 0.9$.]{
		\begin{picture}(1,0.22)
			\put(0,0.045){\includegraphics[height=0.13\textwidth]
				{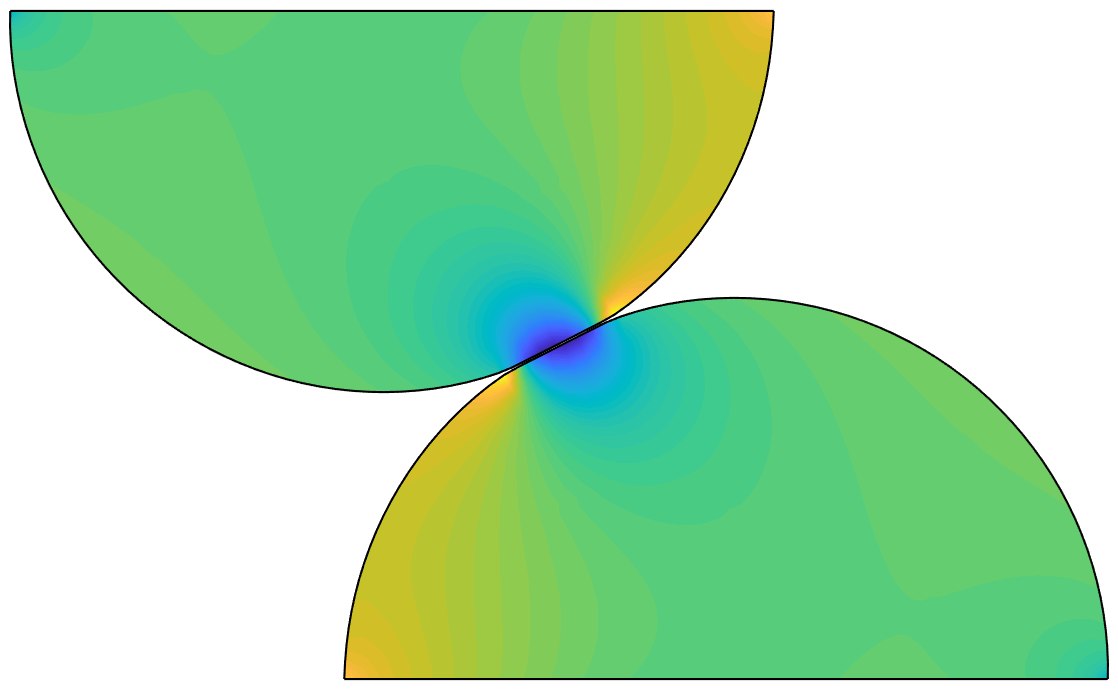}}
			\put(0.21,0.045){\includegraphics[height=0.13\textwidth]
				{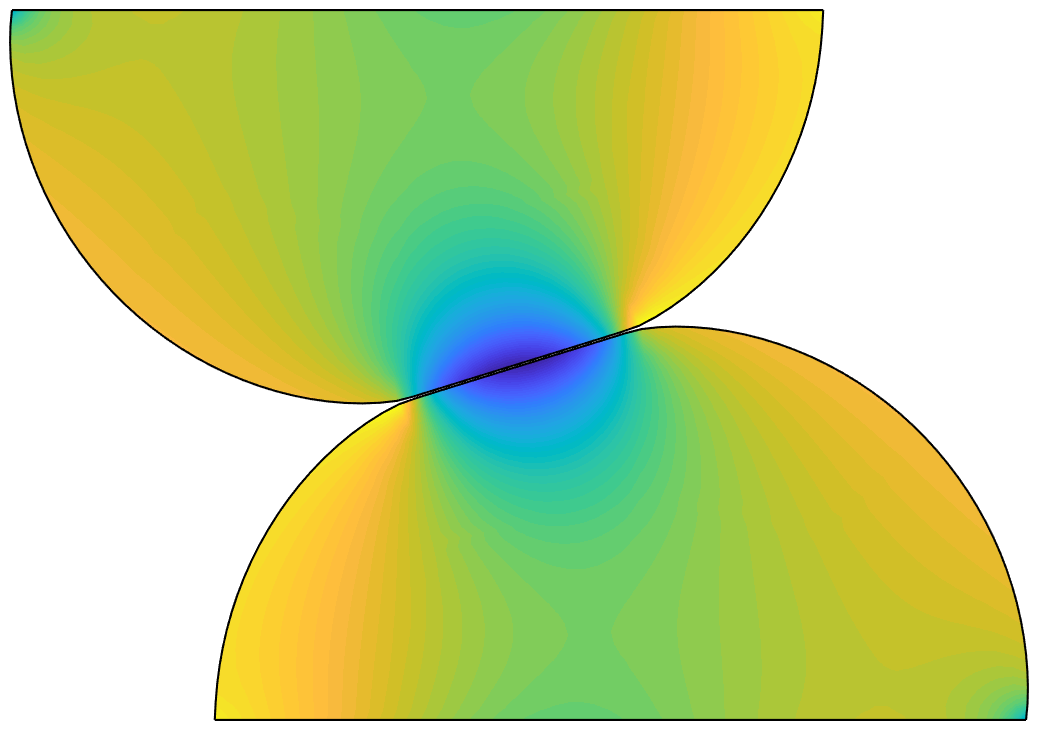}}
			\put(0.42,0.045){\includegraphics[height=0.13\textwidth]
				{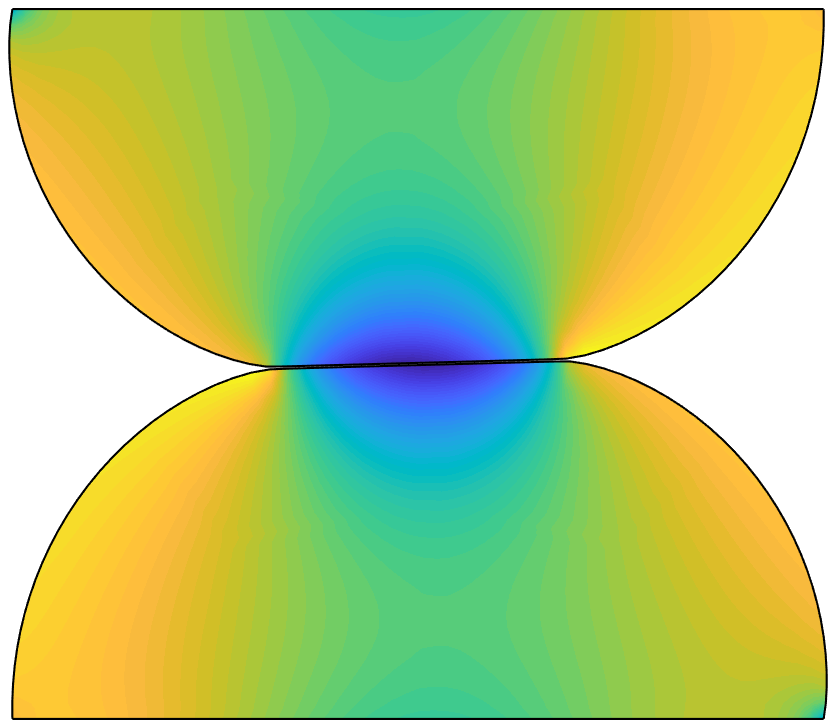}}
			\put(0.595,0.045){\includegraphics[height=0.13\textwidth]
				{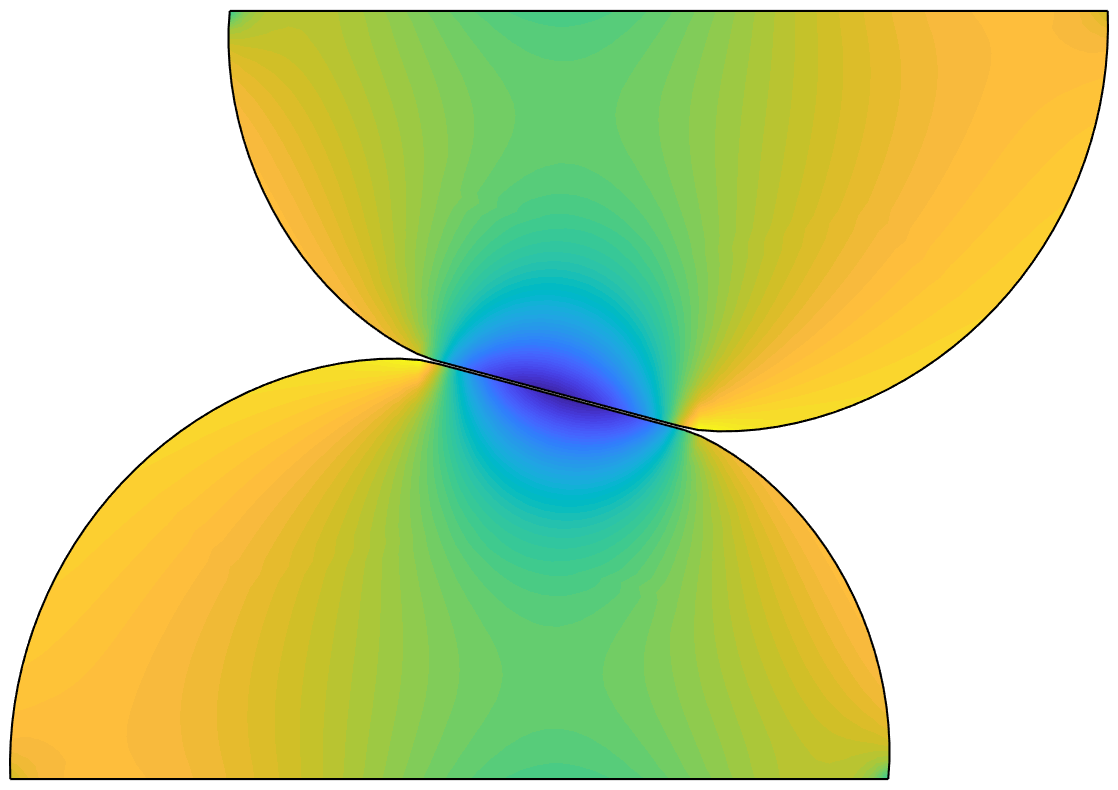}}
			\put(0.78,0.045){\includegraphics[height=0.13\textwidth]
				{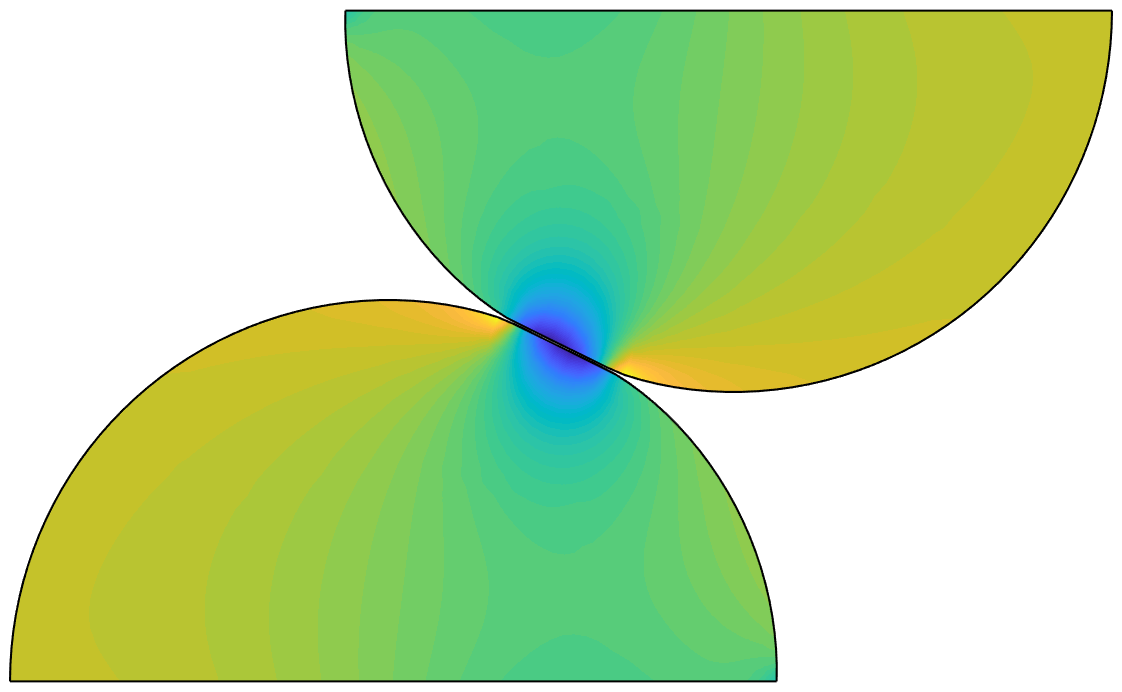}}
			\put(0.03,0.005){\includegraphics[width=0.16\textwidth]
				{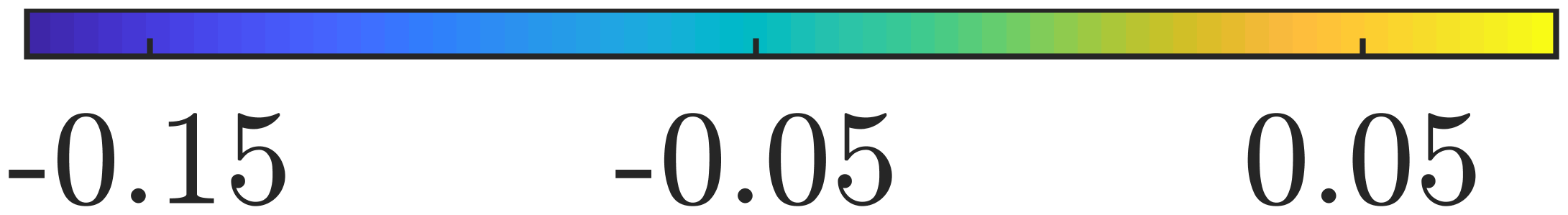}}
			\put(0.23,0.005){\includegraphics[width=0.16\textwidth]
				{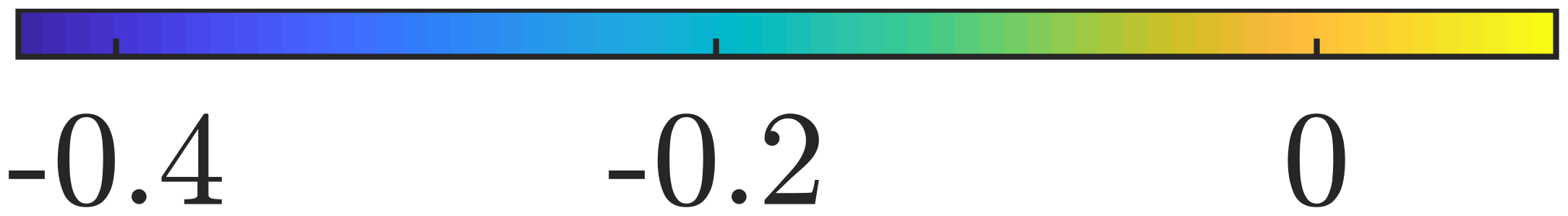}}
			\put(0.415,0.005){\includegraphics[width=0.16\textwidth]
				{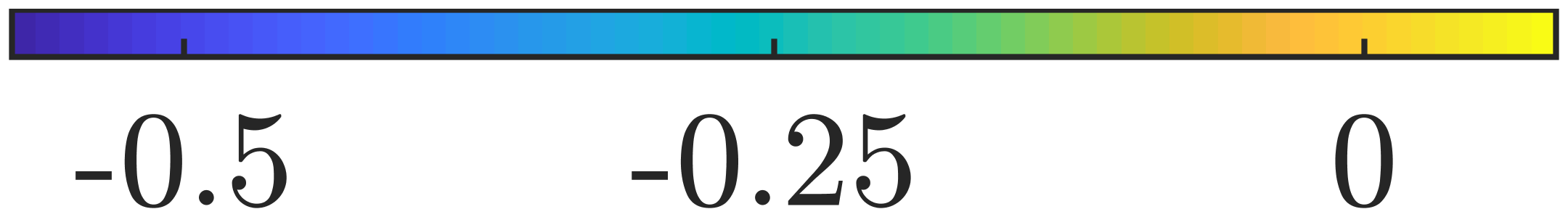}}
			\put(0.60,0.005){\includegraphics[width=0.16\textwidth]
				{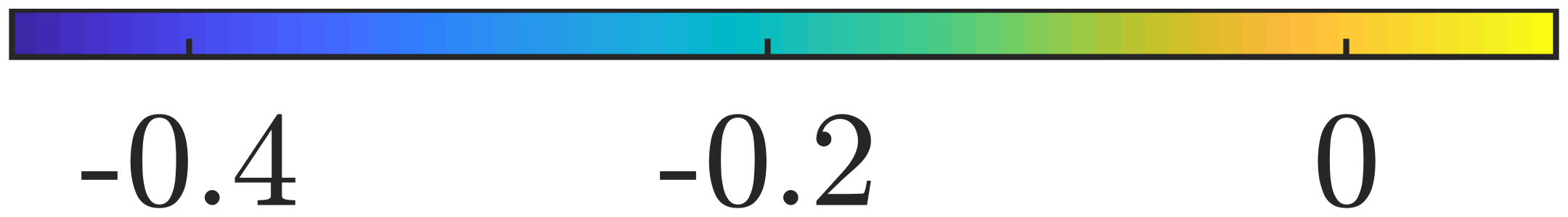}}
			\put(0.80,0.005){\includegraphics[width=0.16\textwidth]
				{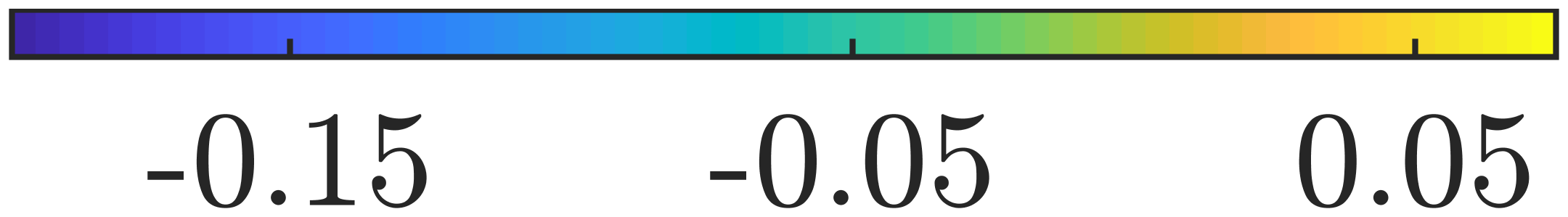}}
			\put(0.027,0.19){\scriptsize $u = 0.125\,R$}
			\put(0.247,0.19){\scriptsize $u = 0.5\,R$}
			\put(0.467,0.19){\scriptsize $u = 1\,R$}
			\put(0.667,0.19){\scriptsize $u = 1.5\,R$}
			\put(0.872,0.19){\scriptsize $u = 1.875\,R$}
		\end{picture}
		\label{f:spheres:xnJ}
	}
	\caption{2D contact of two deformable cylinders: (a)~Setup; (b) --
		(d)~deformation during sliding for (b)~frictionless adhesion or (c) \&
		(d) models~\EA\ and \DI\ with $\gcut = \gmax$, respectively; note that
		the scaling of the colors (showing $\tr\,\bsig\,/\,E$) is adjusted for
		each deformation state.}
	\label{f:spheres}
\end{figure}

We consider two identical, deformable cylinders sliding along each other in
plane strain (see \cref{f:spheres:setup} and \cref{t:spheres:para}). Their
vertical positions are fixed such that their nominal overlap is $25\,\%$ of
their radius,~$R$. We then impose a horizontal displacement~$u$ from~$0$
to~$2\,R$. For each cylinder we use approximately $1,500$ elements and Hermite
enrichment \citep{sauer11ijnme} at the contact surfaces. As discussed in
\cref{s:model:adh}, the volume change at the surfaces is approximated by
$J_\ell \approx \Jcl$ in \cref{e:Tn}. Since in this example (and for our choice
of parameters)~$\Jcl$ has only little influence \citep{mergelPhD}, we proceed
with $\Jcl \approx 1$.
\begin{table}[h]
	\centering
	\begin{tabular}{c@{\qquad}c@{\qquad}c@{\qquad}c@{\qquad}c}
		\hline \noalign{\smallskip}
		$E$ & $\nu$ & $\AH$ & $r_0$ & $R$ \\
		\noalign{\smallskip} \hline \noalign{\smallskip}
		$1\,\mrG\mrP\mra$ & $0.3$ & $2.54\cdot 10^{-20}\,\mrJ$
		& $0.4\,\mrn\mrm$ & $40\,\mrn\mrm$ \\
		\noalign{\smallskip} \hline
	\end{tabular}
	\caption{2D contact of two deformable cylinders: Model parameters.}
	\label{t:spheres:para}
\end{table}

\Cref{f:spheres:xn0,f:spheres:xnD,f:spheres:xnJ} show the two cylinders sliding
across each other for three different adhesion/friction models. Although the
(two-half-pass) friction algorithm does not balance the contact traction
explicitly \citep{sauer15ijnme}, for each deformation state, the stress fields
of both cylinders are point-symmetric up to machine precision (not shown).

\begin{figure}[ht]
	\centering
	\subfigure[Horizontal force$^*$ for model~\EA.]{
		\includegraphics[width=0.47\textwidth]{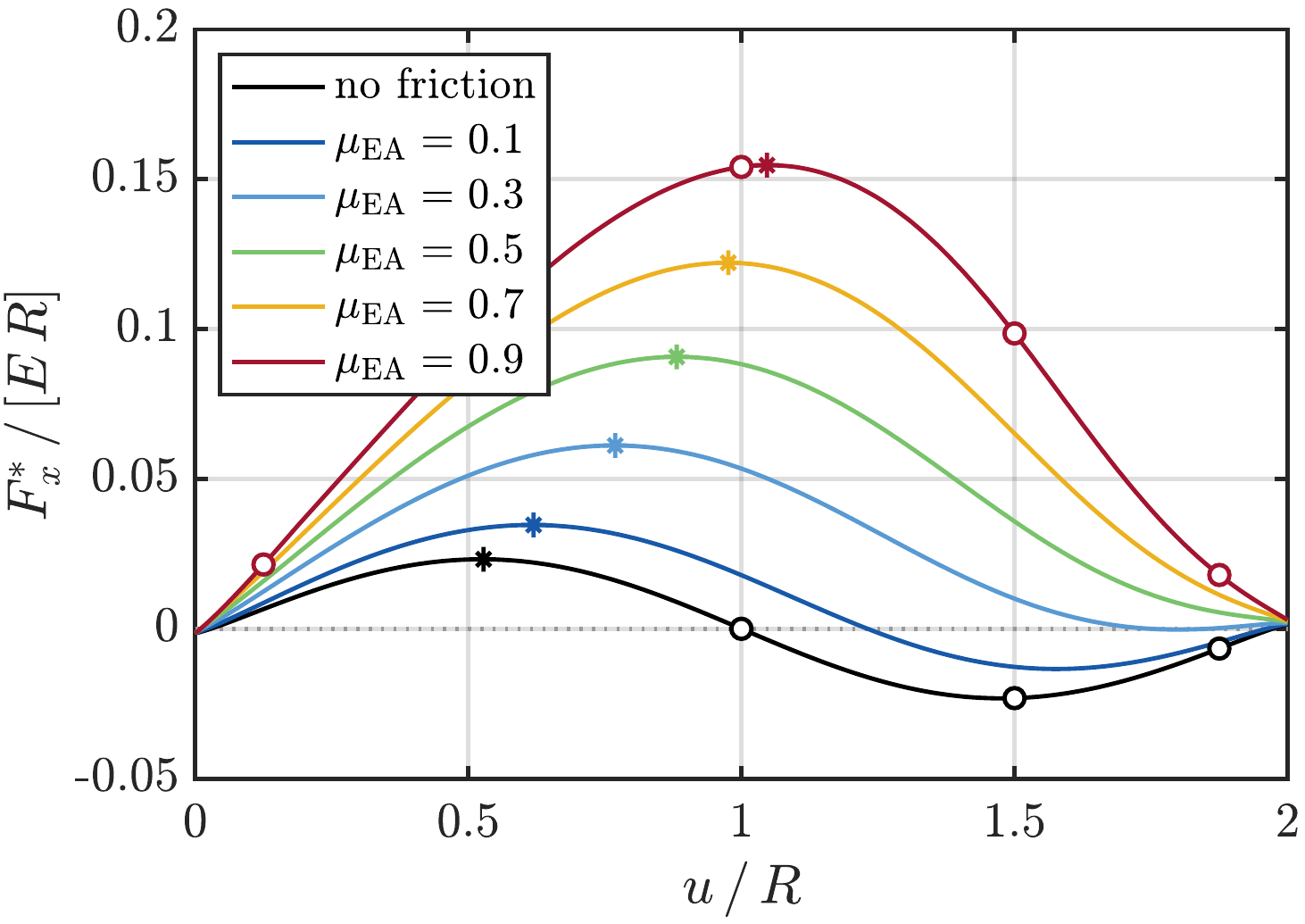}
		\label{f:spheres:PDx}
	}\hspace*{1ex}
	\subfigure[Horizontal force$^*$ for model~\DI\ up to $\mu_\DIe = 0.9$.]{
		\includegraphics[width=0.47\textwidth]{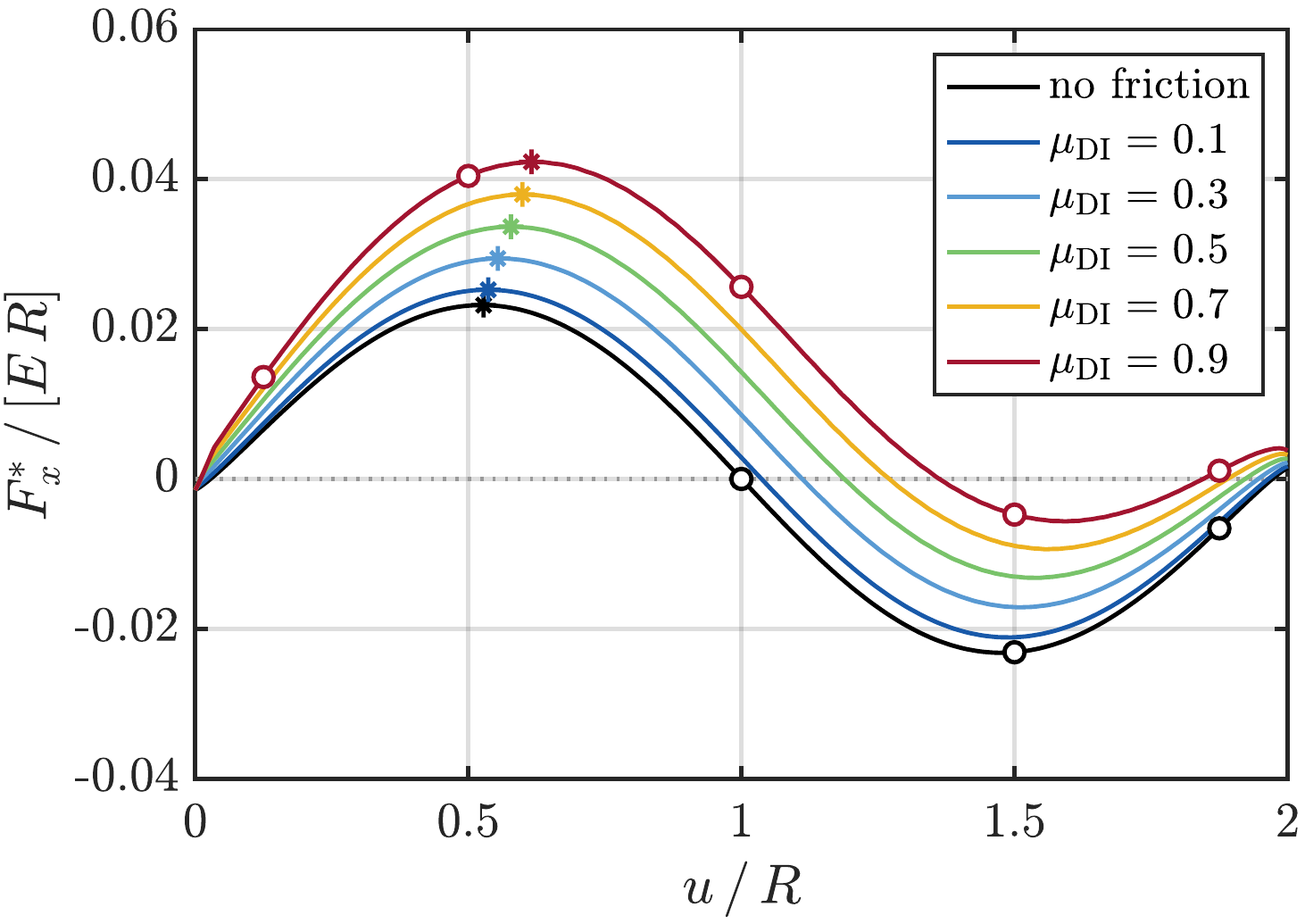}
		\label{f:spheres:PJx}
	}
	\subfigure[Comparison of models~\EA\ and \DI.]{
		\includegraphics[width=0.47\textwidth]{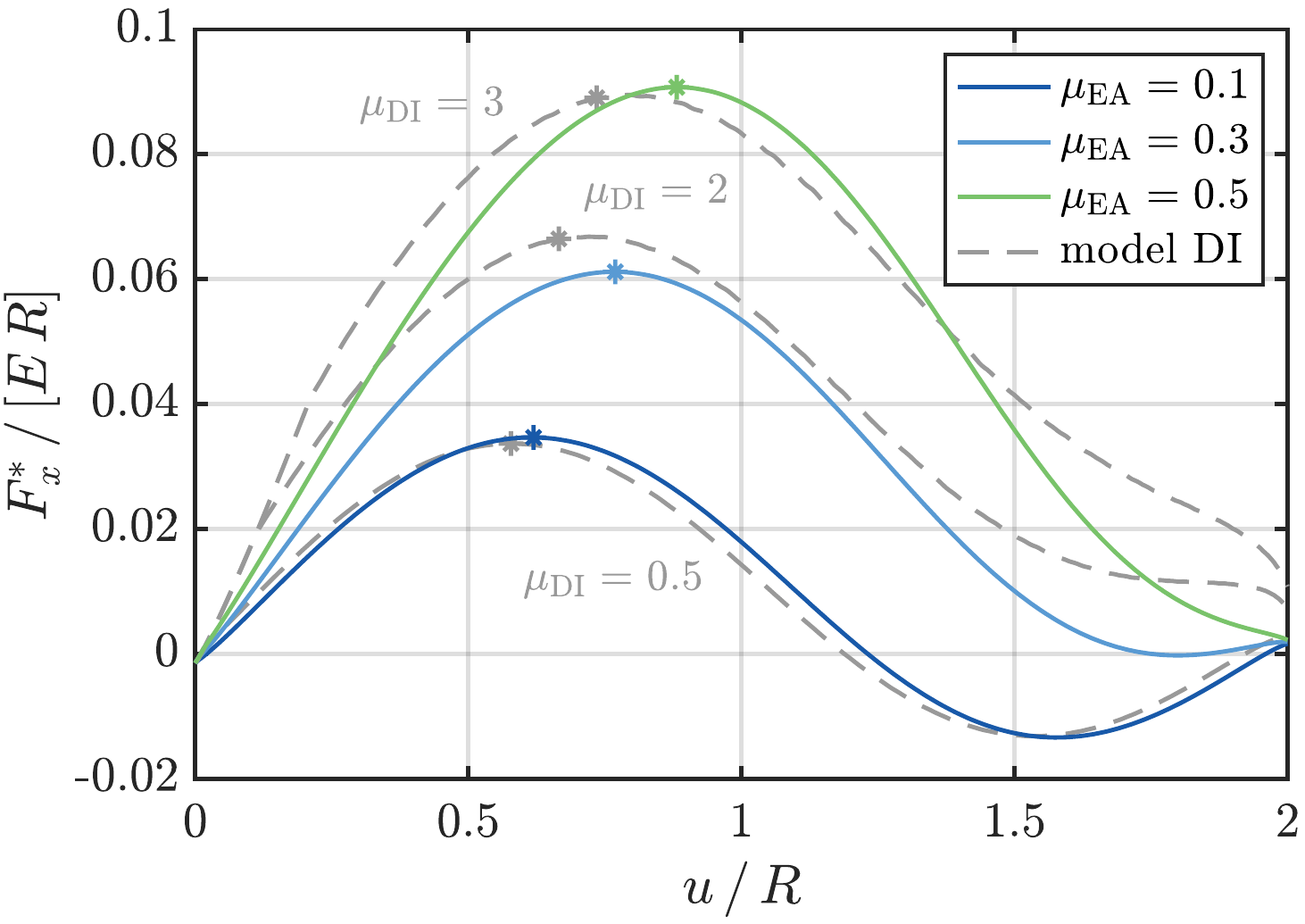}
		\label{f:spheres:PDJx}
	}\hspace*{1ex}
	\subfigure[Normal stress at trailing corner.]{
		\includegraphics[width=0.47\textwidth]{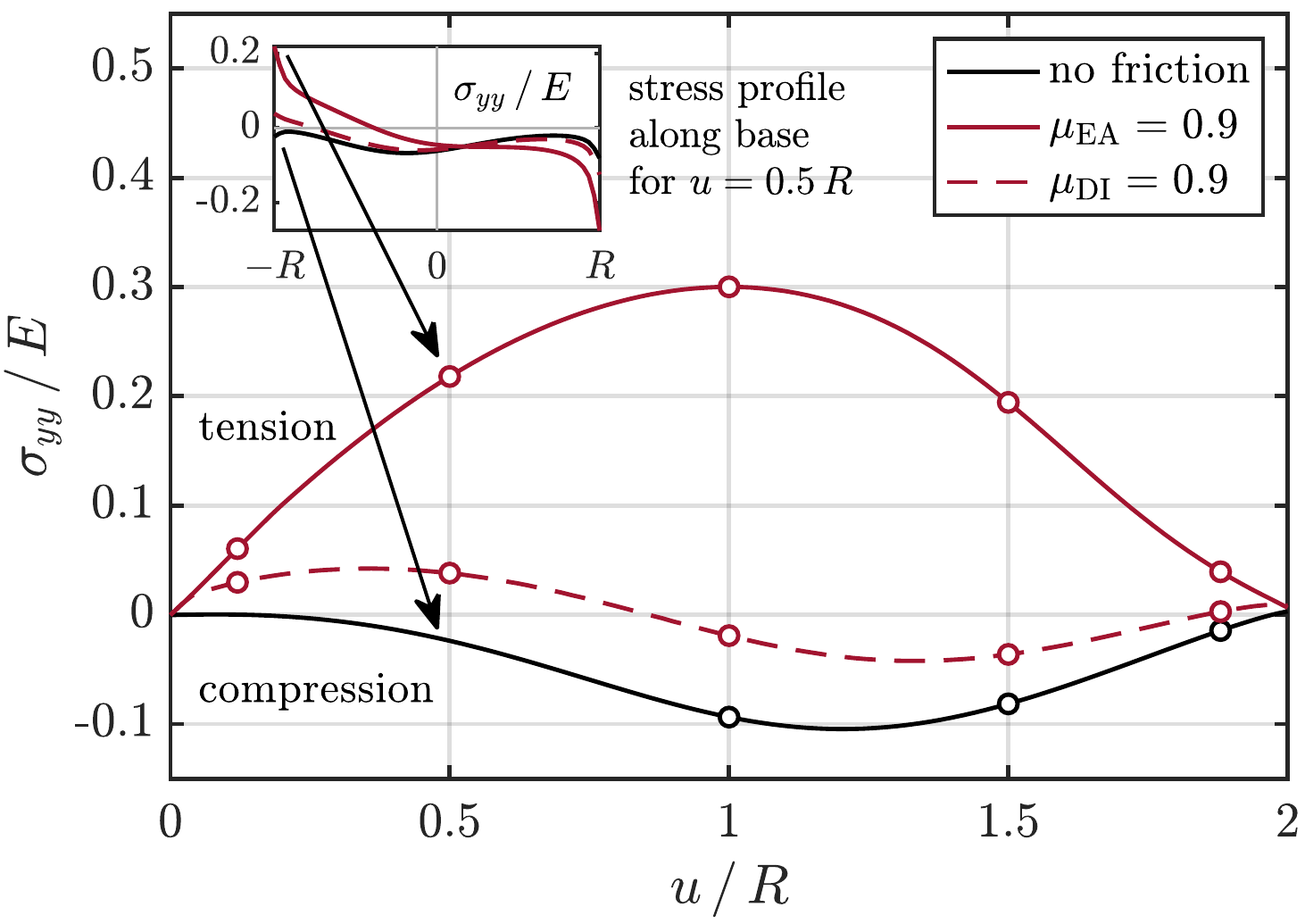}
		\label{f:spheres:sigyy0DJ}
	}
	\caption{2D contact of two deformable cylinders: (a) -- (c)~Horizontal
		forces$^*$ ($^*$: per out-of-plane width) acting on each body for
		models~\EA\ and \DI, both with $\gcut = \gmax$; asterisks indicate the
		horizontal displacement at which snap through would occur in a
		horizontally load-controlled setup; (d)~normal
		stress~$\sigma_{yy}\,/\,E$ at the (left) trailing corner of the bottom
		cylinder for the cases from \cref{f:spheres:xn0} -- 
		\labelcref{f:spheres:xnJ};
		the inset in Fig.~(d) shows~$\sigma_{yy}\,/\,E$ along the base of the
		lower cylinder for a displacement of $u = 0.5\,R$; the white dots in
		Figs.~(a) -- (d) mark the configurations shown in \cref{f:spheres:xn0}
		-- \labelcref{f:spheres:xnJ}, respectively.}
	\label{f:spheres:PJD}
\end{figure}

\Cref{f:spheres:xn0} corresponds to the frictionless, i.e.~purely adhesive,
case. The corresponding horizontal and vertical forces (per out-of-plane
width), acting on the lower cylinder, are shown as black curves in
\cref{f:spheres:PDx,f:spheres:PJx}. Even in the frictionless case, there exists
a finite horizontal force between the cylinders, caused by the geometrical
setup. This force is positive (negative) for $u < R$ ($u > R$), where contact
occurs on the trailing (leading) sides of the two asperities (see also
\cref{f:spheres:setup}). Thus, this setup cannot be reduced to an equivalent
cylinder-plane contact (as commonly done for two cylinders~\citep{johnson85}),
because this would not capture any horizontal forces.

\Cref{f:spheres:xnD,f:spheres:xnJ} show analog snapshots for models \EA\ and
\DI\ with a large friction parameter ($\mu_\EAe = \mu_\DIe = 0.9$). In contrast
to the previous example (where models~\EA\ and \DI\ where almost
undistinguishable), the deformations of the cylinders considerably differ for
both models, see e.g.~$u = R$. We thus investigate also the horizontal forces
(\cref{f:spheres:PDx,f:spheres:PJx}), while varying both~$\mu_\EAe$
and~$\mu_\DIe$. For $\mu_\EAe = \mu_\DIe$, model~\EA\ systematically produces
considerably larger forces than model~\DI. This is explained by the fact that
the normal contact tractions are strongly compressible (unlike in the strip
example), in particular when the two cylinders pass the apex of each other.
Such large normal tractions automatically increase the sliding
threshold~$\Ttmax$ in model~\EA, according to \cref{e:TslideEA}. In contrast,
model~\DI\ depends on the normal traction only implicitly, via a changing
contact area. Compared to the horizontal force, the vertical force (not shown)
changes less with increasing~$\mu_\EAe$ or~$\mu_\DIe$, especially for $\mu_\DIe
\le 0.9$.

In compression-dominated problems like this setup, the friction
parameters~$\mu_\EAe$ and~$\mu_\DIe$ cannot be related to each other via
\cref{e:strip:EJeq} anymore. We thus qualitatively compare cases that provide
similar maximum horizontal forces instead. One of these cases corresponds
e.g.~to $\mu_\EAe = 0.5$ and $\mu_\DIe = 3$ (\cref{f:spheres:PDJx}). The shapes
of the two corresponding curves are rather similar, suggesting that the
evolution of the horizontal force mainly results from the sum of the tractions,
rather than from their distribution, which differs in the two models. The main
deviation can be seen just before detachment of the cylinders, where model~\DI\
systematically predicts larger horizontal forces as well as a more abrupt force
drop.

The results stated above should be put more into perspective. If the cylinders
were driven by a horizontal load instead of a displacement, a snap-through
instability would occur at the maximum of the $F_x^*$~curves, marked with
asterisks in \crefrange{f:spheres:PDx}{f:spheres:PDJx}. Note that in real
tribological tests, finite stiffness springs are used to drive the system (see
also \cite{papangelo20}), so that snap-through may occur even when the springs
themselves are displacement-driven. Thus, in practice the duration of contact
may be much smaller than predicted in
\crefrange{f:spheres:PDx}{f:spheres:PDJx}. Finally note that increasing
friction reduces both the possibility of negative forces and the tendency for a
snap-through before the (midpoints of the) cylinders have moved across each
other ($u = R$).

The interaction between two cylinders, as considered here, has recently gained
renewed attention in the context of wear modeling. In particular, Molinari and
co-workers performed atomistic simulations to investigate the conditions for
formation of debris when two circular asperities collide tangentially. They
found that, depending on the size of the formed junction, such asperities can
either deform plastically, or create debris via crack propagation
\citep{aghababaei16}. In the latter case, a crack nucleates at the trailing
corner of each asperity, where it connects to its underlying bulk
(\cref{f:spheres:setup}). While the contact geometry is similar to ours, the
constitutive laws used by \cite{aghababaei16} are considerably different. They
especially consider a perfectly adhering interface, at which particles coming
into contact remain attached to each other as they would be in the bulk. We
suggest that our friction models could be used to investigate the effect of a
contact interface with a more realistic behavior, including finite adhesion and
friction. Note that our simulation setup could be readily modified to include
also an elastic bulk below the asperity, as done in \cite{aghababaei16}.

To illustrate this, \cref{f:spheres:sigyy0DJ} shows the evolution of the normal
stress at the (left) trailing corner of the bottom cylinder, where the crack
nucleation starts in \cite{aghababaei16}. Here, the cases from
\crefrange{f:spheres:xn0}{f:spheres:xnJ} are considered. For high friction, as
in case~\labelcref{f:spheres:xnD} with model~\EA, the stress is always tensile
(positive), with a maximum reached when the two cylinders pass each other ($u
\approx R$). Assuming a critical stress at which de-bonding between the
asperity and its rigid base initiates, our simulations may allow to investigate
for which parameters and geometrical properties this de-bonding would occur.
The outcome of such an investigation may not be trivial, because both the
stress profile along the base (see the inset in \cref{f:spheres:sigyy0DJ}) and
the stress evolution at the trailing corner can differ significantly for
different models.


\subsection{3D peeling of a tape-like membrane} \label{s:tape}

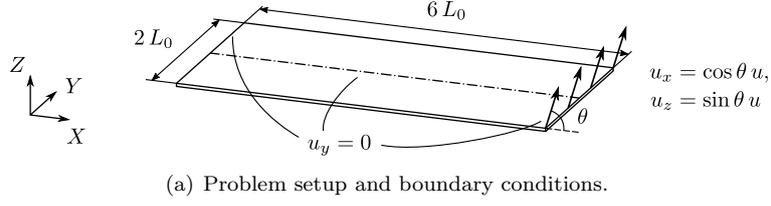
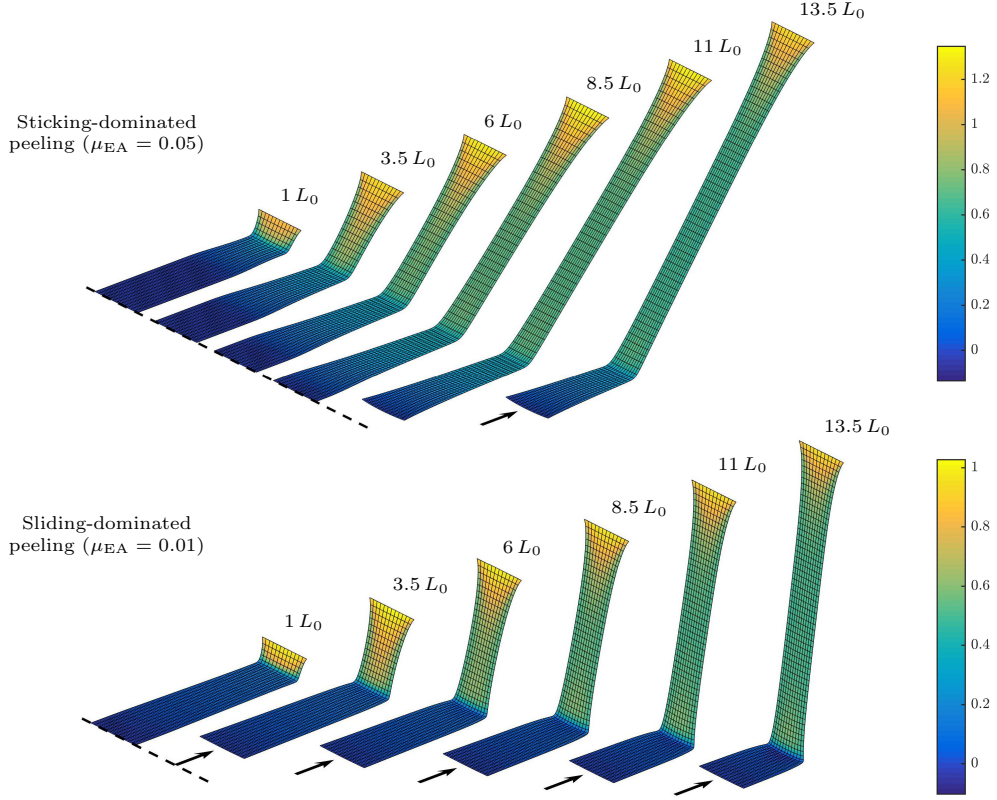
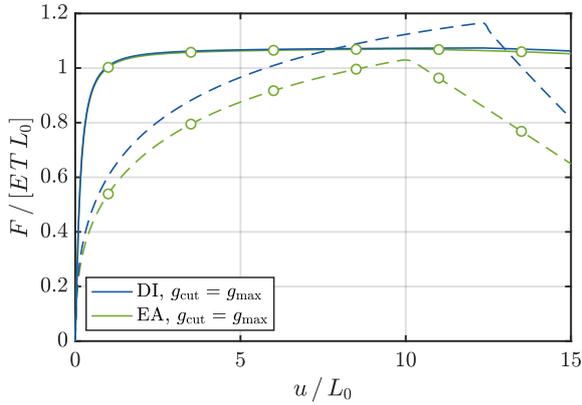
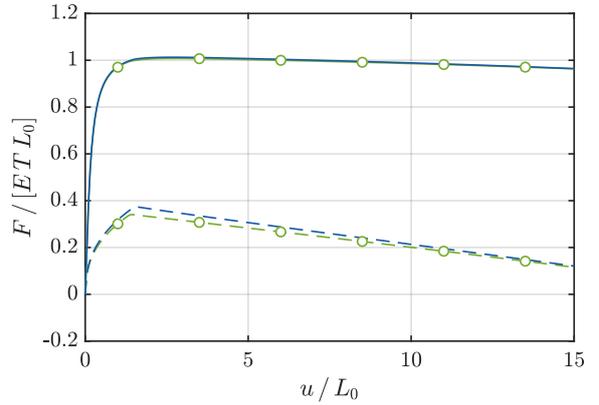
\begin{figure}[p]
	\centering
	\subfigure[Problem setup and boundary conditions.]{
		\includegraphics[width=0.62\textwidth]{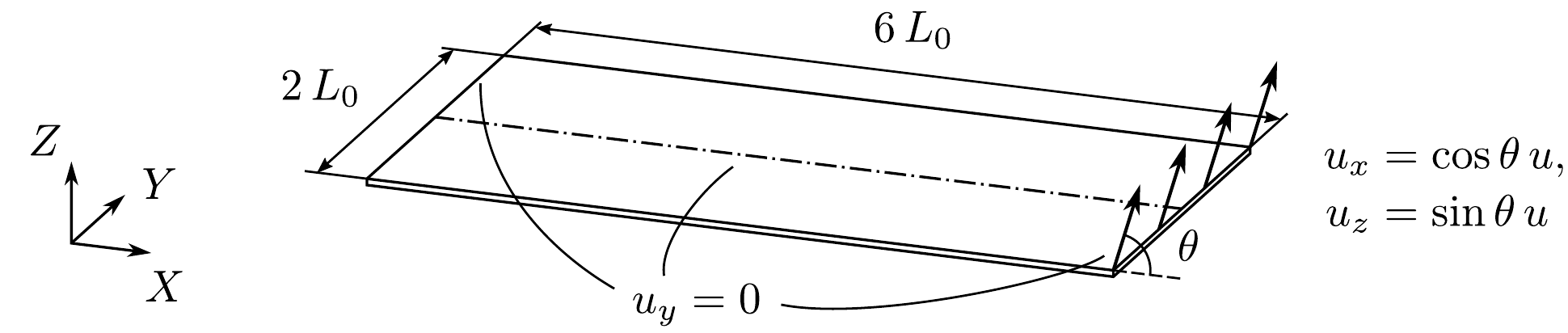}
		\label{f:tape:setup}
	}
	\subfigure{
		\unitlength\textwidth
		\begin{picture}(1,0.32)
			\put(0.17,-0.02){\includegraphics[width=0.60\textwidth]
				{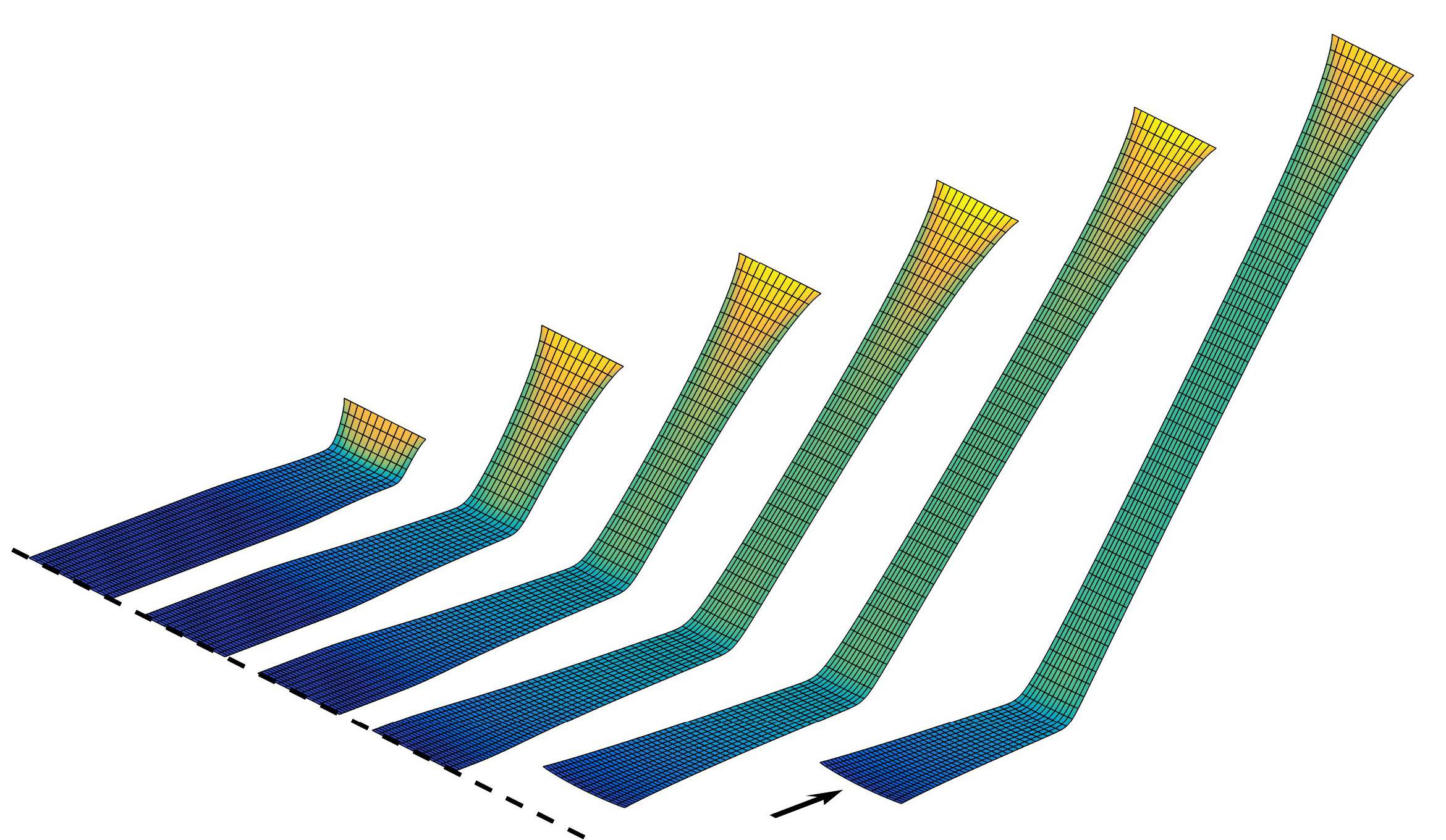}}
			\put(0.85,0.02){\includegraphics[height=0.28\textwidth]
				{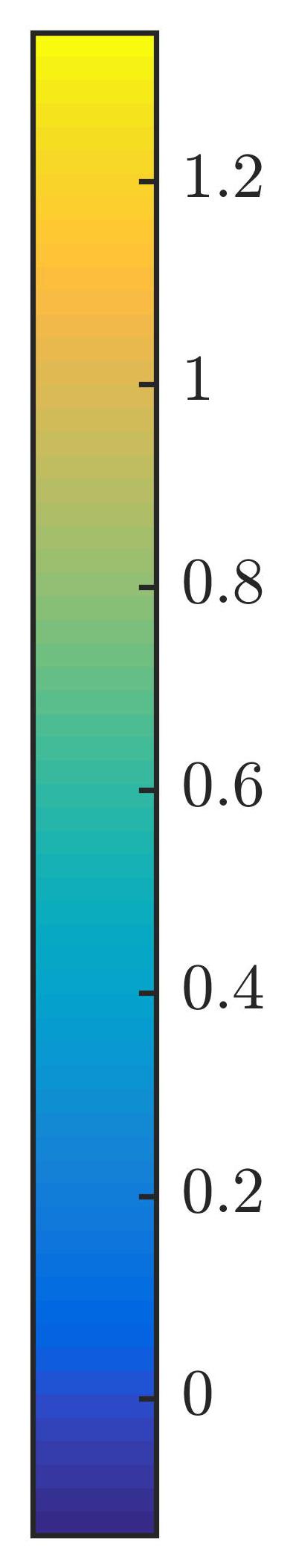}}
			\put(0.1,0.22){\scriptsize \begin{tabular}{c}
				Sticking-dominated \\ peeling ($\mu_\EAe = 0.05$)
				\end{tabular}}
			\put(0.33,0.17){\scriptsize $1\,L_0$}
			\put(0.41,0.20){\scriptsize $3.5\,L_0$}
			\put(0.493,0.23){\scriptsize $6\,L_0$}
			\put(0.575,0.26){\scriptsize $8.5\,L_0$}
			\put(0.66,0.29){\scriptsize $11\,L_0$}
			\put(0.745,0.32){\scriptsize $13.5\,L_0$}
		\end{picture}
	}
	\addtocounter{subfigure}{-1}
	\subfigure[Detachment at $\theta = 45^\circ$ dominated by either sticking
		(top) or sliding (bottom) for model~\EA\ and $\gcut = \gmax$;
		the~colors show the logarithmic area stretch~$\ln J_\mrc$; the figure
		shows both halves of the tape.]{
		\unitlength\textwidth
		\begin{picture}(1,0.30)
			\put(0.17,0.01){\includegraphics[width=0.62\textwidth]
				{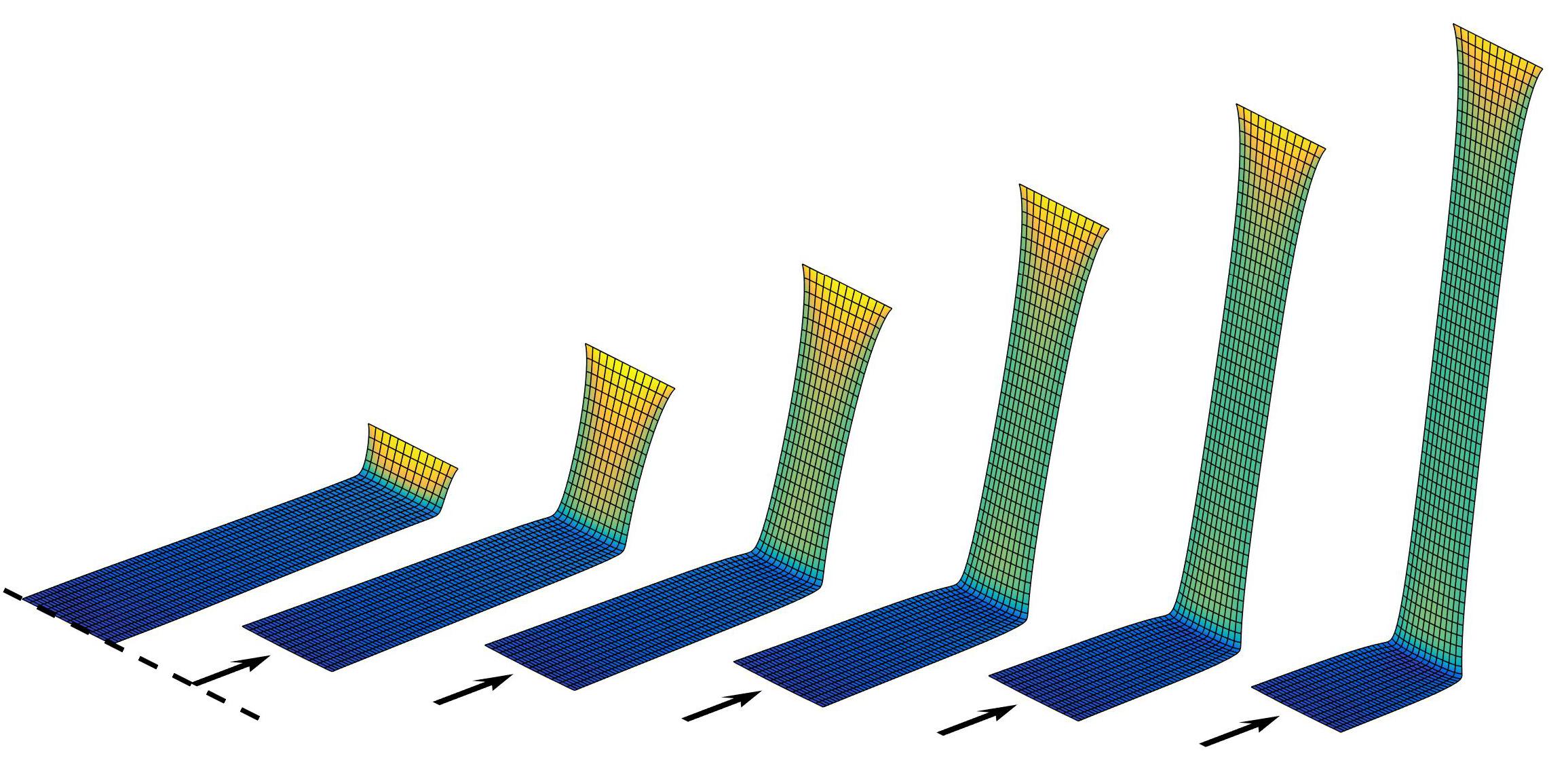}}
			\put(0.85,0.01){\includegraphics[height=0.28\textwidth]
 				{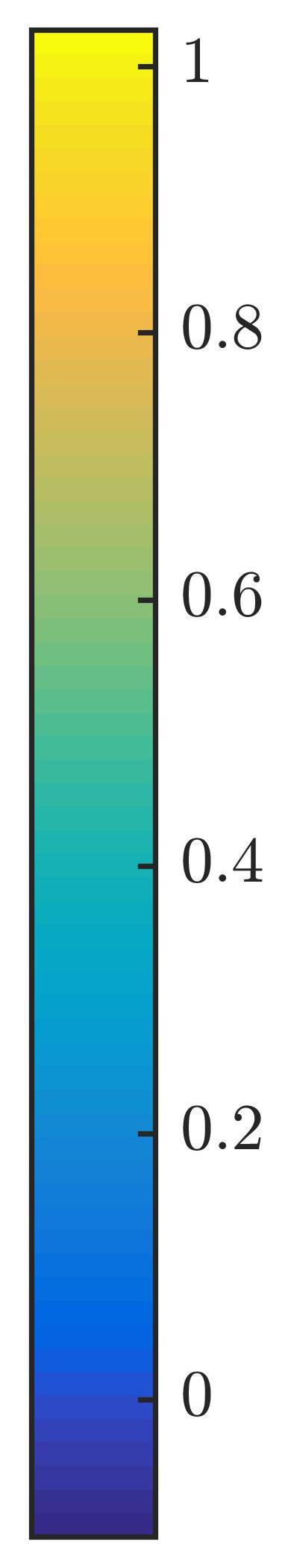}}
			\put(0.1,0.22){\scriptsize \begin{tabular}{c}
				Sliding-dominated \\ peeling ($\mu_\EAe = 0.01$)
				\end{tabular}}
			\put(0.333,0.15){\scriptsize $1\,L_0$}
			\put(0.42,0.18){\scriptsize $3.5\,L_0$}
			\put(0.508,0.21){\scriptsize $6\,L_0$}
			\put(0.595,0.242){\scriptsize $8.5\,L_0$}
			\put(0.68,0.277){\scriptsize $11\,L_0$}
			\put(0.765,0.307){\scriptsize $13.5\,L_0$}
		\end{picture}
		\label{f:tape:x}
	}
	\subfigure[Sticking-dominated peeling: $\mu_\EAe = \mu_\DIe = 0.05$.]{
		\includegraphics[width=0.46\textwidth]{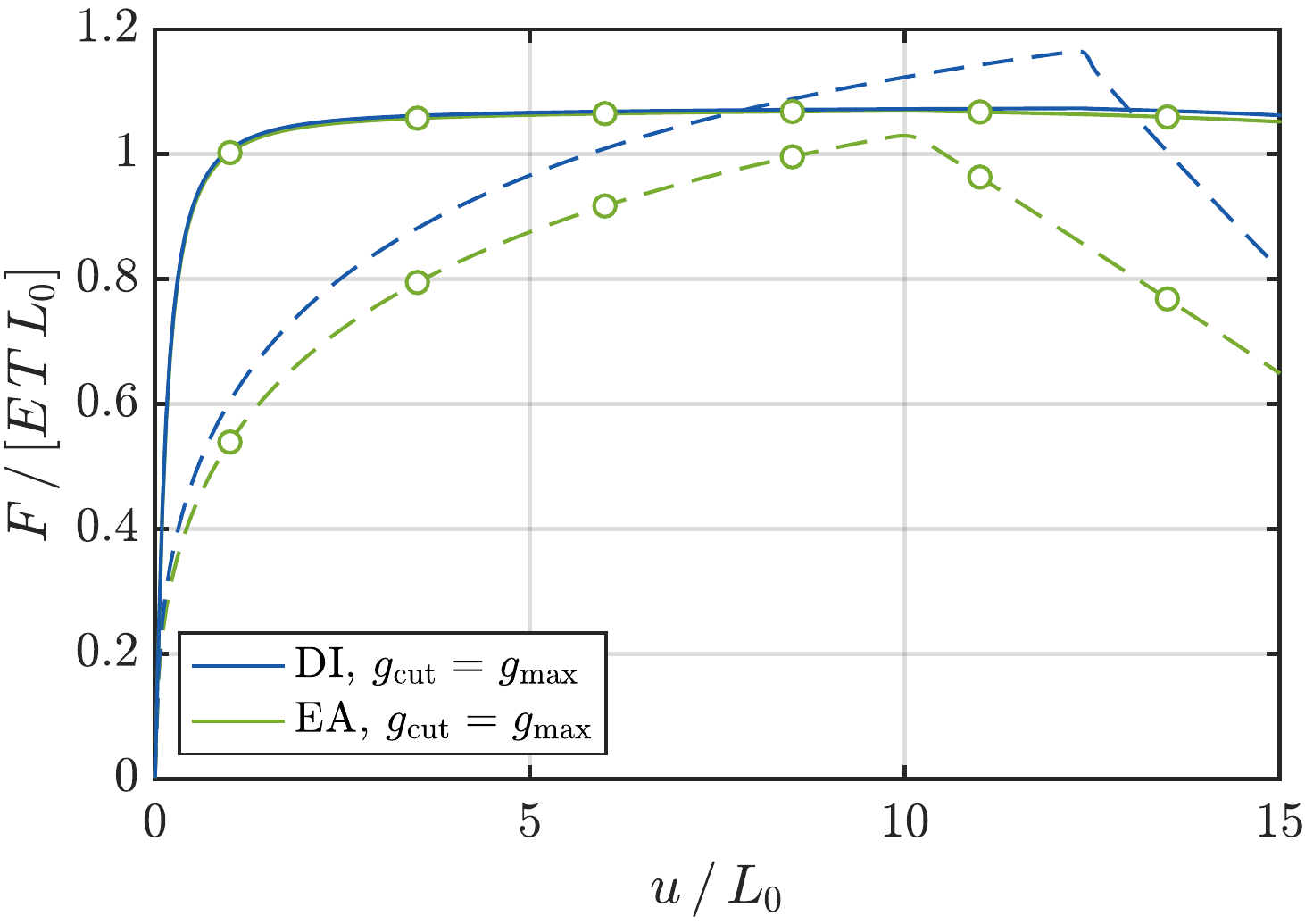}
		\label{f:tape:Pstick}
	}\hspace*{1ex}
	\subfigure[Sliding-dominated peeling: $\mu_\EAe = \mu_\DIe = 0.01$.]{
		\includegraphics[width=0.46\textwidth]{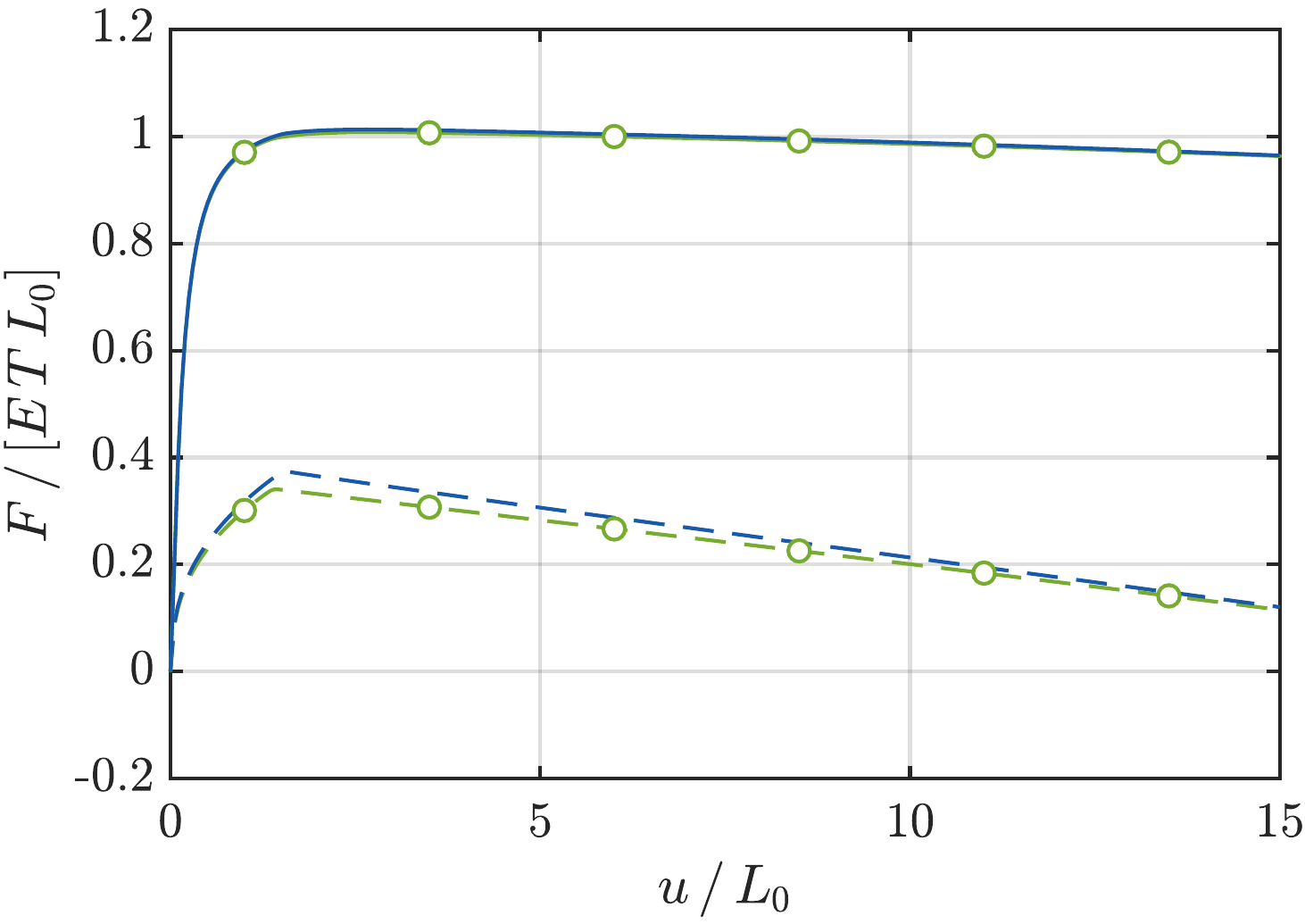}
		\label{f:tape:Pslide}
	}
	\caption{3D peeling of a membrane: (a)~Setup; (b)~detachment at
		$\theta = 45^\circ$ for strong and weak friction; (c)~\&~(d)~forces in
		horizontal (dashed lines) and vertical (solid lines) directions for
		models~\EA\ and \DI, the dots mark the configurations of Fig.~(b).}
	\label{f:tape}
\end{figure}

We now study 3D peeling of a soft and compliant tape
(\cref{f:tape:setup,t:tape:para}), which is modeled as a membrane
\citep{sauer14cmame}. Like other pure membrane models, this formulation only
captures in-plane stresses, but does not account for any bending moments.
%
%
\begin{table}[h]
	\centering
	\begin{tabular}
		{c@{\qquad}c@{\qquad}c@{\qquad}c@{\qquad}c}
		\hline \noalign{\smallskip}
		$E\cdot T$ & $\nu$ & $\AH$ & $r_0$ & $L_0$ \\
		\noalign{\smallskip} \hline \noalign{\smallskip}
		$0.02\,\mrN/\mrm$ & $0.3$ & $10^{-20}\,\mrJ$
		& $0.4\,\mrn\mrm$ & $1\,\mrn\mrm$ \\
		\noalign{\smallskip} \hline
	\end{tabular}
	\caption{3D peeling of a membrane: Model parameters; $T$ denotes the
		thickness of the tape.}
	\label{t:tape:para}
\end{table}

Initially, the entire tape is pre-stretched and attached to a rigid and
motionless substrate. The (uniform) pre-stretch of the tape is chosen
as~$\lambda = 1.001$. This is sufficient to avoid any local in-plane
compression during peeling (which cannot be resisted by a pure membrane). The
frictional sticking to the substrate prevents the tape from directly shrinking
back to its un-stretched configuration. Afterwards, we start peeling off one
side of the tape by prescribing a displacement along the angle~$\theta =
45^\circ$. Note that this angle corresponds to the direction of motion, and not
to the angle between the tape and the substrate at the peeling front.

Since (pure) membranes do not have any bending stiffness, the normal gap from
\cref{e:gn} between the tape and the plane always satisfies $\gn \ge \gequ$
during peeling (in contrast to the behavior observed in \cref{s:strip}). From
\cref{f:Tn} then follows that the normal contact traction is purely
adhesive/tensile within the entire contact area. As a consequence, this
particular setup cannot be investigated with friction models that yield a
tangential resistance only under compression. Such models include both
model~\EA\ with $\scut = 0$ (corresponding to the classical Amontons--Coulomb
law for friction) and the model of \cite{deng12}. Apart from that, even though
it would be possible to model tangential sticking by means of a cohesive zone
model, the membrane would slide without any frictional resistance after
debonding.

The membrane is described by an incompressible Neo-Hookean material model (see
\cite{sauer14cmame}) with the parameters given in \cref{t:tape:para}. To
exploit the symmetry of the tape shown in \cref{f:tape:setup}, we discretize
only one half of it by means of $540$ quadratic NURBS (N2.1) elements
\citep{sauer14cmame}, and apply suitable boundary conditions along the center
line.

\Cref{f:tape:x} shows the peeling process once for a large and once for a small
friction parameter, leading either to strong sticking or immediate sliding. As
can be seen in \cref{f:tape:Pstick,f:tape:Pslide}, in both cases the forces in
the vertical direction are very close to each other for models~\EA\ and \DI.
The horizontal forces, however, are similar only for small friction. This is
caused by the flexibility of the tape (see also \cref{f:tape:x} top), for which
the surface stretch~$\Jck$ is not negligible anymore. It also explains the
differences between the behavior observed here and the results from
\cref{s:strip:forces}. Note that the kink in \cref{f:tape:Pstick,f:tape:Pslide}
appears at the onset of full sliding, also as observed in \cref{s:strip:forces}.

As this example demonstrates, we are able to model adhesive friction even for
very soft structures exhibiting a small or negligible bending stiffness. These
include adhesive tapes as well as adhesive pads of insects. Even though the
bending stiffness of such adhesive pads is sufficiently large at small length
scales (in order to prevent self-sticking and resulting entanglement), at
larger length scales, the bending stiffness becomes negligible, as it scales
with thickness cubed. In summary, the proposed computational model is able to
describe sliding friction during peeling also for a negligible bending
stiffness, which would not be possible with conventional friction models or
cohesive zone models.


\section{Onset of frictional sliding for elastomer-like contact} \label{s:cap}

\begin{figure}[h]
	\centering
	\subfigure[Problem setup.]{
		\unitlength\textwidth
		\begin{picture}(0.43,0.28)
		\put(0.02,0.02){\includegraphics[width=0.4\textwidth]{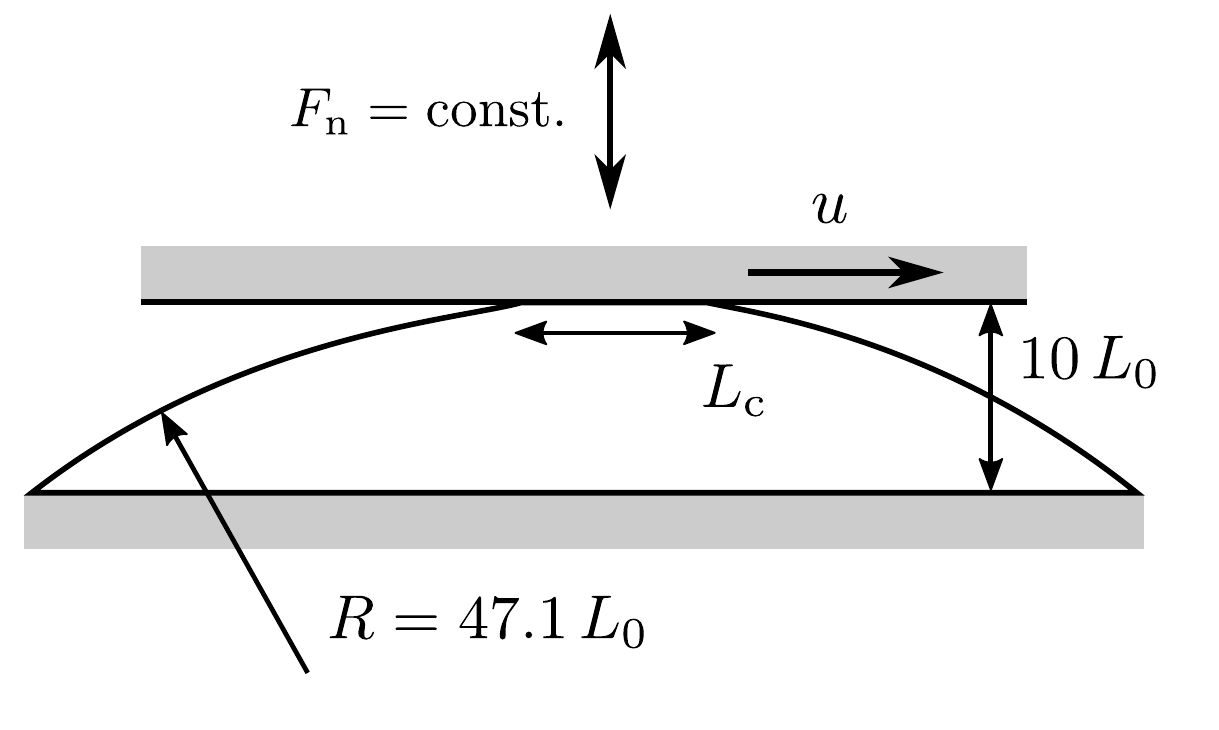}}
		\end{picture}
		\label{f:cap:setup}
	}
	\subfigure[Sliding under zero normal load.]{
		\unitlength\textwidth
		\begin{picture}(0.51,0.26)
		\put(0.017,0.13){\includegraphics[width=0.462\textwidth]
			{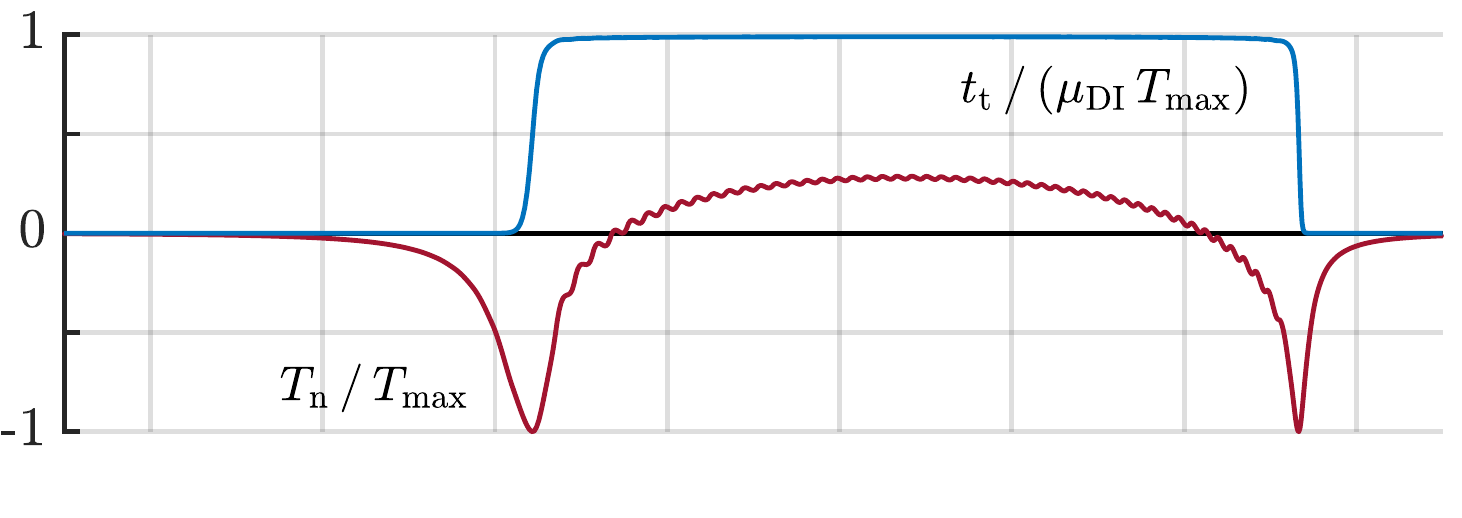}}
		\put(0.035,0.055){\includegraphics[width=0.44\textwidth]
			{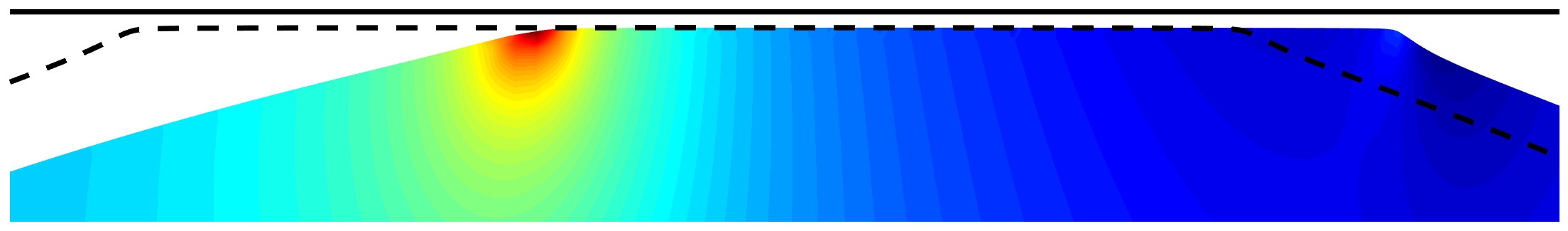}}
		\put(0.035,0.125){\includegraphics[width=0.44\textwidth]
			{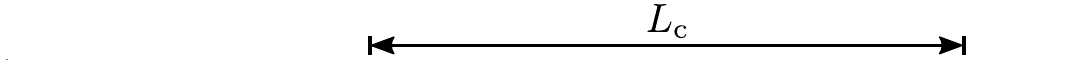}}
		\put(0.035,0){\includegraphics[width=0.44\textwidth]
			{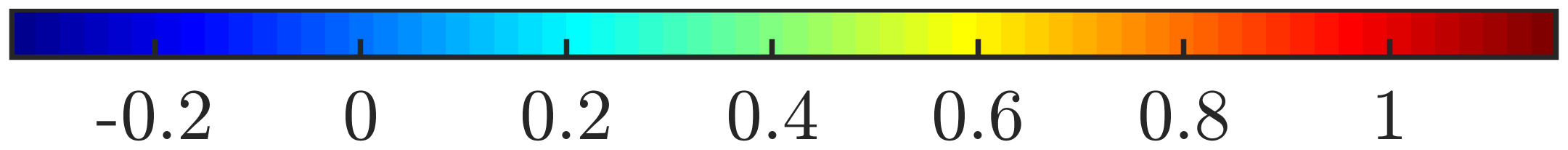}}
		\end{picture}
		\label{f:cap:xn}
	}
	\caption{Adhesive friction of a 2D cap: (a)~Setup (the height of
		$10\,L_0$ refers to the undeformed cap before contact);
		(b)~contact tractions (top panel) as well as deformation and stress
		distribution (bottom panel) at the interface during full sliding for
		$\Fn = 0$ with model~\DI, $\gcut = \gmax$, and $\mu_\DIe = 1$; the
		vertical axis of the deformation plot is stretched by a factor of~$2$;
		the colors show the first invariant of the Cauchy stress,
		$\tr\,\bsig\,/\,E$; the dashed line indicates the contour of the cap
		before horizontal sliding.}
\end{figure}

Let us finally investigate adhesive friction between a smooth elastomer cap and
a rigid substrate, as considered experimentally e.g.~in \cite{sahli18} and
\cite{mergel19jadh}, theoretically e.g.~in \cite{papangelo19}, and numerically
considering linear elasticity \citep{khajehsalehani19}. To this end, we bring a
rigid plate into contact with a deformable, cylindrical cap by applying a
constant normal force~$\Fn$ (\cref{f:cap:setup}). Due to adhesion between the
cap and the plate, this force can be compressive, zero, or even slightly
tensile. Keeping~$\Fn$ constant, we then move the rigid plate horizontally,
while keeping the lower boundary of the cap fixed. As demonstrated in a short
preliminary study~\citep{mergel19jadh}, model~\DI\ is well-suited to
investigate this specific setup.

We here consider 2D plane-strain conditions, nonlinear and nearly
incompressible Neo-Hookean material behavior \citep{bonet97}, and the model
parameters from \cref{t:cap:para}. The cap itself consists of 42,300 linear
(Q1) elements with quadratic NURBS enrichment (Q1N2.1 elements) on the surface
\citep{corbett14}. In order to prevent volumetric locking arising from the
near-incompressibility, we consider reduced integration for the hydrostatic
(also denoted dilatational) part of the stress tensor. See also \cite{hughes00}
for further details. To facilitate a comparison with 3D experimental and
theoretical results, in the following we will denote the (unspecified)
out-of-plane width of the cap by~$W$.
\begin{table}[h]
	\centering
	\begin{tabular}
		{c@{\qquad}c@{\qquad}c@{\qquad}c@{\qquad}c}
		\hline \noalign{\smallskip}
		$E$ & $\nu$ & $\Tnmax$ & $W_\adh$ & $L_0$ \\
		\noalign{\smallskip} \hline \noalign{\smallskip}
		$2\,\mrM\mrP\mra$ & $0.49$ & $0.33\,\mrM\mrP\mra$
		& $0.027\,\mrJ/\mrm^2$ & $1\,\mu\mrm$ \\
		\noalign{\smallskip} \hline
	\end{tabular}
	\caption{Adhesive friction of a 2D cap: Model parameters.}
	\label{t:cap:para}
\end{table}


\subsection{Adhesive friction between cap and substrate} \label{s:cap:adh}

\begin{figure}[h]
	\centering
	\subfigure[Tangential force vs.~horizontal displacement.]{
		\includegraphics[width=0.47\textwidth]{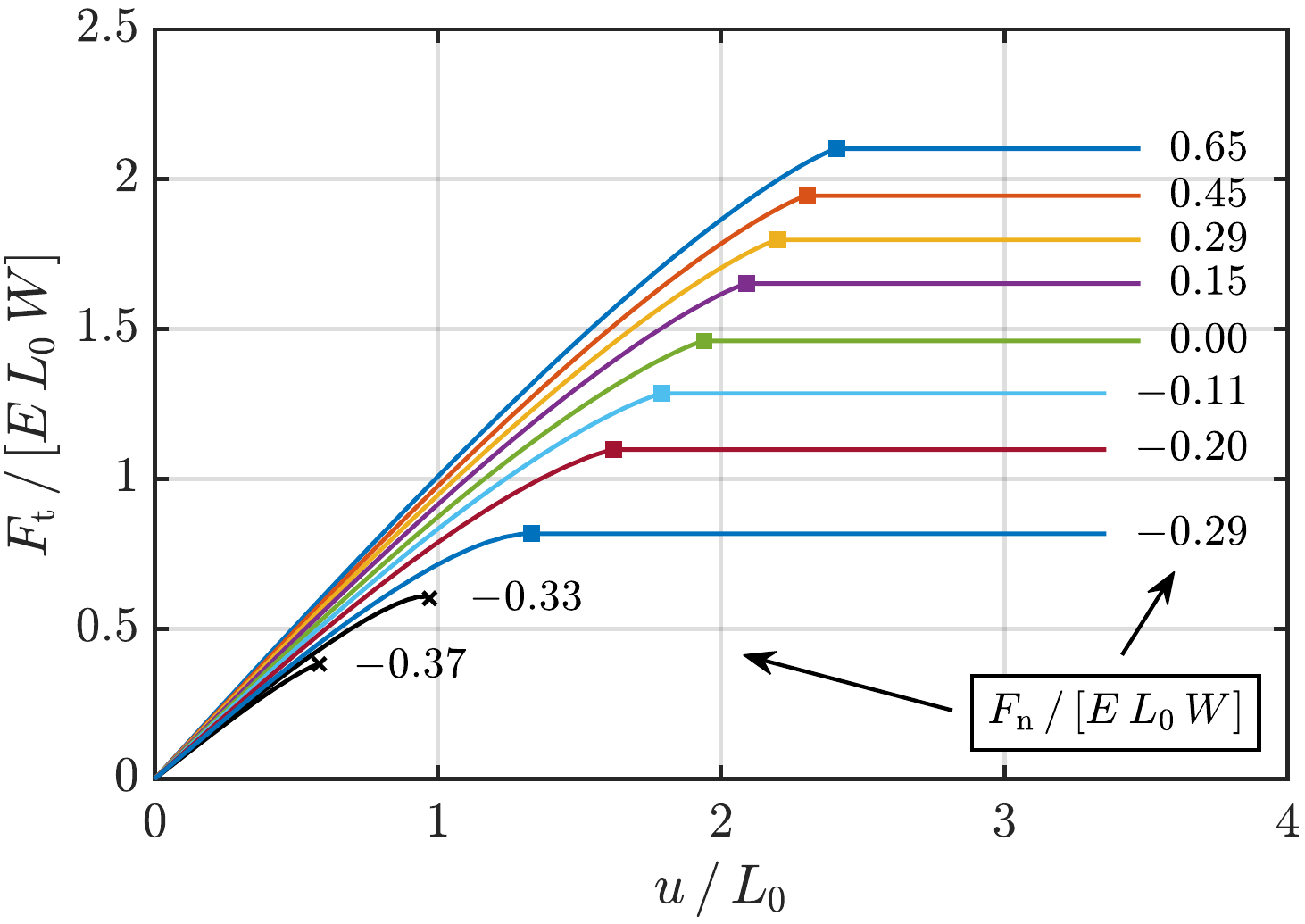}
		\label{f:cap:Ft}
	}\subfigure[Contact length vs.~horizontal displacement.]{
		\includegraphics[width=0.47\textwidth]{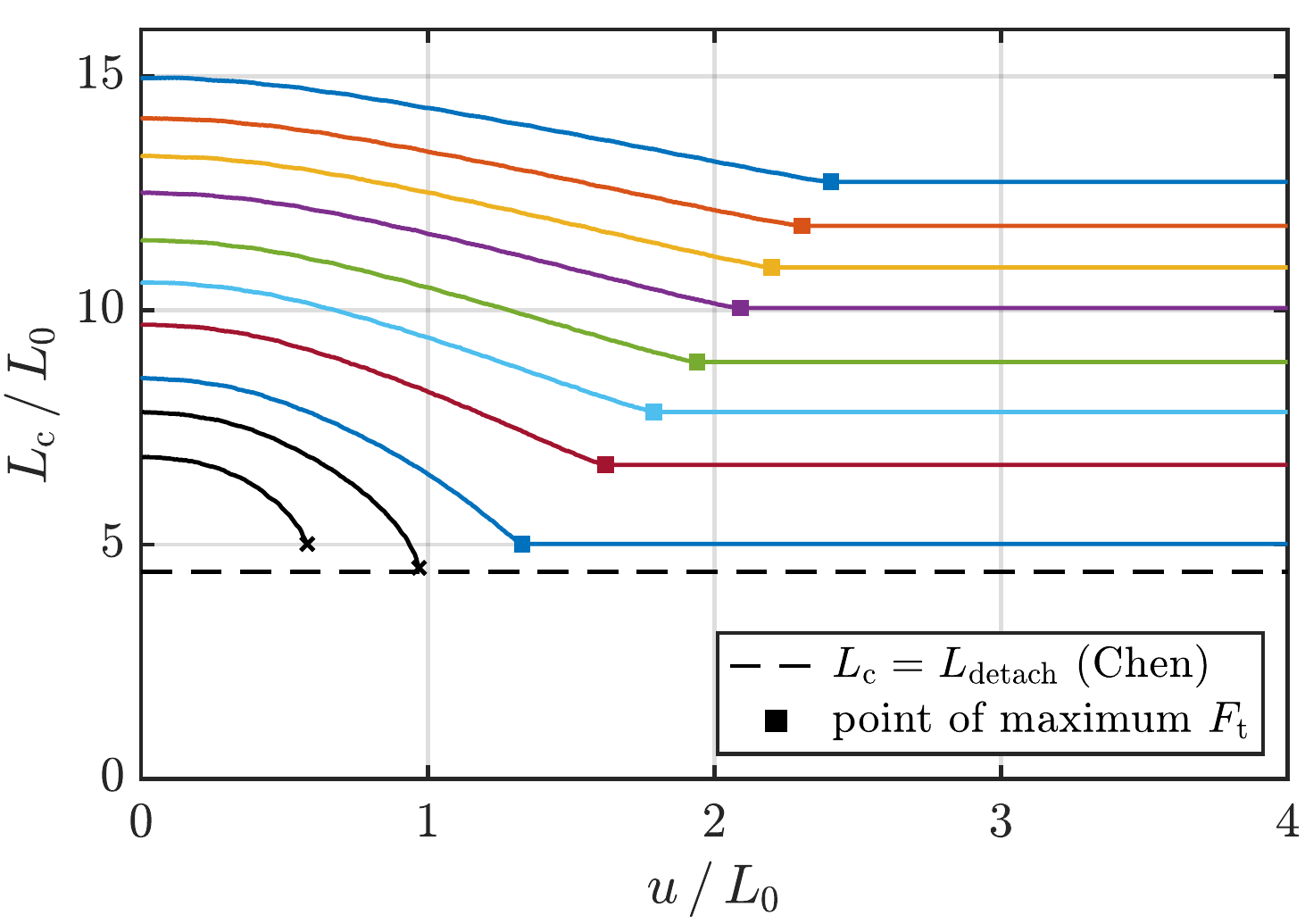}
		\label{f:cap:Lc}
	}
	\subfigure[Contact length vs.~tangential force.]{
		\includegraphics[width=0.47\textwidth]{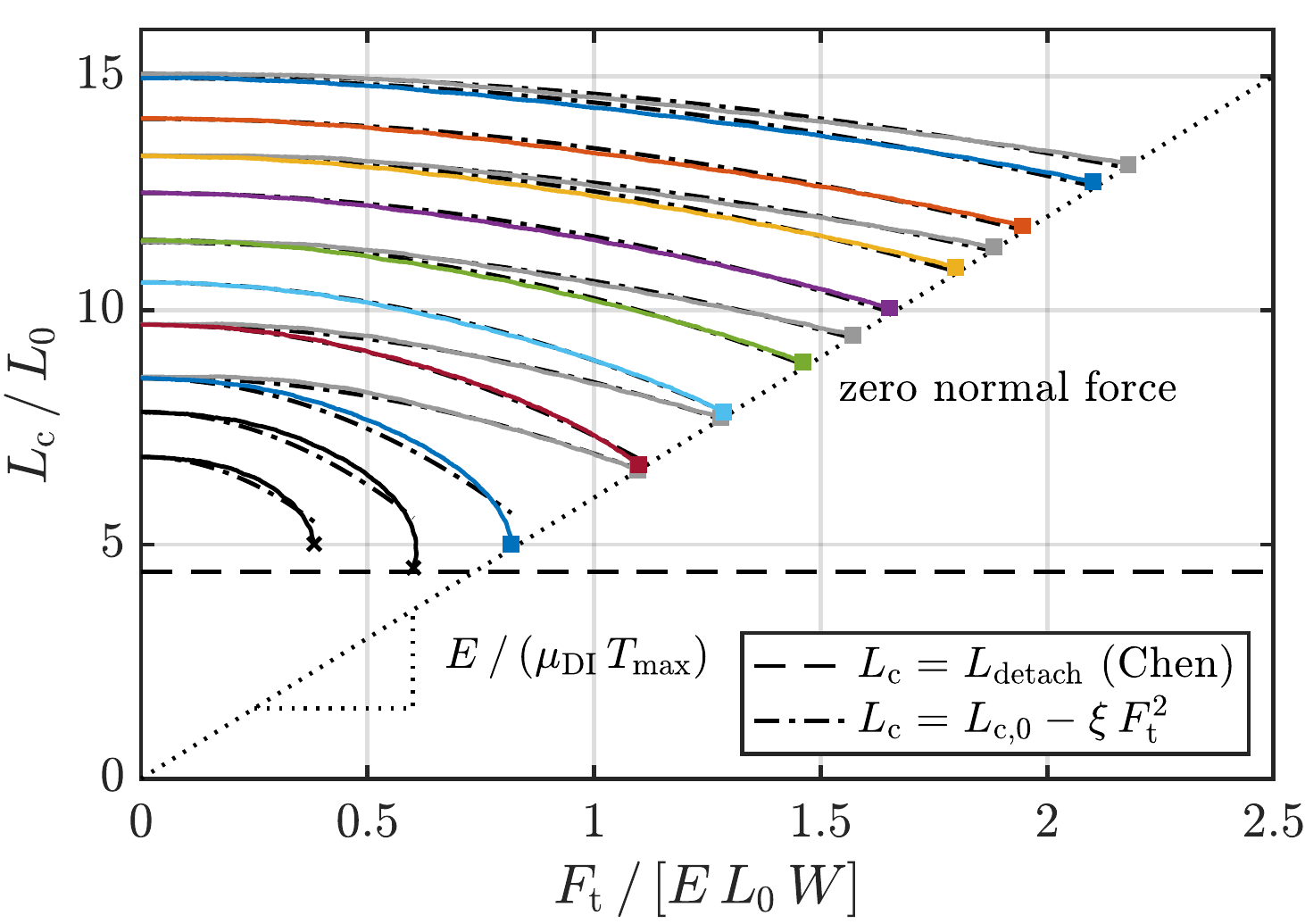}
		\label{f:cap:LcOverFt}
	}\subfigure[Parameter~$\xi$ vs.~initial contact length.]{
		\includegraphics[width=0.47\textwidth]{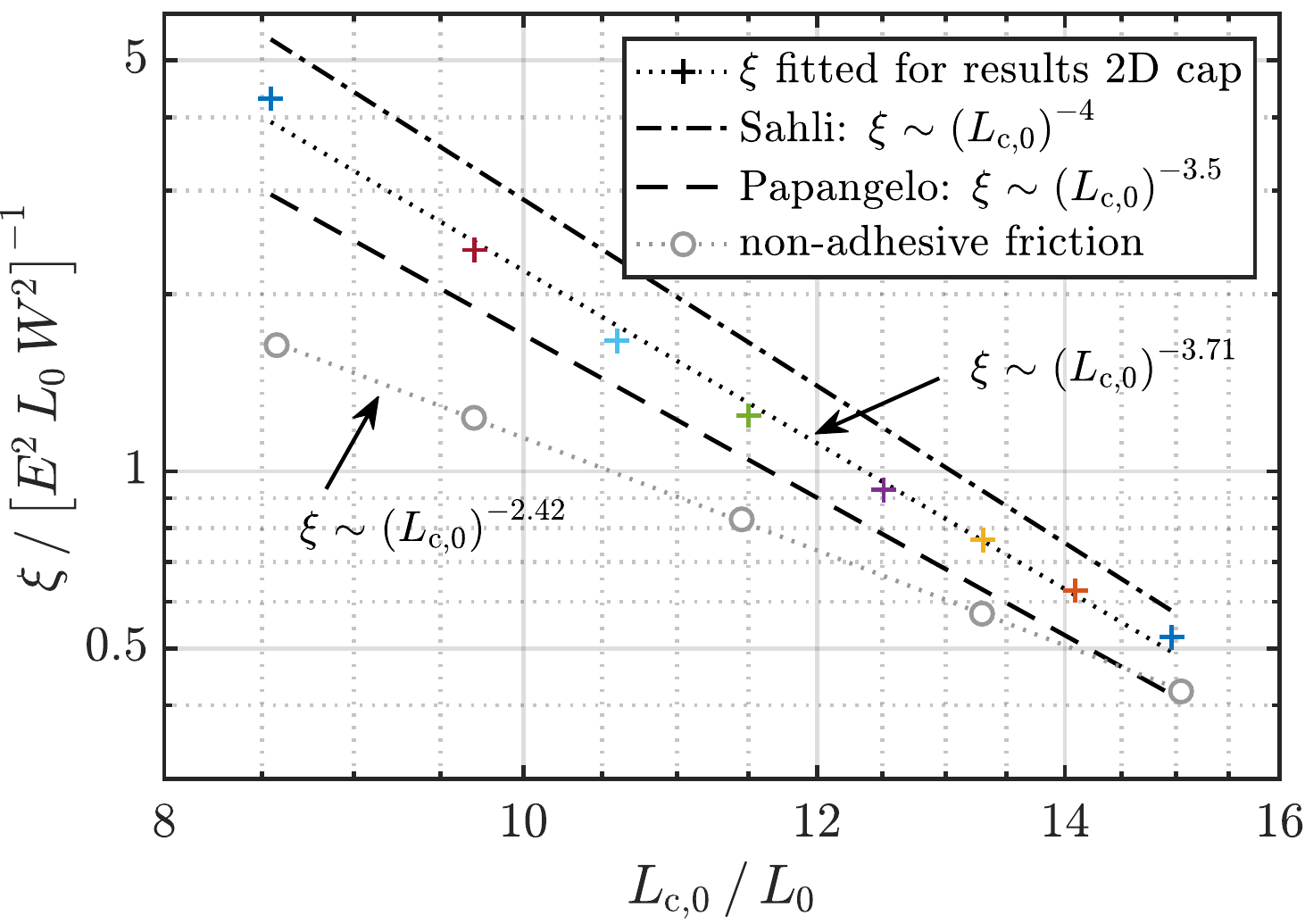}
		\label{f:cap:xiOverLc0}
	}
	\caption{Adhesive friction of a 2D cap: (a) \& (b)~Tangential force and
		contact length, obtained for model~\DI\ ($\gcut = \gmax$,
		$\mu_\DIe = 1$) and different normal forces; the two solid black
		lines indicate contact vanishing under sufficient shear; (c)~contact
		length in dependence of the tangential force; the five gray lines
		correspond to non-adhesive friction (see \cref{s:cap:nonadh});
		(d)~parameter~$\xi$ from \cref{e:Lcfct} fitted for all the cases from
		Fig.~(c) with (non-vanishing) adhesive and non-adhesive contact.}
	\label{f:cap}
\end{figure}

\Cref{f:cap:xn} illustrates the contact interface deformation for sliding at
$\Fn = 0$, including the stress distribution in the bulk of the cap as well as
the corresponding contact tractions. The negative peaks in the normal traction
indicate adhesive stresses at the two contact edges. \Cref{f:cap:Ft} shows the
tangential force versus the prescribed displacement for different normal
forces, while \cref{f:cap:Lc} depicts the corresponding contact length~$\Lc$
(see also \cref{f:cap:setup}). The combination of both, i.e., the contact
length in dependence of the tangential force, is shown in
\cref{f:cap:LcOverFt}. We find that for all investigated normal loads
(compressive, zero, and tensile), the contact length is a decreasing function
of the tangential force. By nature of model~\DI, this decrease stops when all
points within the contact have reached the frictional shear strength $\mu_\DIe
\, \Tnmax$ (see the squares). This explains why all curves in
\cref{f:cap:LcOverFt} end at the dotted line.

For the two shortest initial contact lengths (black solid lines in
\crefrange{f:cap:Ft}{f:cap:LcOverFt}), Newton's method applied in the
(quasi-static) computation stops converging when the contact length decreases
down to a certain, common value. A likely explanation is that for these cases,
the tensile normal force is sufficiently large so that the cap snaps from the
substrate before full tangential sliding. This behavior can also be observed in
experiments \citep{waters10, mergel19jadh}. To test our hypothesis, we consider
the generalized plane-strain JKR model by \cite{chen08b}, and compute the
critical length~$L_\mathrm{detach}$ for which the cap detaches from the
substrate. Inserting the parameters from \cref{t:cap:para} results in
$L_\mathrm{detach} = 4.42\,L_0$, which is very close to the smallest possible
lengths observed in our simulations (see the dashed black lines and the black
crosses in \cref{f:cap:Lc,f:cap:LcOverFt}). This good agreement suggests that
the lower limit for the contact length is indeed related to a (physically
unstable) contact separation.

Let us now compare our results with experiments on shearing of smooth elastomer
spheres on glass plates. As explained in \cite{mergel19jadh}, these comparisons
are only qualitative due to length scale differences between computations
($\mu\mrm$ range, related to the adhesion model by \cite{sauer09cmame}), and
experiments ($\mrm\mrm$ range). In such experiments, the initially circular
contact area shrinks to an ellipse during the onset of sliding
\citep{mergel19jadh, sahli18, sahli19}. \cite{sahli19} demonstrated
experimentally that (i)~the width of the contact area \emph{perpendicular} to
the direction of sliding nearly remains constant; and (ii)~the length~$\Lc$
\emph{parallel} to the direction of sliding is well-captured by the quadratic
fit
\begin{equation}
	\Lc (\Ft) = \Lcz - \xi \, \Ft^2. \label{e:Lcfct}
\end{equation}
In \cref{e:Lcfct}, $\Lcz$ is the initial contact length (at $\Ft = 0$),
and~$\xi$ is a parameter that is independent from~$\Ft$, but dependent on the
applied normal load, or equivalently the initial contact length.

As the dash-dotted lines in \cref{f:cap:LcOverFt} indicate, \cref{e:Lcfct} is
also well-suited to fit the curves from our numerical results, using the
least-squares method. Although the caps in the experiments by \cite{sahli19}
and those in our simulations have different dimensions (spherical in the
experiments vs.~cylindrical in the simulations), we observe consistent
behavior. This motivates us investigating whether the dependence of the
parameter~$\xi$ on the initial contact length/area is similar, too.
\cite{sahli19} observed for their experiments $\xi(\Acz) \sim
{A_{\mrc,0}^{-2}}$, with $\Acz$ being their (circular) initial contact area.
Regarding the initial contact length~$\Lcz$, this translates with $\Acz \sim
L_{\mrc,0}^2$ into $\xi(\Lcz) \sim L_{\mrc,0}^{-4}$. For comparison,
\cref{f:cap:xiOverLc0} shows the parameter~$\xi$ fitted for our numerical
results in dependence of~$\Lcz$. As can be seen, these data are also linear
with logarithmic scales. Our simulations yield $\xi(\Lcz) \sim
L_{\mrc,0}^\beta$ with $\beta = -3.71$, which is in rather good agreement with
the $-4$ inferred from \cite{sahli19}.

Note that the experimental results by \cite{sahli18, sahli19} also match well
with the fracture-mechanics models by \cite{papangelo19} and
\cite{papangelo19b}. Those models describe the evolution of the contact area of
adhesive sphere--plane contact under shear, assuming either a circular or an
elliptic contact area, respectively. For an initially circular area and for
small~$\Ft$, \cite{papangelo19} predict the relation $\Ac(\Ft) = \Acz - \alpha
\, \Ft^2$ with $\alpha(\Acz) \sim A_{\mrc,0}^{-5/4}$. This equation can be
related to the parameter~$\xi$ from \cref{e:Lcfct} as follows: Motivated by
their experiments, \cite{sahli19} assume that the contact area shrinks only
along the shear direction and has an elliptic shape afterwards. With these two
assumptions, the authors show that the exponents of~$\alpha$ and~$\xi$ differ
by the value $-1/2$, so that one expects $\xi(\Acz) \sim A_{\mrc,0}^{-7/4}$.
This, in turn, translates into $\xi(\Lcz) \sim L_{\mrc,0}^{-7/2}$, which gives
an exponent close to the $-3.71$ fitted for our numerical results
(\cref{f:cap:xiOverLc0}). In summary, our exponent~$\beta$ lies in between the
two values observed experimentally and determined theoretically.

\begin{figure}[ht]
	\centering
	\subfigure[Contact length vs.~tangential force (logarithmic).]{
		\includegraphics[width=0.47\textwidth]{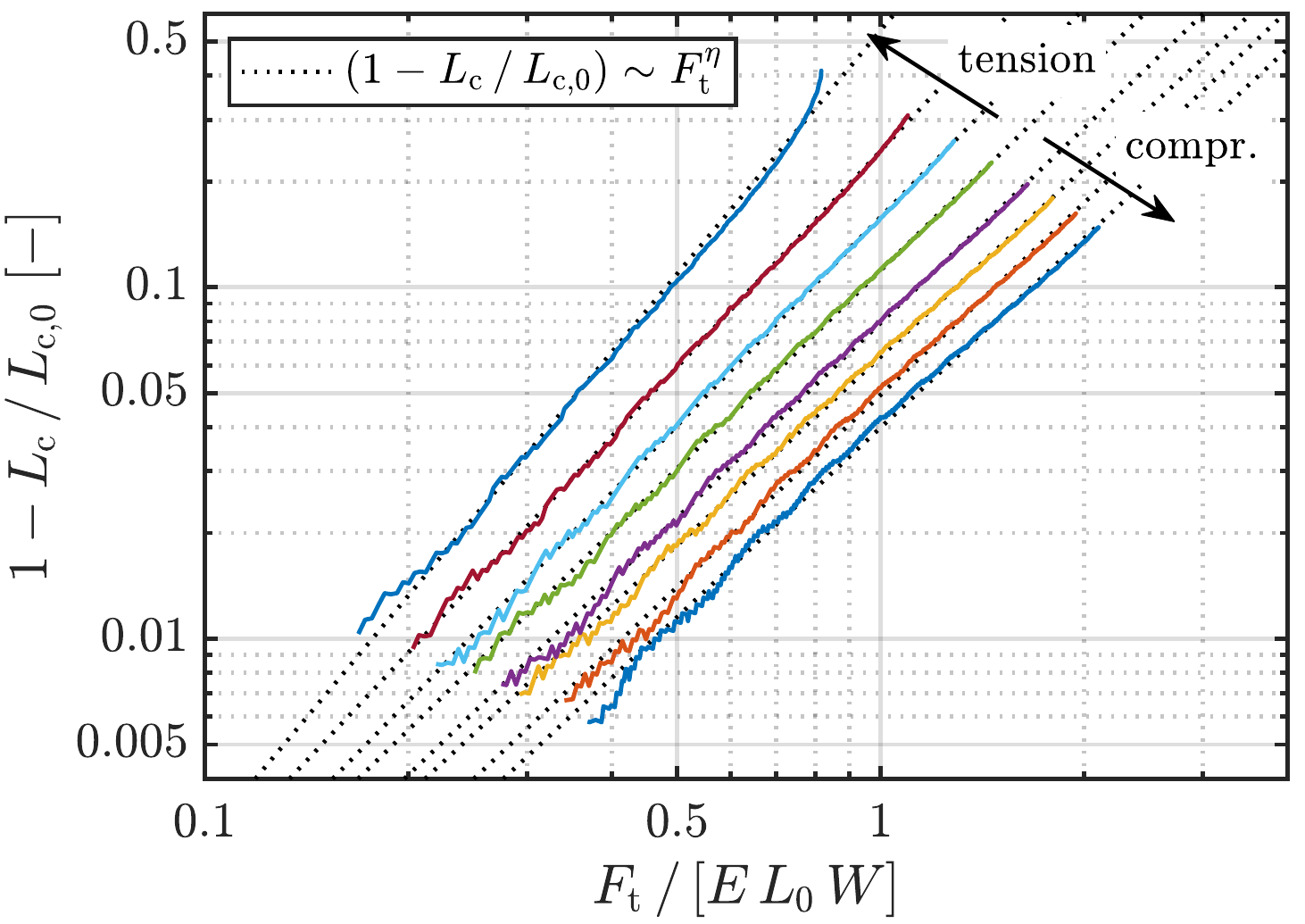}
		\label{f:cap:LogLcOverFt}
	}\subfigure[Parameter~$\eta$ from Fig.~(a) vs.~normal force.]{
		\includegraphics[width=0.47\textwidth]{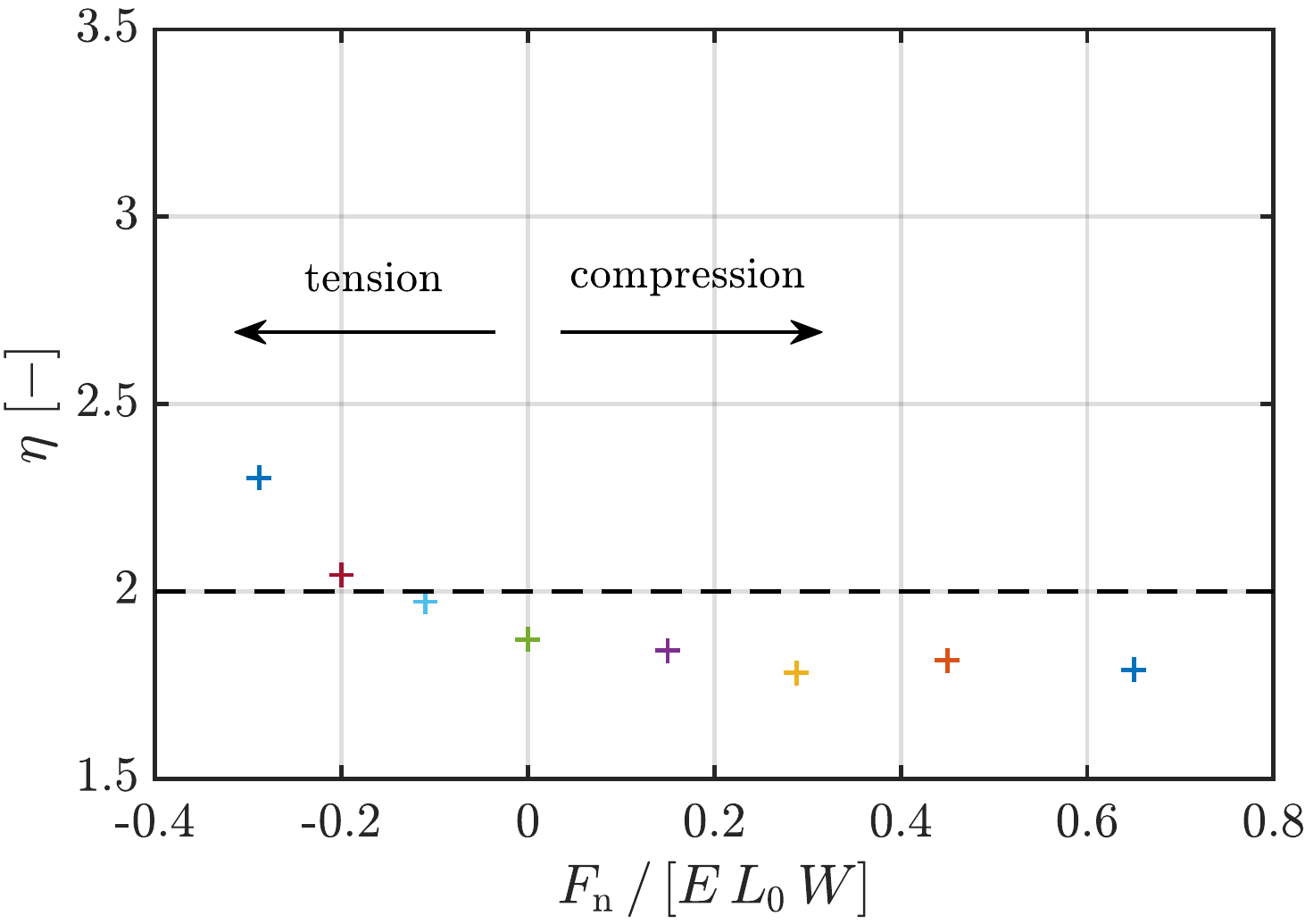}
		\label{f:cap:LogEtaOverFn}
	}
	\caption{Adhesive friction of a 2D cap: (a)~Logarithmic representation of
		the contact length in dependence of the tangential force as considered
		in \cite{papangelo19b}; the figure shows the adhesive cases from
		\cref{f:cap} with stable (non-vanishing) contact; (b)~parameter~$\eta$
		(fitted from~$(1-\Lc\,/\,\Lcz) \sim \Ft^\eta$, and illustrated in
		Fig.~(a) as dotted lines) versus the normal force.}
		\label{f:cap2}
\end{figure}

Motivated by this good agreement, we investigate another finding by
\cite{papangelo19b}: The authors conclude that the contact area does not always
decrease as $\Ft^2$, but rather as $\Ft^\eta$, where $\eta$ is an exponent
depending also on the normal load.~$\eta$ is found to be close to~2 only for
sufficiently large normal loads (such as those applied in \cite{sahli18,
sahli19}), while increasing with smaller loads. To test whether our simulations
produce a similar behavior, in \cref{f:cap:LogLcOverFt} we plot the evolution
of $(1 - \Lc\,/\,\Lcz)$ as a function of $\Ft$. We here consider the adhesive
cases from \cref{f:cap}, except those for which the contact area vanishes
before full sliding. For each normal force, the curves are linear in
logarithmic scales, indicating that indeed $(1 - \Lc\,/\,\Lcz) \sim \Ft^\eta$.
The least-square fitting of~$\eta$ to the numerical results is shown as dashed
lines in \cref{f:cap:LogLcOverFt}.

\Cref{f:cap:LogEtaOverFn} shows the evolution of the fitted exponent~$\eta$ as
a function of the normal force~$\Fn$. Although~$\eta$ monotonously decreases
with increasing~$\Fn$, it is close to~2 for all considered normal loads,
explaining the proper match of the quadratic fits in \cref{f:cap:LcOverFt}.
Overall, both plots in \cref{f:cap2} are in good qualitative agreement with
those of Fig.~10 in \cite{papangelo19b}.

To summarize, even though we have different contact dimensions here, our 2D
simulations agree very well with both experiments \citep{sahli18, sahli19} and
theoretical models \citep{papangelo19, papangelo19b} for adhesive friction of
smooth elastomer spheres and glass plates.


\subsection{Compressible vs.~nearly incompressible material} \label{s:cap:compr}

Since most experiments on the shear-induced reduction of the contact area
include elastomers \citep{savkoor77, waters10, sahli18, mergel19jadh}, all
simulations from \cref{f:cap,f:cap2} are based on nearly incompressible
material. To test whether our findings also extend to compressible materials,
we repeat the simulations discussed in the previous section with $\nu = 0.4$
(instead of $0.49$). These new results are shown in the supplementary material
as Fig.~S1. Below, we just summarize both common points and differences with
respect to the nearly incompressible case.

Overall, the qualitative behavior does not change, meaning that the main
features of the onset of sliding of Hertzian contact is not specific to
incompressible materials. Only small quantitative differences can be observed.
In the compressible case: (i)~both the initial contact length and the final
tangential forces are slightly larger; (ii)~the exponent $\beta$ is $-3.54$
rather than $-3.71$; and (iii) exponent~$\eta$ approaches slightly smaller
values, so that the quadratic fits of the $\Lc$/$\Ft$ curves are slightly less
accurate.


\subsection{Adhesive vs.~non-adhesive friction} \label{s:cap:nonadh}

In the linear elastic, fracture-based models of \cite{papangelo19} and
\cite{papangelo19b}, the presence of adhesion is a necessary condition for a
reduction of the contact area. This could suggest that the same was true also
for our simulations in \cref{s:cap:adh}, which include finite deformations and
adhesion/friction described explicitly rather than being lumped in a
phenomenological mode-mixity function. With our numerical model, we are able to
test this hypothesis by considering five additional cases without adhesion. To
this end, we simply set the normal traction~$\Tn$ in \cref{e:Tn} to zero for
$\gn > \gequ$. The normal forces are chosen such that the initial contact
lengths are close to five adhesive cases (while the latter require a
considerably smaller normal load).

As the five gray curves in \cref{f:cap:LcOverFt} show, the contact length
decreases under tangential shear even in the absence of adhesion. This
observation alone implies that, in contrast to the prevailing view in the
literature, adhesion is not necessary to generate a shear-induced decrease of
the contact length. The curves still seem to be well-fitted by the quadratic
decay~\labelcref{e:Lcfct}, but the decrease in~$\Lc$ is lower than for the
adhesive cases. As a consequence, the reduction parameter~$\xi$ also becomes
smaller for the adhesionless cases (\cref{f:cap:xiOverLc0}), indicating that
adhesion enhances the amount of area reduction.

Assuming that~$\xi$ still follows a power law of the form $\xi(\Lcz) \sim
L_{\mrc,0}^{\beta}$, its exponent is found to be $\beta = -2.42$ (instead of
$-3.71$ with adhesion). This difference suggests that adhesion influences the
area reduction especially for small initial contact lengths or, equivalently,
for small normal loads. The enhancement due to adhesion, expressed by the
factor $(\xi_\mathrm{adh} - \xi_\mathrm{non-adh}) / \xi_\mathrm{non-adh}$, is
e.g.~$+163\,\%$ for $\Lcz \approx 8.5\,L_0$, but only $+24\,\%$ for $\Lcz
\approx 15\,L_0$ (\cref{f:cap:xiOverLc0}). For this specific setup, we thus
expect adhesion to play a negligible role for an initial contact wider than
approximately~$17\,L_0$. Finding a general prediction for~$\Lcz$ beyond which
adhesion is not relevant anymore is an interesting question left for future
studies.

If we (in theory) considered contact between an incompressible, linearly
elastic material and a rigid half-space, we would expect the normal and
tangential displacements in the bulk to remain uncoupled~\citep{johnson85}. In
that case, the tangential shear would not have an influence on the contact
length. From this we can conclude that in the adhesionless case, the reduction
of the contact length is expected to mainly result from the nonlinear
deformation of the cap. This assumption has recently been confirmed in 3D
simulations for adhesionless friction between a hyperelastic sphere and a rigid
plane under high normal loads~\citep{lengiewicz20}. That paper identifies
various elementary mechanisms that all contribute to changes in the contact
area. Investigating those mechanisms with our model, and comparing their
influence with the 3D~case, is again a promising topic for future work, but
beyond the scope of this paper.


\section{Conclusion} \label{s:concl}

This work provides a computational framework to incorporate theoretical models
for local adhesion and friction into 3D, large-deformation simulations. These
models include the two theoretical continuum models~\DI\ and \EA, which were
recently proposed by \cite{mergel19jadh}. Such models are applicable to natural
and technical systems, in which friction is either considerably influenced or
purely dominated by adhesive effects. Up to a certain distance of the adhering
surfaces, both models~\DI\ and \EA\ capture friction even for zero or tensile
contact tractions. This distinguishes them from existing approaches, and is
motivated by soft bio-adhesive pads that are able to generate friction even
under tensile normal loads.

While \cite{mergel19jadh} contains the motivation of the models~\DI\ and \EA,
e.g. in terms of new experimental results, in the first part of this article we
derive the corresponding model equations necessary for a 3D, large-deformation
finite element (FE) formulation. We then describe the algorithmic treatment of
sticking and sliding friction, using the unbiased friction algorithm by
\cite{sauer15ijnme}. Finally, we state the resulting contact FE forces as well
as their tangent matrices for two deformable solids in 3D.

In the second part, we investigate various application examples to illustrate
the capabilities of our simulation framework, as well as the physical
properties of models~\DI\ and \EA. The considered systems include flexible
tapes with a negligible bending stiffness, for which sliding friction cannot be
modeled with existing approaches like cohesive zone or other friction models.
Overall, although qualitative, our results provide interesting perspectives for
future studies, e.g.~in the field of wear, or for the design of adhesives. It
would also be promising to apply our peeling strip model to horizontal peeling,
as reported experimentally for elastomers e.g.~in \cite{ponce15}. Besides, our
models are suitable to computationally investigate peeling of gecko spatulae
\citep{mergelPhD, gouravaraju20a, gouravaraju20b}, or ``friction hairs'' at the
feet of insects \citep{mergelPhD}. They can be further combined with a
geometrically exact beam formulation \citep{sauer14finel}, as done in
\cite{mergelPhD}.

In the third part of this article, we finally study in more detail an example
that is inspired by the onset of sliding of Hertzian elastomer--glass contact.
To this end, we investigate friction of a soft cap and a rigid plate to compare
our results both with experimental findings \citep{sahli18, sahli19} and
theoretical investigations \citep{chen08b, papangelo19, papangelo19b}. As these
results demonstrate, our computational model is able to capture remarkably well
the qualitative behavior of smooth elastomer--glass interfaces. Our findings
suggest that the shear-induced area reduction of such interfaces is not
specific to (nearly) incompressible materials. They also suggest that adhesion
is not a necessary ingredient to model this phenomenon, but enhances an effect
due to finite, non-linear material deformations. These results contribute to
the current debate on the shear-induced area reduction in elastomer contact
\citep{sahli18, sahli19, menga18, menga19, mergel19jadh, papangelo19,
papangelo19b, khajehsalehani19, scheibert20, mcmeeking20, wang20, lengiewicz20}.


\appendix


\section{Tangent matrices for adhesive and frictional contact} \label{a:kc}

This section summarizes the tangent matrices that are required for the
linearization of the contact terms in the governing equations
(\cref{s:comput:FE}). For details on the derivation of these matrices we refer
to \cite{sauer13cmame, sauer15ijnme} and \cite{mergelPhD}.

As becomes apparent in \cref{f:contkin}, the normal gap~$\gn$, and hence also
the elemental contact force $\fck^e = \fnk^e - \ftk^e$, depend on the surfaces
of both interacting bodies, $k,\,\ell = 1,\,2$, $k \neq \ell$. For
linearization we thus need the following two tangent matrices
\begin{equation}
	\kckk^e = \pa{\fck^e}{\uke}, \qquad \kckl^e = \pa{\fck^e}{\ule},
	\label{e:kckke}
\end{equation}
where the vector $\uke$ contains the nodal displacements for element
$\Gamc{k}$, and $\ule$ is the displacement vector of all elements $\Gamc{\ell}$
that are affected by $\Gamc{k}$. In analogy to \cref{e:fc}, the tangent
matrices in \cref{e:kckke} can be decomposed into the contributions $\kckk^e =
\knkk^e - \ktkk^e$ and $\kckl^e = \knkl^e - \ktkl^e$.


\subsection{Tangent matrices for normal (adhesive and repulsive) contact}

For normal contact (\cref{s:model:adh}) with $\theta_k \equiv 1$ and $J_\ell
\approx \Jcl$, the first matrix is given by
\begin{equation}
	\knkk^e = - \int_{\Gamc{\zk}} \Nk^\mrT \, \pa{\bTnk}{\xk} \, \Nk~\dA_k,
	\label{e:knkke}
\end{equation}
where
\begin{equation}
	\pa{\bTnk}{\xk} = \frac{\Tn'}{\Jcle} \, \np \otimes \np + \frac{\Tn}{\Jcle}
	\, \pa{\np}{\xk} - \frac{\Tn}{{(\Jcle)}^2} \, \np \otimes \pa{\Jcle}{\xk},
\end{equation}
and
\begin{align}
	\Tn'			& = \pa{\Tn}{\gn} = - \frac{\AH}{2\pi r_0^4}
						\left[{ \frac{1}{5} {\Big(\frac{r_0}{\gn}\Big)}^{10}
						- {\Big(\frac{r_0}{\gn}\Big)}^4}\right],
						\label{e:dTndgn} \\
	\pa{\np}{\xk}	& = \frac{1}{\gn} \left[{ \bone - \np \otimes \np
						- c^\alphbet_\mrp \, \apco{\alpha} \otimes \apco{\beta}
						}\right]. \label{e:dnpdxk}
\end{align}
The scalar $\Jcle$ is the surface stretch at projection point~$\xp$: $\Jcle =
\left\|{ \apco{1}\times\apco{2}} \right\| / \left\|{ \Apco{1}\times\Apco{2}
}\right\|$ (see also \cref{e:Jcl}). The derivative of $\Jcle$ with respect to
$\xk$ can be computed from
\begin{equation}
	\pa{\Jcle}{\xk} = J_{\mrc\ell,\alpha}^e \, c^\alphbet_\mrp \, \apco{\beta},
	\qquad J_{\mrc\ell,\alpha}^e := \pa{\Jcle}{\xip{}} = \Jcle
	\left[{ \apcon{\beta} \cdot \pa{\apco{\beta}}{\xi^\alpha}
	- \Apcon{\beta} \cdot \pa{\Apco{\beta}}{\xi^\alpha} }\right].
	\label{e:dJcldxk}
\end{equation}
In \cref{e:dnpdxk}, $\bone$ denotes the identity tensor, $\apco{\alpha}$ and
$\np$ are given by \cref{e:apco,e:np}, and $c^\alphbet_\mrp$ are the components
of the matrix
\begin{equation}
	\big[c^\alphbet_\mrp\big] = {\Big[ a_\alphbet^\mrp - \gn \, \big( \np \cdot
	\apco{\alpha,\beta} \big)\Big]}^{-1}, \qquad \apco{\alpha,\beta}
	= \left.{ \pa{\ba_\alpha(\bxi)}{\xi^\beta}
	}\right|_{\boldsymbol{\xi}\,=\,\boldsymbol{\xi}_\mrp}\!;
\end{equation}
see \cref{e:apcontra}. The second tangent matrix, containing the derivatives
with respect to the neighboring nodes~$\ule$, is determined from
\begin{equation}
	\knkl^e = - \int_{\Gamc{\zk}} \Nk^\mrT \, \pa{\bTnk}{\ule}~\dA_k
	\label{e:knkle}
\end{equation}
and
\begin{equation}
	\pa{\bTnk}{\ule} = - \pa{\bTnk}{\xk} \, \mN_\ell - \frac{\Tn}{\Jcle}
	\left[{ c^\alphbet_\mrp \, \apco{\alpha} \otimes \np + \np \otimes
	\apcon{\beta} + \gn \, \frac{J_{\mrc\ell,\alpha}^e}{\Jcle} \,
	c^\alphbet_\mrp \, \np \otimes \np }\right] \mN_{\ell,\beta}.
	\label{e:dTndul}
\end{equation}
In \cref{e:dTndul}, $\mN_{\ell,\beta}$ contains the partial derivatives of the
nodal shape functions~$\mN_\ell$ with respect to coordinate~$\xi^\beta$.


\subsection{Tangent matrices for tangential (sticking and sliding) contact}
\label{a:kc:kt}

For a friction model stated in the \emph{reference} configuration (like
model~\EA), the associated tangent matrices have a form that is very similar to
\cref{e:knkke,e:knkle}:
\begin{equation}
	\ktkk^e = - \int_{\Gamc{\zk}} \Nk^\mrT \, \pa{\bTtk}{\xk} \, \Nk~\dA_k,
	\qquad \ktkl^e = - \int_{\Gamc{\zk}} \Nk^\mrT \, \pa{\bTtk}{\ule}~\dA_k.
	\label{e:ktmat}
\end{equation}
In contrast, for a model in the \emph{current} configuration (like model~\DI),
we must additionally linearize the surface stretch appearing in $\da_k =
\Jcke~\dA_k$; this leads to a second term in~$\ktkk^e$,
\begin{align}
	\ktkk^e	& = - \int_\Gamc{k} \Nk^\mrT \, \pa{\bttk}{\xk} \, \Nk~\da_k
				- \int_\Gamc{k} \Nk^\mrT \, \bttk \otimes \ba_k^\alpha \,
				\mN_{k,\alpha}~\da_k, \\
	\ktkl^e	& = - \int_\Gamc{k} \Nk^\mrT \, \pa{\bttk}{\ule}~\da_k,
				\label{e:ktklspat}
\end{align}
see e.g.~\cite{sauer15ijnme}. The partial derivatives appearing in
\crefrange{e:ktmat}{e:ktklspat} are specified in the following for both
sticking and sliding friction. Unless stated otherwise, all quantities are
evaluated at the current pseudo-time step, $t_{n+1}$; see \cref{s:comput:algo}.
Like in that section, we first discuss the case for which the traction is
defined in the current configuration.


\subsubsection{Sticking friction} \label{a:kc:ktstick}

During sticking, the tangential traction corresponds to the trial
value~$\bttrial$; see \cref{e:ttrial,e:ttnewstick}. The derivatives of this
trial value with respect to both $\xk$ and $\ule$ are given by
\begin{align}
	\pa{\bttrial}{\xk}	& = \epsT \, c^\alphbet_\mrp \, \apco{\alpha} \otimes
							\apco{\beta}, \label{e:dttrialdxk} \\
	\pa{\bttrial}{\ule} & = - \pa{\bttrial}{\xk} \, \mN_\ell(\bxip{})
							+ \epsT \left[{ \mN_\ell(\bxip{})
							- \mN_\ell(\bxis{n}) }\right] + \epsT\,\gn \,
							c^\alphbet_\mrp \, \apco{\alpha} \otimes \np \,
							\mN_{\ell,\beta} (\bxip{}). \label{e:dttrialdule}
\end{align}
Regarding a model in the reference configuration, one obtains the same two
expressions for~$\partial\bTtrial/\partial\xk$
and~$\partial\bTtrial/\partial\ule$, but with a penalty parameter that refers
to the \emph{reference} area, $\dA_k$.


\subsubsection{Sliding friction} \label{a:kc:ktslide}

During sliding, the contact traction satisfies \cref{e:ttslide}. After
introducing
\begin{equation}
	\bp_\mrt = \frac{\ttmax}{\|\bttrial\|} \, \big[ \bone - \nt \otimes
	\nt \big], \qquad \nt = \frac{\bttrial}{\|\bttrial\|},
\end{equation}
and using \cref{e:dttrialdxk,e:dttrialdule}, we obtain
\begin{align}
	\pa{\bttk}{\xk}		& = \bp_\mrt \, \pa{\bttrial}{\xk} + \nt \otimes
							\pa{\ttmax}{\xk}, \label{e:dttdxk} \\
	\pa{\bttk}{\ule}	& = \bp_\mrt \, \pa{\bttrial}{\ule} + \nt \otimes
							\pa{\ttmax}{\ule}.
\end{align}
In analogy, we obtain for the reference configuration
\begin{align}
	\pa{\bTtk}{\xk}		& = \bP_\mrt \, \pa{\bTtrial}{\xk} + \Nt \otimes
							\pa{\Ttmax}{\xk}, \\
	\pa{\bTtk}{\ule}	& = \bP_\mrt \, \pa{\bTtrial}{\ule} + \Nt \otimes
							\pa{\Ttmax}{\ule}, \label{e:dTtdul}
\end{align}
where
\begin{equation}
	\bP_\mrt = \frac{\Ttmax}{\|\bTtrial\|} \, \big[ \bone - \Nt \otimes
	\Nt \big], \qquad \Nt = \frac{\bTtrial}{\|\bTtrial\|}.
\end{equation}
The derivatives in \crefrange{e:dttdxk}{e:dTtdul} are either stated in
\labelcref{a:kc:ktstick}, or they are specified next.


\subsubsection{Partial derivatives of the sliding threshold}

\begin{enumerate}[1)]
\item Model~\DI\ (\cref{s:model:DI}): The partial derivatives of~$\ttmax$ from
\cref{e:tslideDI} are given by
\begin{equation}
	\pa{\ttmax}{\xk} = \ttmax'(\gn) \, \np, \qquad
	\pa{\ttmax}{\ule} = -\ttmax'(\gn) \, \mN_\ell^\mrT \, \np,
\end{equation}
where
\begin{equation}
	\ttmax'(\gn) = \pa{\ttmax}{\gn} = \frac{\tau_\DIe \, k_\DIe}
	{1 + \mre^{\, k_\DIe (\gn - \gcut)}}
	\left[{ \frac{1}{1 + \mre^{\, k_\DIe (\gn - \gcut)}} - 1 }\right].
\end{equation}
\item Model~\EA\ (\cref{s:model:EA}): From \cref{e:TslideEA} follows that
\begin{align}
	\pa{\Ttmax}{\xk}	& = \frac{\mu_\EAe}{\Jcle} \, \Tn'(\gn) \, \np
							- \frac{\Ttmax}{\Jcle} \, \pa{\Jcle}{\xk}, \\
	\pa{\Ttmax}{\ule}	& = - \frac{\mu_\EAe}{\Jcle} \, \Tn'(\gn) \,
							\mN_\ell^\mrT \, \np - \frac{\Ttmax}{\Jcle} \,
							\pa{\Jcle}{\ule}.
\end{align}
The surface stretch~$\Jcle$ and its derivatives are given in
\cref{e:Jcl,e:dJcldxk} and
\begin{equation}
	{\left[{ \pa{\Jcle}{\ule} }\right]}^\mrT = - J_{\mrc\ell,\alpha}^e \,
	c^\alphbet_\mrp \, {\big[\apco{\beta}\big]}^\mrT \, \mN_\ell(\bxip{})
	+ {\left[{ \Jcle \, \apcon{\beta} + \gn \, J_{\mrc\ell,\alpha}^e \,
	c^\alphbet_\mrp \, \np }\right]}^\mrT \, \mN_{\ell,\beta} (\bxip{}).
\end{equation}
\end{enumerate}


\section*{Acknowledgements}

This work was initially supported by the German Research Foundation (DFG) under
grants SA1822/5-1 and GSC111. JS acknowledges support from the Institut Carnot
Ing{\'e}nierie @ Lyon and from LABEX MANUTECH-SISE (ANR-10-LABX-0075) of the
Universit{\'e} de Lyon, within the program ``Investissements d'Avenir''
(ANR-11-IDEX-0007) operated by the French National Research Agency (ANR). We
further thank Dr.~Thang X.~Duong (RWTH Aachen University) and Antonio
Papangelo, Ph.D.~(Politecnico di Bari) for helpful comments.


\renewcommand{\thefigure}{S\arabic{figure}}
\setcounter{figure}{0}

\begin{figure}[p]
	\centering
	\subfigure[Tangential force vs.~horizontal displacement.]{
		\includegraphics[width=0.45\textwidth]{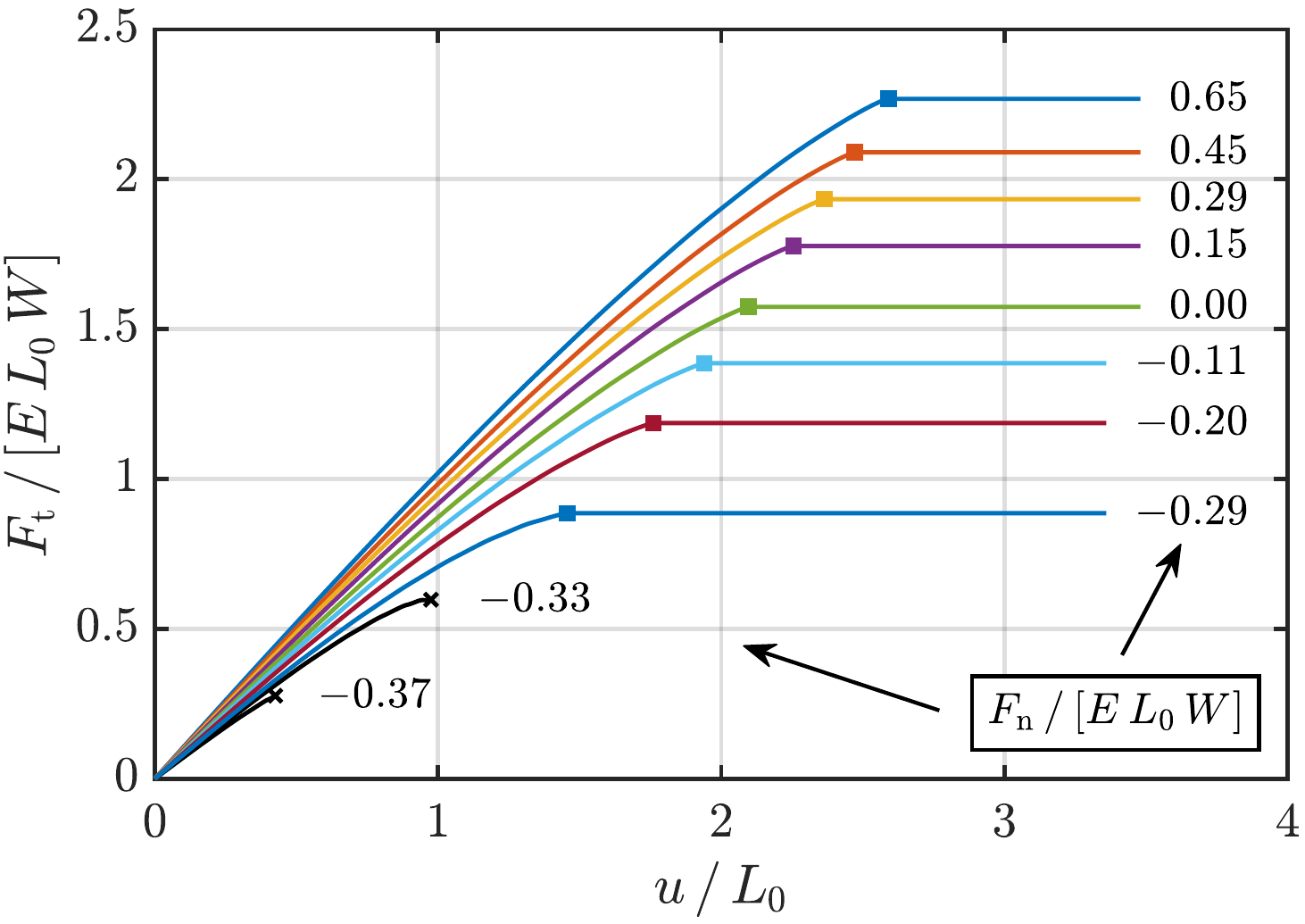}
	}\hspace*{1ex}
	\subfigure[Contact length vs.~horizontal displacement.]{
		\includegraphics[width=0.45\textwidth]{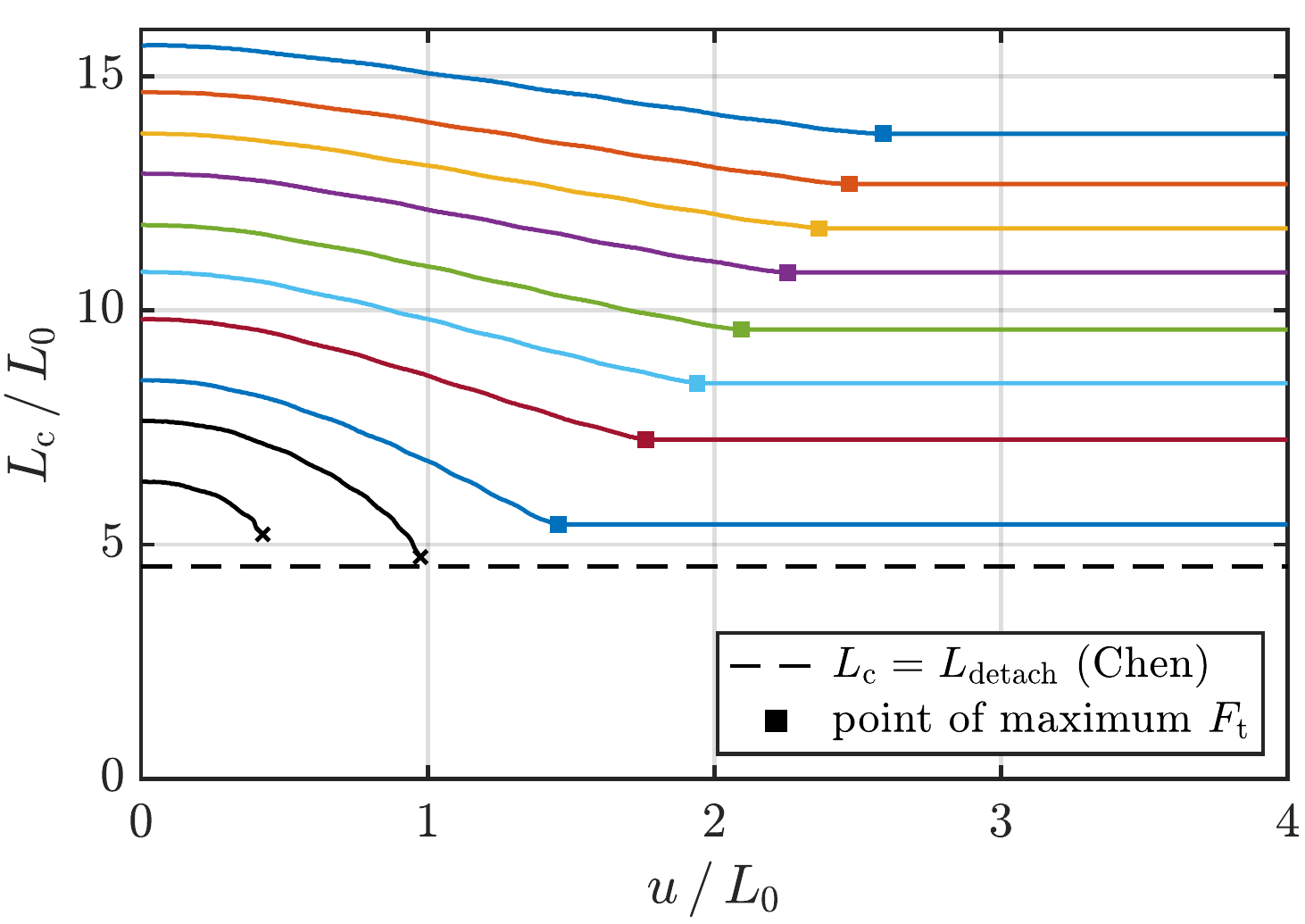}
	}
	\subfigure[Contact length vs.~tangential force.]{
		\includegraphics[width=0.45\textwidth]{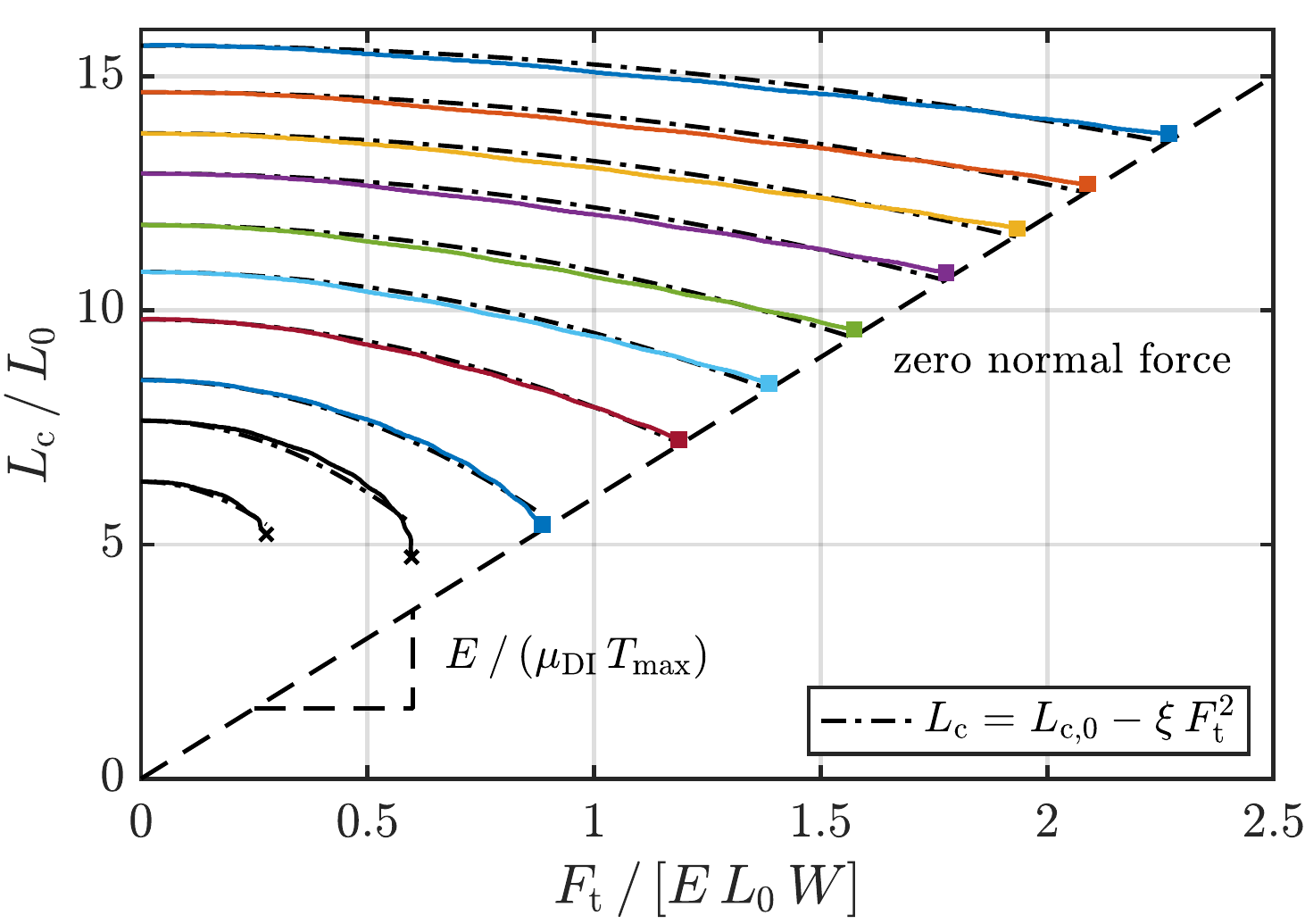}
	}\hspace*{1ex}
	\subfigure[Parameter~$\xi$ vs.~initial contact length.]{
		\includegraphics[width=0.45\textwidth]{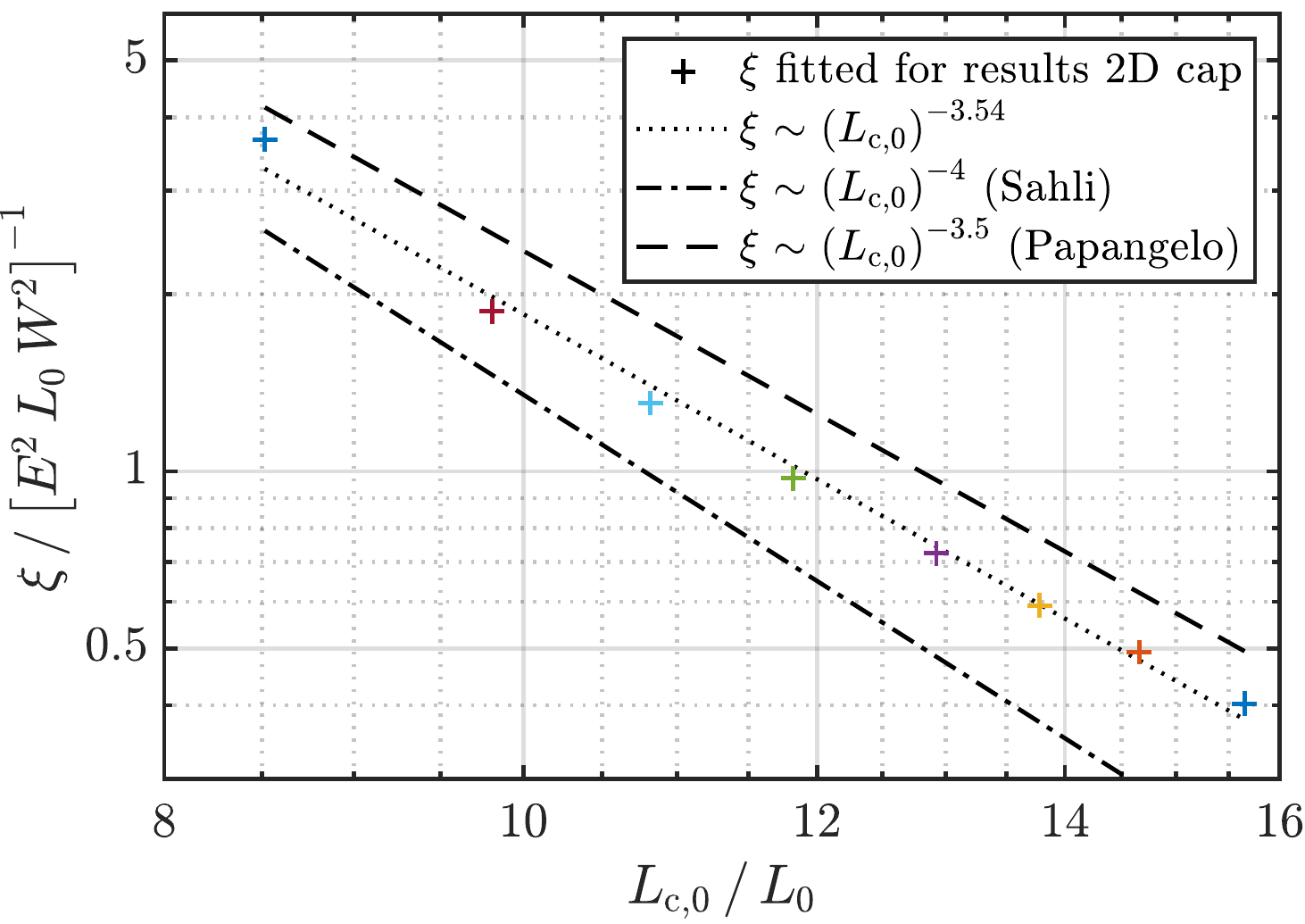}
	}
	\subfigure[Contact length vs.~tangential force (logarithmic).]{
		\includegraphics[width=0.45\textwidth]{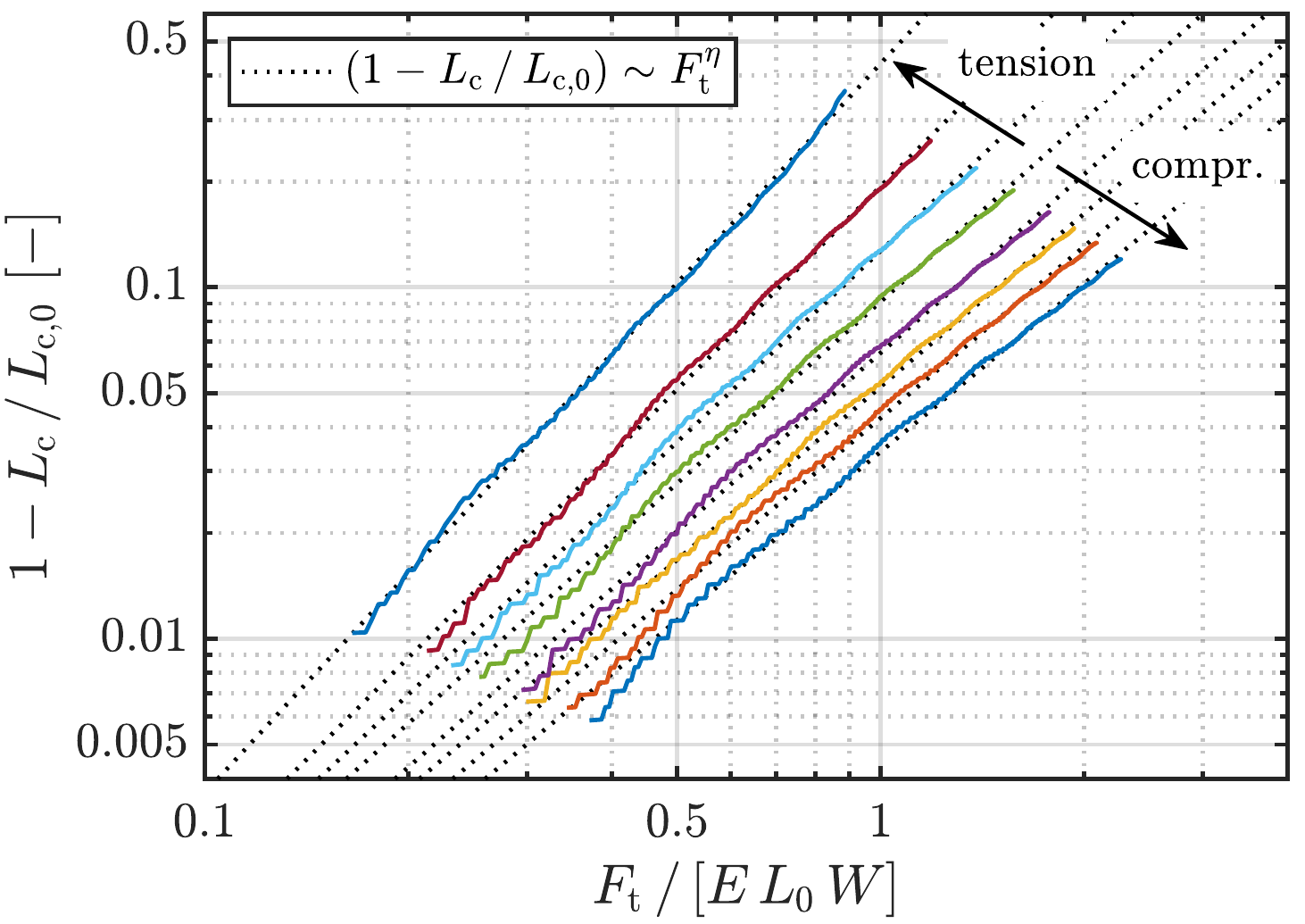}
	}\hspace*{1ex}
	\subfigure[Parameter~$\eta$ from Fig.~(e) vs.~normal force.]{
		\includegraphics[width=0.45\textwidth]{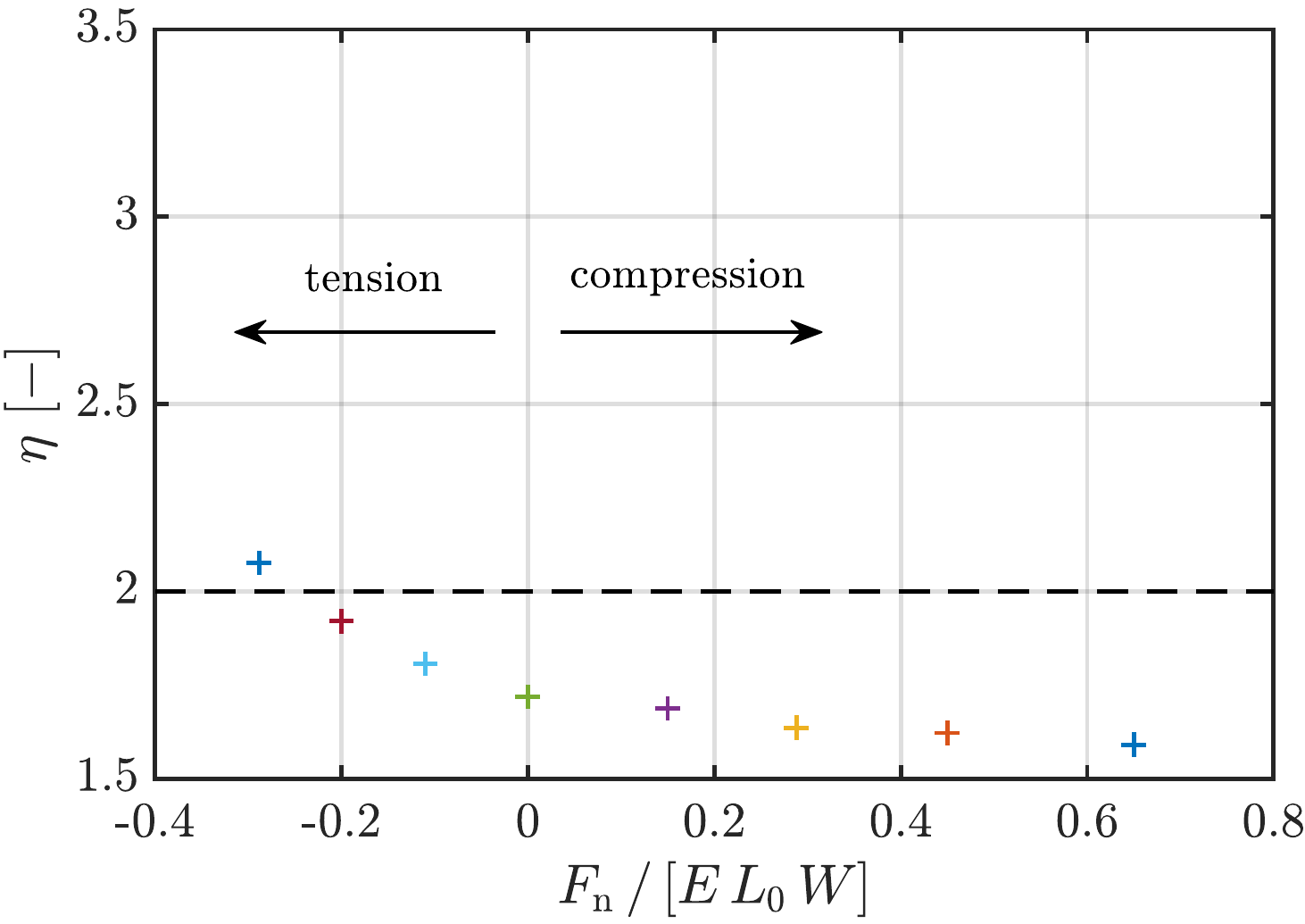}
	}
	\caption{Supplementary material -- adhesive friction of a 2D cap:
		Test cases of Sect.~5.1 (panels analogous to Figs.~13 and~14), but for
		compressible material ($\nu = 0.4$ instead of $\nu = 0.49$).}
\end{figure}

\newpage


\bibliography{JournalsAbbr,Bibliography}


\end{document}